\documentclass[aps,twocolumn,prl,reprint,superscriptaddress,longbibliography]{revtex4-2}

\usepackage{amsmath,amssymb,bm}
\usepackage{graphicx,placeins}
\usepackage{booktabs,xurl,xcolor}
\usepackage{etoolbox}
\usepackage[labelstoglobalaux]{bibunits}
\usepackage[hypertexnames=false]{hyperref}

\newcommand{\Tr}{\operatorname{Tr}}
\newcommand{\Var}{\operatorname{Var}}
\newcommand{\Cov}{\operatorname{Cov}}
\newcommand{\Reop}{\operatorname{Re}}

\newcommand{\ket}[1]{\lvert#1\rangle}
\newcommand{\bra}[1]{\langle#1\rvert}
\newcommand{\norm}[1]{\left\|#1\right\|}

\newcommand{\JQ}{\mathcal J_Q}
\newcommand{\JC}{\mathcal J_C}
\newcommand{\heading}[1]{\emph{#1}---\ignorespaces}
\newcounter{mainthm}
\newenvironment{mainthm}{\par\refstepcounter{mainthm}\textit{Theorem \themainthm---}\ \ignorespaces}{\par}

\newcommand{\diag}{\operatorname{diag}}

\newcommand{\EE}{\mathbb E}

\newtheorem{theorem}{Theorem}

\newtheorem{lemma}{Lemma}

\newenvironment{maintheorem}[1]{\renewcommand{\thetheorem}{#1}\begin{theorem}[main text]}{\end{theorem}}

\edef\suppEmergencyStretch{\the\emergencystretch}
\allowdisplaybreaks[3]
\hypersetup{
	colorlinks = true,
	urlcolor = teal,
	citecolor = teal,
	linkcolor = teal
}

\begin{document}

\makeatletter
\let\savedAddToContents\addtocontents
\long\def\addtocontents#1#2{\ifstrequal{#1}{toc}{}{\savedAddToContents{#1}{#2}}}
\makeatother

\title{Restoring Heisenberg Scaling in Hamiltonian Estimation despite Noise Uncertainty}

\author{Yan Luo}
\affiliation{Institute of Quantum Precision Measurement, State Key Laboratory of Radio Frequency Heterogeneous Integration, College of Physics and Optoelectronic Engineering, Shenzhen University, Shenzhen, China}

\author{Lingna Wang}
\affiliation{Department of Mechanical and Automation Engineering, The Chinese University of Hong Kong, Shatin, Hong Kong SAR, China}

\author{Hongzhen Chen}
\email{hzchen@szu.edu.cn}
\affiliation{Institute of Quantum Precision Measurement, State Key Laboratory of Radio Frequency Heterogeneous Integration, College of Physics and Optoelectronic Engineering, Shenzhen University, Shenzhen, China}

\begin{abstract}
	Noise degrades quantum coherence and limits the precision of quantum sensors. Quantum control can counteract these effects, but existing strategies typically assume that the noise is known. In practice, noise is often only partially characterized, so its parameters must be estimated jointly with the Hamiltonian parameters of interest. Here we show that uncertainty in the noise parameters can destroy otherwise attainable Heisenberg scaling, even when coherence is preserved. For general Markovian dynamics under arbitrary quantum control, we derive a necessary and sufficient condition for each Hamiltonian parameter to retain this scaling and determine its optimal asymptotic precision when the condition holds.
	When the condition fails, we propose a protocol that uses a suitable noisy reference system to restore this scaling, and prove that a reference insensitive to the Hamiltonian parameters but subject to an independent copy of the sensor's noise recovers the optimal precision attainable with known noise.
	Our results reveal how incomplete knowledge of the noise affects the fundamental limits of quantum metrology, with broad implications for achieving quantum-enhanced precision in realistic environments.
\end{abstract}
{\maketitle}

Quantum metrology exploits coherence and entanglement to surpass the standard quantum limit (SQL) and, in ideal settings, achieve Heisenberg scaling, with the estimation variance decreasing as the inverse square of the interrogation time or the number of probes~\cite{Giovannetti2004Measurements,Giovannetti2006Metrology,Giovannetti2011Advances,Degen2017Sensing,Pezze2018Metrology}. However, this quantum advantage is vulnerable to noise, which can destroy coherence and entanglement and degrade the precision scaling~\cite{Huelga1997Frequency,Demkowicz2012Elusive,Chin2012NonMarkovian,Kolodynski2013Efficient,Chaves2013Noisy,Smirne2016NonMarkovian,Kurdzialek2023Adaptiveness,Escher2011Framework}. Quantum control and quantum error correction can suppress the effects of noise while preserving signal sensitivity and, in suitable models, restore Heisenberg scaling~\cite{Demkowicz2017Adaptive,Zhou2018Heisenberg,Gorecki2020Optimal,Duer2014QEC,Arrad2014Sensing,Kessler2014QEC,HerreraMarti2015Magnetometer,Unden2016Repetitive,Kwon2026Autonomous,Sekatski2017FastControl}.

These strategies typically rely on prior knowledge of the noise, as do the corresponding precision limits. In practice, however, noise parameters such as strengths, directions, or correlations are often unknown and must be estimated jointly with the Hamiltonian parameters of interest~\cite{Szczykulska2016Multiparameter,Suzuki2020Nuisance,Kwon2026Unbiased,Crowley2014Tradeoff,Bezerra2025PhaseLoss,Vidrighin2014Joint,Wang2026Bell,Roccia2018Visibility}. Studies in quantum sensing~\cite{Chen2024Informative,Das2025Scalings,Kwon2026VirtualPurification,Brady2026Correlated,Gardner2025Lindblad,Brady2026Multiparameter,Abbasgholinejad2026Function,Tanggara2026Sequential,Gong2026Scrambling} and noise characterization~\cite{Alvarez2011Spectrum,Norris2016Spectroscopy,Omkar2015Characterization,Romanov2026Learning} have addressed aspects of this problem. Yet the precision limits for jointly estimating Hamiltonian and noise parameters under general Markovian dynamics remain unknown.

In this Letter, we first illustrate in a spin model how uncertainty in the noise direction propagates into the frequency estimate, limiting its precision to the SQL even though coherence is preserved. We then prove that this mechanism extends to general Markovian dynamics under arbitrary quantum control and derive a necessary and sufficient condition for each Hamiltonian parameter to retain Heisenberg scaling, together with its optimal asymptotic precision.
Finally, we propose encoding the sensor jointly with a reference system and determine when and how well it restores Heisenberg scaling. In particular, a reference insensitive to the Hamiltonian parameters but subject to an independent copy of the sensor's noise recovers the optimal precision attainable with known noise.

Consider the estimation of parameters $\bm\theta=(\theta_1,\theta_2,\ldots)$ encoded in a quantum state $\rho_{\bm\theta}$. A measurement $\{M_y\}$ gives outcome probabilities $p(y|\bm\theta)=\Tr(\rho_{\bm\theta}M_y)$. For locally unbiased estimators $\hat{\bm\theta}$ constructed from $\nu$ independent repetitions, the covariance matrix $\Cov(\hat{\bm\theta})$ obeys the classical and quantum Cram\'er--Rao bounds~\cite{Braunstein1994Statistical,Liu2020QFIM}
\begin{equation}\label{eq:qcrb}
	\Cov(\hat{\bm\theta})\succeq\frac{1}{\nu}\JC^{-1}\succeq\frac{1}{\nu}\JQ^{-1}.
\end{equation}
Here $[\JC]_{jk}=\sum_y\partial_jp(y|\bm\theta)\,\partial_kp(y|\bm\theta)/p(y|\bm\theta)$ is the classical Fisher information matrix (CFIM), with $\partial_j=\partial/\partial\theta_j$, and $[\JQ]_{jk}=\Reop\Tr(\rho_{\bm\theta}L_jL_k)$ is the quantum Fisher information matrix (QFIM), where the symmetric logarithmic derivatives (SLDs) $L_j$ satisfy $\partial_j\rho_{\bm\theta}=(\rho_{\bm\theta}L_j+L_j\rho_{\bm\theta})/2$. For a single parameter, the QFIM reduces to the quantum Fisher information (QFI) $J_Q$, which a measurement in the eigenbasis of the SLD attains locally. For each parameter $\theta_i$, Eq.~\eqref{eq:qcrb} gives
\begin{equation}
	\Var(\hat\theta_i)\geq\frac{1}{\nu}[\JQ^{-1}]_{ii}\geq\frac{1}{\nu[\JQ]_{ii}},
	\label{eq:nuisance_bound}
\end{equation}
where the second inequality is strict whenever $\JQ$ has nonzero off-diagonal elements between $\theta_i$ and the other unknown parameters.

We first study a spin-$1/2$ sensor in a magnetic field along the $z$ axis (Fig.~\ref{fig:noise_direction}), with Hamiltonian $H=\omega\sigma_z/2$, where the frequency $\omega$ is the parameter to be estimated. Without noise, the initial superposition $(\ket{0}+\ket{1})/\sqrt{2}$ accumulates a relative phase $\omega t$ over an interrogation time $t$, so the QFI for $\omega$ is $J_Q=t^2$, corresponding to Heisenberg scaling. Noise can destroy this coherence and reduce the QFI. We model transverse magnetic-field fluctuations by the Markovian master equation
\begin{equation}\label{eq:toy}
	\frac{d\rho}{dt}=-i\frac{\omega}{2}\left[\sigma_z,\rho\right]
	+\kappa\left(\sigma_\phi\rho\sigma_\phi-\rho\right).
\end{equation}
Here $\kappa>0$ is the known noise rate, and $\sigma_\phi=\cos\phi\,\sigma_x+\sin\phi\,\sigma_y$ specifies the noise direction. Without control, the noise limits the QFI for $\omega$ to $J_Q\leq t/\kappa$, corresponding to SQL scaling~\cite{Albarelli2018Monitoring,SupplementalMaterial}.

Quantum error correction (QEC) can restore Heisenberg scaling by correcting spin flips while preserving the accumulated phase~\cite{Kessler2014QEC,Arrad2014Sensing}. With the sensor as the first qubit and a noiseless ancilla as the second, we use the code $\ket{C_0}=\ket{00}$, $\ket{C_1}=\ket{11}$ and prepare $(\ket{C_0}+\ket{C_1})/\sqrt{2}$. A parity measurement detects each spin flip, after which the fixed recovery $\sigma_x$ applies the net rotation $\sigma_x\sigma_\phi=e^{i\phi\sigma_z}$ to the sensor, giving the two codewords opposite phases $\pm\phi$. In the rapid-correction limit, the number $N$ of detected errors is Poisson distributed with mean $\kappa t$~\cite{Dalibard1992WaveFunction}. Recording $N$ in a classical register gives the output state $\rho^{\mathrm{out}}=\sum_N p_N\ket{N}\bra{N}\otimes\ket{\psi_N}\bra{\psi_N}$, with
\begin{equation}
	\ket{\psi_N}=\frac{1}{\sqrt2}\bigl(\ket{C_0}+e^{i\Phi_N}\ket{C_1}\bigr),\quad \Phi_N=\omega t-2N\phi,
	\label{eq:toystate}
\end{equation}
where the $p_N$ are independent of $\omega$ and $\phi$. For each recorded $N$, the logical state thus remains coherent. 

For a known noise direction $\phi_0$, compensating the phase $-2N\phi_0$ with the recorded count restores the noiseless logical state, whose QFI $t^2$ yields $\Var(\hat\omega)=1/(\nu t^2)$.

\begin{figure}[t]
	\centering
	\includegraphics[width=0.95\columnwidth]{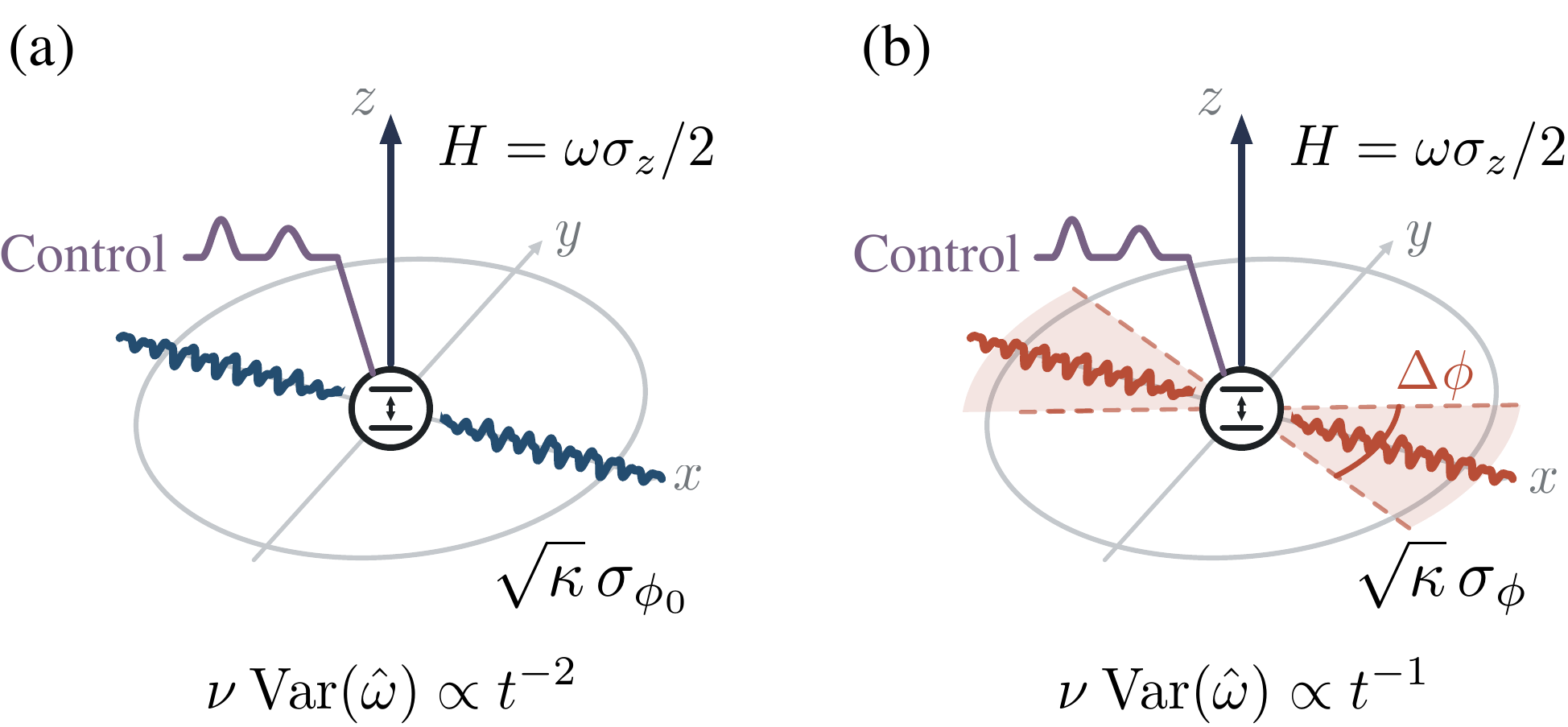}
	\caption{(a) Estimation of $\omega$ with known noise direction $\phi_0$. (b) Joint estimation of $\omega$ and $\phi$. Shading schematically indicates the standard deviation $\Delta\phi$ of the estimated noise direction. The labels give the optimal long-time variance scalings under arbitrary quantum control.}
	\label{fig:noise_direction}
\end{figure}

We now consider joint estimation of $\omega$ and an unknown noise direction $\phi$, for which the error phase $-2N\phi$ can no longer be compensated. The QFI of $\ket{\psi_N}$ with respect to $\Phi_N$ is 1, so by the chain rule its QFIM with respect to $(\omega,\phi)$ is $\nabla\Phi_N\nabla\Phi_N^{\mathsf T}$, where $\nabla\Phi_N=(\partial_\omega\Phi_N,\partial_\phi\Phi_N)^{\mathsf T}=(t,-2N)^{\mathsf T}$. Since the counts are recorded and $p_N$ is independent of the parameters, the QFIM per interrogation is the average of $\nabla\Phi_N\nabla\Phi_N^{\mathsf T}$ over $N$,
\begin{equation}
	\JQ=\left\langle\begin{pmatrix}t^2&-2tN\\-2tN&4N^2\end{pmatrix}\right\rangle_N=\begin{pmatrix}t^2&-2\kappa t^2\\-2\kappa t^2&4\kappa t(1+\kappa t)\end{pmatrix}.
	\label{eq:toyqfim}
\end{equation}
The off-diagonal elements do not vanish: because $\langle N\rangle=\kappa t$, errors shift the mean phase $\langle\Phi_N\rangle=(\omega-2\kappa\phi)t$ exactly as a frequency offset $-2\kappa\phi$ would. Only the count fluctuations, with variance $\kappa t$, resolve this ambiguity between $\omega$ and $\phi$, so that
\begin{equation}
	\Var(\hat\omega)\geq\frac{1}{\nu}[\JQ^{-1}]_{\omega\omega}=\frac{1}{\nu t^2}+\frac{\kappa}{\nu t},
	\label{eq:toyvariance}
\end{equation}
and $\Var(\hat\phi)\geq[\JQ^{-1}]_{\phi\phi}/\nu=1/(4\nu\kappa t)$~\cite{Tanggara2026Sequential,SupplementalMaterial}. Measuring the logical phase in a basis that depends on the recorded $N$ attains both bounds. The second term in Eq.~\eqref{eq:toyvariance}, from uncertainty in the noise direction, dominates at long times, so both variances scale as $1/(\nu t)$ even though coherence is preserved. Calibrating $\phi$ first, with the same total time, does not improve this term~\cite{SupplementalMaterial}. Is the SQL scaling in Eq.~\eqref{eq:toyvariance} unavoidable under arbitrary quantum control, and can Heisenberg scaling be restored?

\heading{General setting}
We consider a quantum sensor evolving according to the Markovian master equation~\cite{Lindblad1976Generators,Gorini1976Semigroups}
\begin{equation}\label{eq:lindblad}
	\frac{d\rho}{dt}=-i[H(\bm\omega),\rho]+\sum_k\mathcal D[E_k(\bm\eta)]\rho,
\end{equation}
where $\mathcal D[E]\rho=E\rho E^\dagger-\{E^\dagger E,\rho\}/2$ is the Lindblad dissipator. Since the noise parameters $\bm\eta=(\eta_1,\ldots,\eta_q)$ are unknown, we estimate them jointly with the Hamiltonian parameters $\bm\omega=(\omega_1,\ldots,\omega_r)$, collecting all parameters in $\bm\theta=(\bm\omega,\bm\eta)$. In the general sequential scheme, each of $\nu$ independent interrogations of duration $t$ allows arbitrary parameter-independent state preparation, control, measurement, and feedback, assisted by noiseless ancillas~\cite{SupplementalMaterial}.

Let $\JQ(t)$ denote the QFIM of the output state of an interrogation of duration $t$. For a single target parameter $\theta_i$, we keep $\theta_i$ unchanged and reparameterize the other unknown parameters as $\theta'_j=\theta_j-v_j\theta_i$ for $j\ne i$~\cite{Gross2021OneFromMany}. With $v_i=1$, the diagonal QFIM entry for $\theta_i$ in the new coordinates is $\bm v^{\mathsf T}\JQ(t)\bm v$. As shown in End Matter, minimizing this entry over $\bm v$ and using Eq.~\eqref{eq:qcrb} gives
\begin{equation}
	\Var(\hat\theta_i)\geq\frac{1}{\nu}[\JQ^{-1}(t)]_{ii}=\frac{1}{\nu\min_{\bm v:v_i=1}\bm v^{\mathsf T}\JQ(t)\bm v}.
	\label{eq:joint_direction_variance}
\end{equation}
The minimum is the information about $\theta_i$ that remains after accounting for the other unknown parameters. This bound is locally attainable for each parameter~\cite{Suzuki2020Nuisance}, whereas the full matrix bound $\Cov(\hat{\bm\theta})\succeq\JQ^{-1}/\nu$ need not be attainable with a common measurement~\cite{Holevo2011Probabilistic,Matsumoto2002PureState,Demkowicz2020Multiparameter,Chen2022Hierarchical,Wang2026MinimalTradeoff,Ragy2016Compatibility}. Minimizing the attainable variance over probe states, controls, and measurements defines the optimal variance $\Var_{\mathrm{opt}}(\hat\theta_i)$.

Our first result determines when Hamiltonian parameters can retain Heisenberg scaling while the noise parameters are estimated jointly. For a single Hamiltonian parameter $\omega_a$ with known noise, Heisenberg scaling requires the Hamiltonian derivative $\partial_{\omega_a}H$ to lie outside the Lindblad span $\mathcal S=\operatorname{span}\{I,E_k,E_k^\dagger,E_k^\dagger E_l\}$, so that the noise can be suppressed without erasing the information about $\omega_a$~\cite{Demkowicz2017Adaptive,Zhou2018Heisenberg}. Joint estimation further requires distinguishing $\partial_{\omega_a}H$ from the coherent contributions of the other unknown parameters. These are the derivatives $\partial_{\omega_{a'}}H$ for the other Hamiltonian parameters and, for the noise parameters $\eta_b$, the operators~\cite{Chen2024Informative,Das2025Scalings}
\begin{equation}\label{eq:Hhat}
	C_{\eta_b}=\frac i2\sum_k\left[E_k^\dagger\partial_{\eta_b}E_k-\left(\partial_{\eta_b}E_k\right)^\dagger E_k\right].
\end{equation}
For $\omega_a$, the relevant span is therefore
\begin{equation}
	\mathcal N_a=\mathcal S+\operatorname{span}\bigl(\{\partial_{\omega_{a'}}H\}_{a'\ne a}\cup\{C_{\eta_b}\}_b\bigr).
	\label{eq:noise_span}
\end{equation}

\begin{mainthm}\label{thm:joint_scaling}
	In the general sequential scheme with unknown noise parameters, a locally unbiased estimator of $\omega_a$ can attain Heisenberg scaling, $\Var(\hat\omega_a)=O(1/(\nu t^2))$, if and only if
	\begin{equation}\label{eq:jointHS}
		\partial_{\omega_a}H\notin\mathcal N_a.
	\end{equation}
	Otherwise, the inverse variance grows at most linearly with the interrogation time, limiting the precision to SQL scaling. If Eq.~\eqref{eq:jointHS} holds, the optimal variance satisfies
	\begin{equation}
		\lim_{t\to\infty}\nu t^2\Var_{\mathrm{opt}}(\hat\omega_a)=\frac{1}{4d_a^2},
		\label{eq:scalaropt}
	\end{equation}
	where the operator-norm distance from $\partial_{\omega_a}H$ to $\mathcal N_a$ is
	\begin{equation}
		d_a=\min_{K=K^\dagger\in\mathcal N_a}\norm{\partial_{\omega_a}H-K}_\infty.
		\label{eq:di}
	\end{equation}
\end{mainthm}

Here $\norm{X}_\infty$ is the largest absolute eigenvalue of $X$. Equivalently, the optimal quadratic coefficient, i.e., the leading coefficient of $t^2$ in $1/[\nu\Var_{\mathrm{opt}}(\hat\omega_a)]$, is $4d_a^2$. The distance $d_a$ can be computed by solving a semidefinite program~\cite{SupplementalMaterial}.

\heading{Proof sketch}
By Eq.~\eqref{eq:joint_direction_variance}, Heisenberg scaling for $\omega_a$ requires the QFI $\bm v^{\mathsf T}\JQ(t)\bm v$ to grow quadratically in $t$ for every reparameterization $\bm v$. This is the QFI for $\omega_a$ in the new coordinates $\theta'_j=\theta_j-v_j\omega_a$: a change $\delta\omega_a$ at fixed $\theta'_j$ shifts all parameters by $\delta\bm\theta=\bm v\,\delta\omega_a$. We show in End Matter that the effect of this shift on the dynamics has a coherent part and a dissipative part. The coherent part is generated by $\partial_{\omega_a}H-K_a$, where
\begin{equation}
	K_a(\bm v,S)=S-\sum_{a'\ne a}v_{\omega_{a'}}\partial_{\omega_{a'}}H-\sum_b v_{\eta_b}C_{\eta_b}.
	\label{eq:target_K_definition}
\end{equation}
The sums are the coherent contributions of the shifts of the other parameters. The Hermitian operator $S\in\mathcal S$ can be chosen freely through the Kraus representation, reflecting that components in the Lindblad span cannot be separated from the noise. As $\bm v$ and $S$ vary, $K_a$ ranges over the Hermitian operators in $\mathcal N_a$. The coherent part can accumulate information quadratically in $t$, whereas the dissipative part, described by a positive semidefinite operator $A_a$ defined through Eq.~\eqref{eq:joint_jump_em} in End Matter, adds information only linearly. Under arbitrary quantum control, together they bound the QFI along $\bm v$~\cite{SupplementalMaterial},
\begin{equation}
	\bm v^{\mathsf T}\JQ(t)\bm v\leq4\left[t\norm{\partial_{\omega_a}H-K_a}_\infty+\sqrt{t\norm{A_a}_\infty}\right]^2.
	\label{eq:target_purification_bound}
\end{equation}

If $\partial_{\omega_a}H\in\mathcal N_a$, some choice of $\bm v$ and $S$ gives $K_a=\partial_{\omega_a}H$, so the coherent part vanishes: the shifts of the other parameters cancel the coherent effect of $\omega_a$ up to terms in the Lindblad span. For this $\bm v$, Eq.~\eqref{eq:target_purification_bound} reduces to $\bm v^{\mathsf T}\JQ(t)\bm v\leq4t\norm{A_a}_\infty$, and Eq.~\eqref{eq:joint_direction_variance} gives $\Var(\hat\omega_a)\geq1/(4\nu t\norm{A_a}_\infty)$, ruling out Heisenberg scaling.

If $\partial_{\omega_a}H\notin\mathcal N_a$, the coherent part cannot be cancelled. Its smallest operator norm over $\bm v$ and $S$ is the distance $d_a$ in Eq.~\eqref{eq:di}. Inserting the minimizing $\bm v$ and $S$ into Eqs.~\eqref{eq:joint_direction_variance} and~\eqref{eq:target_purification_bound} gives $\liminf_{t\to\infty}\nu t^2\Var(\hat\omega_a)\geq1/(4d_a^2)$.

To attain this bound, we choose two codewords $\ket{C_0}$ and $\ket{C_1}$ with
\begin{equation}
	PKP\propto P\enspace\forall K\in\mathcal N_a,\quad\langle\partial_{\omega_a}H\rangle_0-\langle\partial_{\omega_a}H\rangle_1=2d_a,
	\label{eq:code_conditions}
\end{equation}
where $P=\ket{C_0}\bra{C_0}+\ket{C_1}\bra{C_1}$ is the projector onto the code space, $\langle O\rangle_i=\bra{C_i}O\ket{C_i}$, and $2d_a$ is the largest difference allowed by the first condition. End Matter gives the construction, which follows Refs.~\cite{Zhou2018Heisenberg,Gorecki2020Optimal,Abbasgholinejad2026Function}. We apply the Hamiltonian control $-H(\tilde{\bm\omega})$ together with rapid error correction, where $\tilde{\bm\theta}=(\tilde{\bm\omega},\tilde{\bm\eta})$ denotes the prior estimates. These estimates can be refined adaptively at negligible cost. To first order in $\delta\bm\theta=\bm\theta-\tilde{\bm\theta}$, the logical state $\rho_C$ then evolves as
\begin{equation}\label{eq:logical_evolution}
	\frac{d\rho_C}{dt}=-id_a\,\delta\omega_a[\sigma_z^C,\rho_C],
\end{equation}
where $\sigma_z^C=\ket{C_0}\bra{C_0}-\ket{C_1}\bra{C_1}$. An equal superposition of the codewords thus accumulates a relative phase $2d_at\,\delta\omega_a$, and measuring $\sigma_y^C$ attains the bound. For several Hamiltonian parameters, splitting the interrogations among their optimal codes retains Heisenberg scaling for each parameter that satisfies Eq.~\eqref{eq:jointHS}, although the optimal coefficients need not be attained simultaneously~\cite{Ragy2016Compatibility,Gorecki2020Optimal,Albarelli2022Probe}. \hfill$\square$

In Theorem~\ref{thm:joint_scaling}, the unknown noise parameters enter only through the operators $C_{\eta_b}$ in $\mathcal N_a$. Relative to known noise with the same unknown Hamiltonian parameters, these operators enlarge the span, which can destroy Heisenberg scaling or reduce $d_a$~\cite{SupplementalMaterial}. In contrast, if all $C_{\eta_b}\in\mathcal S$, the span $\mathcal N_a$ and hence each optimal quadratic coefficient are the same as for known noise. This includes unknown rates and correlations of fixed noise operators $E_l^{(0)}$, that is, $E_k=\sum_lM_{lk}(\bm\eta)E_l^{(0)}$ with invertible $M(\bm\eta)$, for which $C_{\eta_b}\in\operatorname{span}\{E_k^\dagger E_l\}\subseteq\mathcal S$.

Theorem~\ref{thm:joint_scaling} extends to noise parameters, with $C_{\eta_b}$ in place of $\partial_{\omega_a}H$ and the Lindblad span enlarged by the Hamiltonian derivatives and the other $C_{\eta_{b'}}$~\cite{SupplementalMaterial}. Rates, with $C_{\eta_b}\in\mathcal{S}$, are thus limited to SQL scaling, consistent with Ref.~\cite{Brady2026Multiparameter}, whereas parameters that rotate rank-deficient noise operators can retain Heisenberg scaling~\cite{Chen2024Informative,Das2025Scalings}. The Supplemental Material also treats parameters entering both $H$ and $E_k$, and bounds the coefficient of $t$ when Heisenberg scaling is lost, a bound attainable for known noise~\cite{Zhou2020Approximate,Abbasgholinejad2026Function}.

Unlike unknown rates and correlations, an unknown noise direction rotates $E$ out of the Lindblad span and can destroy Heisenberg scaling, as in the spin model. There, $E=\sqrt\kappa\,\sigma_\phi$ and $\mathcal S=\operatorname{span}\{I,\sigma_\phi\}$, but $\partial_\phi E=\sqrt\kappa\,\sigma_{\phi+\pi/2}$ lies outside $\mathcal S$ and contributes $C_\phi=-\kappa\sigma_z=-2\kappa\,\partial_\omega H$. Jointly estimating $\phi$ therefore places $\partial_\omega H$ in $\mathcal N_\omega$, so $d_\omega=0$ and Heisenberg scaling is lost. Physically, changing the noise direction $\phi$ has the same coherent effect as shifting the frequency $\omega$, which is why the mean phase $\langle\Phi_N\rangle=(\omega-2\kappa\phi)t$ cannot separate the two. Along $\bm v=(1,1/(2\kappa))$, which leaves $\omega-2\kappa\phi$ unchanged, the coherent part vanishes and the dissipative part reduces to $A_\omega=v_\phi^2(\partial_\phi E)^\dagger\partial_\phi E=I/(4\kappa)$, so Eqs.~\eqref{eq:joint_direction_variance} and~\eqref{eq:target_purification_bound} give, under arbitrary control,
\begin{equation}\label{eq:toyallcontrol}
	\Var(\hat\omega)\geq\frac{\kappa}{\nu t}.
\end{equation}
The QEC protocol above attains this bound and the optimal covariance matrix in the long-time limit~\cite{SupplementalMaterial}.

Since Theorem~\ref{thm:joint_scaling} optimizes over all controls, no strategy acting on the sensor alone can undo this loss. We now show that a noisy reference system can restore the precision lost to unknown noise parameters, including Heisenberg scaling. 

Consider a reference $R$ that evolves alongside the sensor $S$ according to Eq.~\eqref{eq:lindblad}, with its own Hamiltonian $H_R(\bm\omega)$ and Lindblad operators $F_l(\bm\eta)$ acting on $R$. These operators may vanish or depend on only some of the parameters, and $\bm\omega$ and $\bm\eta$ may include unknown parameters specific to $R$. To compare the two systems, we use Eq.~\eqref{eq:target_K_definition} to define the smallest norm of the coherent part along $\bm v$, with $v_{\omega_a}=1$, as
\begin{equation}
	d_a(\bm v)=\min_{S=S^\dagger\in\mathcal{S}}\Big\|\sum_{a'}v_{\omega_{a'}}\partial_{\omega_{a'}}H+\sum_bv_{\eta_b}C_{\eta_b}-S\Big\|_\infty,
	\label{eq:dav}
\end{equation}
so that $d_a=\min_{\bm v}d_a(\bm v)$. We write $d_a^S(\bm v)$ for the sensor and $d_a^R(\bm v)$ for the reference, the latter with $H_R$, $F_l$, and the Lindblad span $\mathcal{S}_R$ of $R$ in place of $H$, $E_k$, and $\mathcal{S}$.

\begin{mainthm}\label{thm:reference_restoration}
	In the general sequential scheme acting jointly on $S$ and $R$, the parameter $\omega_a$ can attain Heisenberg scaling if and only if no $\bm v$ gives $d_a^S(\bm v)=d_a^R(\bm v)=0$. If so, the optimal variance satisfies $\lim_{t\to\infty}\nu t^2\Var_{\mathrm{opt}}(\hat\omega_a)=1/[4(d_a^{SR})^2]$, where $d_a$ for the joint system reduces to
	\begin{equation}
		d_a^{SR}=\min_{\bm v}\big[d_a^S(\bm v)+d_a^R(\bm v)\big].
		\label{eq:dSR}
	\end{equation}
\end{mainthm}

The distance $d_a^{SR}$ can likewise be computed by solving a semidefinite program~\cite{SupplementalMaterial}.

\heading{Proof sketch}
For the joint system, the coherent contributions $\partial_{\omega_{a'}}H$ and $C_{\eta_b}$ in Eq.~\eqref{eq:dav} are sums of those of $S$ and $R$, and the Lindblad span contains $\mathcal{S}\otimes I$ and $I\otimes\mathcal{S}_R$.
By the triangle inequality, the smallest norm of the coherent part of a joint shift along $\bm v$ is at most $d_a^S(\bm v)+d_a^R(\bm v)$, so Eqs.~\eqref{eq:joint_direction_variance} and~\eqref{eq:target_purification_bound} bound the optimal coefficient by $4(d_a^{SR})^2$.

To attain this bound, we encode each system in a code that is required only to correct its own noise: $\ket{C_0}$ and $\ket{C_1}$ satisfy the first condition in Eq.~\eqref{eq:code_conditions} with $\mathcal{S}$ in place of $\mathcal{N}_a$, and $\ket{C_0^R}$ and $\ket{C_1^R}$ satisfy it with $\mathcal{S}_R$.
The joint codewords are $\ket{C_0^{SR}}=\ket{C_0C_1^R}$ and $\ket{C_1^{SR}}=\ket{C_1C_0^R}$, for which $\bra{C_i^{SR}}A\otimes B\ket{C_j^{SR}}=\bra{C_i}A\ket{C_j}\bra{C_{1-i}^R}B\ket{C_{1-j}^R}$.
Each operator spanning the joint Lindblad span has this form, with $A\in\mathcal{S}$ and $B\in\mathcal{S}_R$, so both factors are proportional to $\delta_{ij}$ and the first condition holds. A joint coherent contribution $X_j\otimes I+I\otimes X_j^R$, with $X_j=\partial_{\omega_{a'}}H$ for $\theta_j=\omega_{a'}$, $X_j=C_{\eta_b}$ for $\theta_j=\eta_b$, and $X_j^R$ likewise for $R$, has no off-diagonal elements because the local codewords are orthogonal.
Its diagonal elements are $\langle X_j\rangle_i+\langle X_j^R\rangle_{1-i}^R$, where $\langle\cdot\rangle_i^R$ is the expectation in $\ket{C_i^R}$ and the reference codewords enter in reverse order. By Eq.~\eqref{eq:corrected_local_generator_em} in End Matter, the logical state thus evolves, to first order, as
\begin{equation}
	\frac{d\rho_C}{dt}=-i\sum_j\delta\theta_j\big(y_j^S-y_j^R\big)[\sigma_z^C,\rho_C],
	\label{eq:jointgen}
\end{equation}
where $y_j^S=(\langle X_j\rangle_0-\langle X_j\rangle_1)/2$ and $y_j^R=(\langle X_j^R\rangle_0^R-\langle X_j^R\rangle_1^R)/2$ are the responses of the two codes to $\theta_j$. If $y_j^S=y_j^R$ for all $\theta_j\neq\omega_a$ and $y_{\omega_a}^S-y_{\omega_a}^R=d_a^{SR}$, Eq.~\eqref{eq:jointgen} reduces to Eq.~\eqref{eq:logical_evolution} with $d_a^{SR}$ in place of $d_a$, and the protocol of Theorem~\ref{thm:joint_scaling} attains the bound.
End Matter constructs the codes that meet these conditions by convex duality, with $\bm v$ as the Lagrange multipliers. \hfill$\square$

If $H_R=0$ and $F_l(\bm\eta)=E_l(\bm\eta)$, the contributions $C_{\eta_b}$ of $S$ and $R$ cancel in the triangle inequality, so $d_a^{SR}$ equals the distance for known noise with the same unknown Hamiltonian parameters, $d_a^{\rm known}=\min_{\bm v}d_a(\bm v)$ with all $v_{\eta_b}=0$.
The joint codewords $\ket{C_0C_1}$ and $\ket{C_1C_0}$ built from the optimal code for known noise attain it, since their responses to each $\eta_b$ cancel: such a reference restores the optimal coefficient attainable with known noise.
If $\eta_b$ is the only other unknown parameter and mimics $\omega_a$, that is, $\partial_{\omega_a}H+\lambda_SC_{\eta_b}\in\mathcal{S}$ and $\partial_{\omega_a}H_R+\lambda_RC_{\eta_b}^R\in\mathcal{S}_R$ for real $\lambda_S$ and $\lambda_R$, Eq.~\eqref{eq:dSR} gives
\begin{equation}
	d_a^{SR}=\min\{a_S,a_R\}\,|\lambda_S-\lambda_R|,
	\label{eq:single}
\end{equation}
with $a_S=\min_{S=S^\dagger\in\mathcal{S}}\|C_{\eta_b}-S\|_\infty$ and $a_R=\min_{S=S^\dagger\in\mathcal{S}_R}\|C_{\eta_b}^R-S\|_\infty$.
A unit shift of $\omega_a$ thus has the same coherent effect on the sensor as a shift $-\lambda_S$ of $\eta_b$. The reference distinguishes the two only if their effects on it differ, which requires $\lambda_R\neq\lambda_S$ and a coherent response to $\eta_b$, $a_R>0$. A noiseless reference, with $a_R=0$, therefore never helps, and neither does a copy of the sensor, with $\lambda_R=\lambda_S$.

\begin{figure}[t]
    \centering
    \includegraphics[width=0.95\linewidth]{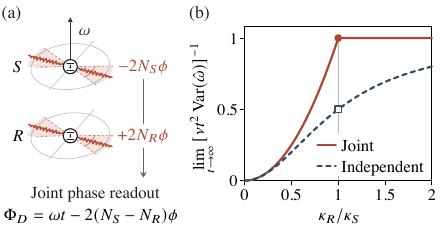}
    \caption{Reference-assisted frequency estimation. (a) Only the sensor $S$ accumulates the signal phase $\omega t$. Corrected flips on $S$ and the reference $R$ shift the joint logical phase in opposite directions. (b) For known flip rates with ratio $r=\kappa_R/\kappa_S$ and a common unknown noise direction, joint encoding in $(\ket{C_0C_1}+\ket{C_1C_0})/\sqrt2$ attains the optimal quadratic coefficient $\min\{1,r^2\}$ (solid), whereas independent encodings in $[(\ket{C_0}+\ket{C_1})/\sqrt2]^{\otimes2}$ give $r^2/(1+r^2)$ (dashed).}
    \label{fig:reference_rates}
\end{figure}

In the spin model, a reference $R$ flipped independently about the same unknown axis at the same rate $\kappa$ as the sensor $S$ thus restores the precision attainable with known noise.
We encode each spin in the code $\ket{C_0}=\ket{00}$, $\ket{C_1}=\ket{11}$ with its own noiseless ancilla, prepare $(\ket{C_0C_1}+\ket{C_1C_0})/\sqrt{2}$, and correct each spin as before.
As in Eq.~\eqref{eq:toystate}, each corrected flip on $S$ shifts the relative phase of the two joint codewords by $-2\phi$. Since $R$ always holds the opposite codeword, each corrected flip on $R$ shifts it by $+2\phi$, as depicted in Fig.~\ref{fig:reference_rates}(a).
Conditional on the independent Poisson counts $N_S$ and $N_R$, with mean $\kappa t$, the logical state is
\begin{equation}\label{eq:reference_spin_phase}
	\ket{\psi_D}=\frac{1}{\sqrt{2}}\big(\ket{C_0C_1}+e^{i\Phi_D}\ket{C_1C_0}\big),
\end{equation}
where $\Phi_D=\omega t-2D\phi$ with $D=N_S-N_R$.
As in Eq.~\eqref{eq:toyqfim}, with $N$ replaced by $D$, which has zero mean and variance $2\kappa t$, the QFIM per interrogation is
\begin{equation}
	\JQ=\left\langle\begin{pmatrix}t^2&-2tD\\-2tD&4D^2\end{pmatrix}\right\rangle_D=\begin{pmatrix}t^2&0\\0&8\kappa t\end{pmatrix}.
	\label{eq:reference_qfim}
\end{equation}
The off-diagonal elements now vanish because $\langle\Phi_D\rangle=\omega t$ no longer depends on $\phi$, so measuring the logical phase in a basis that depends on $D$ estimates $\omega$ with the variance $1/(\nu t^2)$ for known noise, and $\phi$ with half the variance for the sensor alone, $1/(8\nu\kappa t)$~\cite{SupplementalMaterial}.

For different rates $\kappa_S$ and $\kappa_R=r\kappa_S$, Eq.~\eqref{eq:single} with $\lambda_S=1/(2\kappa_S)$, $\lambda_R=0$, and $a_{S,R}=\kappa_{S,R}$ gives the coefficient $4(d_\omega^{SR})^2=\min\{1,r^2\}$, whereas independent encodings give only $r^2/(1+r^2)$~\cite{SupplementalMaterial}, as shown in Fig.~\ref{fig:reference_rates}(b).
To attain $\min\{1,r^2\}$, we expose the faster-flipping spin only for a time $\min\{rt,t/r\}$ and then swap its state into a further noiseless ancilla. Both counts then have the same mean, so $\phi$ again drops out of $\langle\Phi_D\rangle$. If the rates are unknown, we instead expose each spin until its $M$th corrected flip, which gives $D=0$~\cite{SupplementalMaterial}.

In a uniform field, the reference can be a clock transition whose frequency is first-order insensitive to the field~\cite{Harty2014TrappedIon}, provided that it is flipped about the same unknown axis as the sensor transition. If a common fluctuating field flips both, the noise is correlated, and Theorem~\ref{thm:joint_scaling} for the joint system gives the same coefficient~\cite{SupplementalMaterial}.

We also show in the Supplemental Material that a single sensor can act as its own reference if its noise strength can be switched by a known factor and its codewords are interchanged at the switch~\cite{SupplementalMaterial}.

\heading{Summary}
We have established fundamental limits for Hamiltonian estimation with unknown noise parameters. Because a change in the noise parameters can shift the accumulated phase in the same way as a change in the Hamiltonian parameters, Heisenberg scaling can be lost even when both signal sensitivity and coherence are preserved. Jointly encoding the sensor with a suitable noisy reference resolves this ambiguity and restores Heisenberg scaling.
In particular, a reference insensitive to the Hamiltonian parameters but subject to an independent copy of the sensor's noise recovers the optimal precision attainable with known noise, for any chosen Hamiltonian parameter.
These results identify the noise parameters whose uncertainty is most detrimental, such as noise directions rather than rates or correlations, and provide explicit remedies.
Future work includes extending these results to non-Markovian noise and to Bayesian estimation, which naturally incorporates partial prior knowledge of the noise.

\begin{acknowledgments}
	\heading{Acknowledgments}
	This work is supported by the National Natural Science Foundation of China (Grants No. 12505024, No. 92476201), Department of Science and Technology of Guangdong Province (Grant No. 2024QN11X234), Guangdong Basic and Applied Basic Research Foundation (Grant No. 2025A1515011441), Shenzhen Science and Technology Program (Grants No. ZDYJ20251211120900001, No. JCYJ20240813141350066).
\end{acknowledgments}

\bibliography{reference}

\onecolumngrid
\begin{center}
	\textbf{End Matter}
\end{center}
\twocolumngrid
\begingroup
\appendix
\setcounter{secnumdepth}{1}
\makeatletter
\renewcommand\section{\@startsection{section}{1}{\parindent}{.8\baselineskip}{-1sp}{\normalfont\normalsize\itshape}}
\def\@runin@to@section#1#2#3{#1#2:\ #3\normalfont---}
\makeatother

\section{Variance bound in joint estimation}\leavevmode\label{sec:joint_direction_variance}
With the target parameter $\theta_i$ placed first in the parameter vector $\bm\theta$, we write the QFIM as
\begin{equation}
	\JQ(t)=\begin{pmatrix}
		a&\bm c^{\mathsf T}\\
		\bm c&B
	\end{pmatrix}.
	\label{eq:qfim_block_em}
\end{equation}
Let $\bm\theta_{-i}$ collect the remaining parameters and define $\bm\theta'_{-i}=\bm\theta_{-i}+B^{-1}\bm c\,\theta_i$, with $B$ and $\bm c$ fixed. The chain rule then gives the reparameterized QFIM for $\bm\theta'=(\theta_i,\bm\theta'_{-i})$ as $\JQ'(t)=\operatorname{diag}(\mathcal I_i,B)$, where $\mathcal I_i=a-\bm c^{\mathsf T}B^{-1}\bm c$. For any vector $\bm v=(1,\bm b^{\mathsf T})^{\mathsf T}$ in this ordering,
\begin{equation}
	\bm v^{\mathsf T}\JQ(t)\bm v=\mathcal I_i+(\bm b+B^{-1}\bm c)^{\mathsf T}B(\bm b+B^{-1}\bm c)\geq\mathcal I_i,
	\label{eq:directional_qfi_decomposition_em}
\end{equation}
where the equality is attained at $\bm b=-B^{-1}\bm c$. Since the reparameterization leaves $\theta_i$ unchanged, $[\JQ^{-1}(t)]_{ii}=[\JQ'(t)^{-1}]_{ii}=1/\mathcal I_i$. The Cram\'er--Rao bound therefore gives Eq.~\eqref{eq:joint_direction_variance}.

\section{Precision bound under arbitrary control}\leavevmode\label{sec:purification_bound}
By Eq.~\eqref{eq:joint_direction_variance}, any $\bm v_i$ with $(\bm v_i)_i=1$ gives $1/[\nu\Var(\hat\theta_i)]\leq\bm v_i^{\mathsf T}\JQ\bm v_i$. The right-hand side is the QFI with respect to $s$ along the line $\bm\theta(s)=\bm\theta_0+s\bm v_i$ through the true values $\bm\theta_0$, on which $\partial_s=\sum_j(\bm v_i)_j\partial_{\theta_j}$.
Purifying the initial state and representing controls, measurements, and feedback by parameter-independent unitaries gives a purification $\ket{\Psi_t}$ of the output state satisfying~\cite{Escher2011Framework,Demkowicz2017Adaptive,Kurdzialek2023Adaptiveness}
\begin{equation}
	\bm v_i^{\mathsf T}\JQ\bm v_i\leq4\norm{\partial_s\Psi_t}^2.
	\label{eq:purification_qfi_em}
\end{equation}

The right-hand side of Eq.~\eqref{eq:purification_qfi_em} depends on the choice of purification, which we fix through the Kraus representation of each short-time channel. Let $\bm K=(K_{\mathrm{ch},0},K_{\mathrm{ch},1},\ldots)^{\mathsf T}$ be the Kraus operators for an interval $\delta t$. A Hermitian matrix $g_i$ defines an equivalent representation $\widetilde{\bm K}(s)=e^{isg_i}\bm K(s)$. For a pure input $\ket{\psi}$ and an orthonormal environment basis $\{\ket{\ell}_E\}$, the purification $\ket{\Psi_s}=\sum_\ell K_{\mathrm{ch},\ell}(s)\ket{\psi}\otimes\ket{\ell}_E$ transforms as $\ket{\widetilde\Psi_s}=(I\otimes e^{isg_i})\ket{\Psi_s}$. At $s=0$, we define $\alpha_{i,\delta t}=(\partial_s\widetilde{\bm K})^\dagger\partial_s\widetilde{\bm K}$ and $\beta_{i,\delta t}=-i(\partial_s\widetilde{\bm K})^\dagger\widetilde{\bm K}$~\cite{Escher2011Framework,Demkowicz2017Adaptive}.

For the master equation in Eq.~\eqref{eq:lindblad}, evolution over an infinitesimal time interval $\delta t$ has Kraus operators $K_{\mathrm{ch},0}=I-(iH+\sum_kE_k^\dagger E_k/2)\delta t+O(\delta t^2)$ and $K_{\mathrm{ch},k}=E_k\sqrt{\delta t}+O(\delta t^{3/2})$. Differentiating along $\bm\theta(s)$ gives $\partial_sH=\sum_a(\bm v_i)_{\omega_a}\partial_{\omega_a}H$ and $\partial_sE_k=\partial_{\bm v_i}E_k=\sum_b(\bm v_i)_{\eta_b}\partial_{\eta_b}E_k$. For this short-time channel, we choose
\begin{equation}
	g_i=\begin{pmatrix}
		a_i\delta t&\sqrt{\delta t}\,\bm b_i^\dagger\\
		\sqrt{\delta t}\,\bm b_i&h_i
	\end{pmatrix},
	\label{eq:kraus_gauge_em}
\end{equation}
with $a_i\in\mathbb R$, $\bm b_i=(b_{i1},b_{i2},\ldots)^{\mathsf T}$, and $h_i=h_i^\dagger$. At $s=0$, $\partial_s\widetilde{\bm K}=\partial_s\bm K+ig_i\bm K$. The short-time expansion gives $\bm K^\dagger g_i\bm K=S_i\delta t+O(\delta t^2)$, where
\begin{equation}
	S_i=a_iI+\sum_k(b_{ik}^*E_k+b_{ik}E_k^\dagger)+\sum_{kl}(h_i)_{kl}E_k^\dagger E_l\in\mathcal S.
	\label{eq:joint_span_em}
\end{equation}
The unitary transformation therefore gives~\cite{Chen2024Informative}
\begin{equation}
	\beta_{i,\delta t}=-i(\partial_s\bm K)^\dagger\bm K-\bm K^\dagger g_i\bm K=\mathcal R_i\delta t+O(\delta t^2),
	\label{eq:kraus_beta_expansion_em}
\end{equation}
where the derivatives of the noise operators combine into the coherent contributions $C_{\eta_b}$ of Eq.~\eqref{eq:Hhat},
\begin{equation}
	\mathcal R_i=\sum_a(\bm v_i)_{\omega_a}\partial_{\omega_a}H+\sum_b(\bm v_i)_{\eta_b}C_{\eta_b}-S_i.
	\label{eq:joint_path_residual}
\end{equation}
The derivatives of $\widetilde{\bm K}$ satisfy $\partial_s\widetilde K_{\mathrm{ch},0}=O(\delta t)$ and $\partial_s\widetilde K_{\mathrm{ch},k}=\sqrt{\delta t}\,F_{i,k}+O(\delta t^{3/2})$ for $k\geq1$, with
\begin{equation}
	F_{i,k}=\partial_{\bm v_i}E_k+ib_{ik}I+i\sum_l(h_i)_{kl}E_l.
	\label{eq:joint_jump_em}
\end{equation}
With the positive semidefinite operator $A_i=\sum_kF_{i,k}^\dagger F_{i,k}$ ($A_a$ for $\theta_i=\omega_a$), we obtain $\alpha_{i,\delta t}=A_i\delta t+O(\delta t^2)$. For $\theta_i=\omega_a$, $\mathcal R_i=\partial_{\omega_a}H-K_a$, with $K_a$ defined in Eq.~\eqref{eq:target_K_definition}. For $\theta_i=\eta_b$, $\mathcal R_i=C_{\eta_b}-K_b$, with $K_b$ defined likewise. Thus $\mathcal R_i$ generates the coherent part and $A_i$ describes the dissipative part of the parameter shift along $\bm v_i$, both fixed by the choice of $\bm v_i$ and $g_i$.

To bound $\norm{\partial_s\Psi_t}$, we represent the channel for each interval $\delta t_j$ by an isometry $V_{\mathrm{ch},j}=\sum_\ell\widetilde K_{\mathrm{ch},\ell,j}\otimes\ket{\ell}_{E_j}$ onto a fresh environment, on which the controls act trivially. We split $\partial_sV_{\mathrm{ch},j}$ into a component $-iV_{\mathrm{ch},j}\beta_{i,j}$ within its image and an orthogonal component $Z_j$, with $Z_j^\dagger Z_j=\alpha_{i,j}-\beta_{i,j}^2$, where $\alpha_{i,j}=\alpha_{i,\delta t_j}$ and $\beta_{i,j}=\beta_{i,\delta t_j}$. In the derivative of the full evolution, the components within the images add coherently across intervals, whereas the orthogonal components are mutually orthogonal and add in quadrature. Since $\beta_{i,j}=\mathcal R_i\delta t_j+O(\delta t_j^2)$ and $\alpha_{i,j}-\beta_{i,j}^2=A_i\delta t_j+O(\delta t_j^2)$, the limit $\max_j\delta t_j\to0$ at fixed $t$ gives~\cite{SupplementalMaterial}
\begin{equation}
	\norm{\partial_s\Psi_t}\leq\int_0^t\sqrt{\Tr(\rho_\tau\mathcal R_i^2)}\,d\tau+\sqrt{\int_0^t\Tr(\rho_\tau A_i)\,d\tau},
	\label{eq:purified_derivative_em}
\end{equation}
where $\rho_\tau$ is the sensor state averaged over measurement records. Bounding both expectation values by operator norms and using Eq.~\eqref{eq:purification_qfi_em} gives Eq.~\eqref{eq:target_purification_bound}. Keeping the state dependence gives a tighter bound in terms of the time-averaged sensor state~\cite{SupplementalMaterial}.

\section{Saturating the precision bound}\leavevmode\label{sec:optimal_coefficient}
We now construct the code described in the proof sketch of Theorem~\ref{thm:joint_scaling}, following Refs.~\cite{Zhou2018Heisenberg,Gorecki2020Optimal,Abbasgholinejad2026Function}, and show that it attains the optimum in Eq.~\eqref{eq:scalaropt}.

We first derive the evolution in a given code space spanned by $\ket{C_0}$ and $\ket{C_1}$, with encoding $V_a=\ket{C_0}\bra{0}+\ket{C_1}\bra{1}$ and projector $P=V_aV_a^\dagger$~\cite{Chen2024Informative}. Suppose that, at $\tilde{\bm\theta}$, the code satisfies the Knill--Laflamme conditions for correcting $\{I,E_k\}$~\cite{Knill1997Codes},
\begin{equation}
	V_a^\dagger E_kV_a=\epsilon_kI_C,\qquad V_a^\dagger E_k^\dagger E_lV_a=\Lambda_{kl}I_C,
	\label{eq:code_correction_conditions_em}
\end{equation}
where $I_C$ is the logical identity.

Let the state at time $\tau$ be $\rho=V_a\rho_CV_a^\dagger$. After an infinitesimal interval $\delta t$, it evolves to $\rho^{\prime}$. A syndrome measurement with projections $\{P,I-P\}$ gives
\begin{equation}
	\rho_p=P\rho^{\prime}P+\delta t\sum_kM_k\rho M_k^\dagger+O(\delta t^2),
	\label{eq:syndrome_state_em}
\end{equation}
where $M_k=(I-P)E_kP$ takes the state out of the code space. At $\tilde{\bm\theta}$, Eq.~\eqref{eq:code_correction_conditions_em} gives $M_k^\dagger M_l=\Gamma_{kl}P$, where $\Gamma_{kl}=\Lambda_{kl}-\epsilon_k^*\epsilon_l$. Choose a fixed unitary $u$ with $u^\dagger\Gamma u=\operatorname{diag}(\chi_k)$. The transformation $E_k\mapsto\sum_lu_{lk}E_l$ leaves the master equation unchanged. Retaining the notation for the transformed operators and coefficients, we have $M_k^\dagger M_l=\chi_k\delta_{kl}P$. The combined syndrome measurement and recovery have Kraus operators $R_0=P$ and $R_k=PM_k^\dagger/\sqrt{\chi_k}$ for $\chi_k>0$. Completing this set gives a trace-preserving map $\mathcal R_{\mathrm{rec}}[X]=\sum_jR_jXR_j^\dagger$ into the code space. We fix the code and recovery at $\tilde{\bm\theta}$ and add Hamiltonian control $-H(\tilde{\bm\omega})$ to the evolution defining $\rho^{\prime}$. One correction cycle gives
\begin{equation}
	\begin{aligned}
		\rho_C(\tau+\delta t)&=V_a^\dagger\mathcal R_{\mathrm{rec}}[\rho^{\prime}]V_a\\
		&=\rho_C(\tau)+\delta t\,\mathcal L_a^C(\bm\theta)(\rho_C)+O(\delta t^2).
	\end{aligned}
	\label{eq:corrected_cycle_em}
\end{equation}
In the rapid-correction limit, $\mathcal L_a^C$ generates the logical evolution, with $\mathcal L_a^C(\tilde{\bm\theta})=0$. To expand this generator around $\tilde{\bm\theta}$, we define $D_{b,k}=\partial_{\eta_b}E_k$ and $Q_b=V_a^\dagger\sum_kE_k^\dagger D_{b,k}V_a$. Using $R_0E_kV_a=\epsilon_kV_a$ and $R_jE_kV_a=\delta_{jk}\sqrt{\chi_k}V_a$ for $\chi_j>0$, we obtain
\begin{equation}
	V_a^\dagger\mathcal R_{\mathrm{rec}}\!\left[\sum_kD_{b,k}\rho E_k^\dagger\right]V_a=Q_b\rho_C.
	\label{eq:recovered_noise_terms_em}
\end{equation}
Only $R_0$ contributes to the derivative of the anticommutator, yielding $-\{Q_b+Q_b^\dagger,\rho_C\}/2$. Adding the Hermitian-conjugate term $\rho_CQ_b^\dagger$ and using Eq.~\eqref{eq:Hhat} gives
\begin{equation}
	\partial_{\eta_b}\mathcal L_a^C(\rho_C)=\left[\frac{Q_b-Q_b^\dagger}{2},\rho_C\right]=-i[V_a^\dagger C_{\eta_b}V_a,\rho_C].
	\label{eq:recovered_noise_derivative_em}
\end{equation}
After error correction, a change in $\eta_b$ thus acts on the logical state only through its coherent contribution $C_{\eta_b}$. Combining this with the Hamiltonian derivatives, the first-order change in the logical generator is
\begin{equation}
	\delta\mathcal L_a^C(\rho_C)=-i\sum_j\delta\theta_j[V_a^\dagger X_jV_a,\rho_C],
	\label{eq:corrected_local_generator_em}
\end{equation}
with $X_j$ the coherent contribution $\partial_{\omega_{a'}}H$ or $C_{\eta_b}$ of $\theta_j$.

We now choose the code to maximize the response to $\omega_a$ and eliminate the first-order responses to the other parameters. By norm duality, the distance in Eq.~\eqref{eq:di} is
\begin{equation}
	d_a=\max_{\substack{Y=Y^\dagger,\;\norm{Y}_1\leq1\\Y\perp\mathcal N_a}}\Tr(Y\partial_{\omega_a}H),
	\label{eq:optimal_code_dual_em}
\end{equation}
where $\norm{Y}_1=\Tr|Y|$ and $Y\perp\mathcal N_a$ means $\Tr(YO)=0$ for all $O\in\mathcal N_a$. For $d_a>0$, an optimizer $Y_\star$ has $\norm{Y_\star}_1=1$, and $I\in\mathcal N_a$ gives $\Tr Y_\star=0$. Thus $\rho_\pm=|Y_\star|\pm Y_\star$ are density matrices. We choose $\ket{C_0}$ and $\ket{C_1}$ as their purifications in orthogonal subspaces of a noiseless ancilla. For $O$ acting on the system, we obtain
\begin{equation}
	V_a^\dagger O V_a=\Tr(|Y_\star|O)I_C+\Tr(Y_\star O)\sigma_z^C,
	\label{eq:optimal_code_compression_em}
\end{equation}
where $\sigma_z^C$ is the logical Pauli operator. Since $Y_\star\perp\mathcal N_a$ and $\Tr(Y_\star\partial_{\omega_a}H)=d_a$, the code satisfies Eq.~\eqref{eq:code_conditions}. In particular, $\mathcal S\subseteq\mathcal N_a$ gives Eq.~\eqref{eq:code_correction_conditions_em}. Equation~\eqref{eq:corrected_local_generator_em} therefore reduces to Eq.~\eqref{eq:logical_evolution}. For the initial state $(\ket{C_0}+\ket{C_1})/\sqrt2$ and a measurement of $\sigma_y^C$ after time $t$, the estimator $\hat\omega_a=\tilde\omega_a+\bar z/(2d_at)$, with $\bar z$ the mean of the $\nu$ outcomes $z=\pm1$, is locally unbiased with variance $1/(4\nu d_a^2t^2)$, attaining the optimum in Eq.~\eqref{eq:scalaropt}. An auxiliary experiment with a nonsingular CFIM can estimate the remaining parameters using a vanishing fraction of interrogations without changing the quadratic coefficient~\cite{SupplementalMaterial}.

\section{Optimal codes for Theorem~2}\leavevmode\label{sec:reference_codes}
For sensor codewords that satisfy the first condition in Eq.~\eqref{eq:code_conditions} with $\mathcal{S}$, the operator $Y_S=\Tr_A(\ket{C_0}\bra{C_0}-\ket{C_1}\bra{C_1})/2$, with $A$ the ancilla, satisfies $Y_S\perp\mathcal{S}$ and $\|Y_S\|_1\leq1$, and gives $y_j^S=\Tr(Y_SX_j)$ in Eq.~\eqref{eq:jointgen}. Conversely, as in Appendix~\ref{sec:optimal_coefficient}, codewords that purify $|Y_S|\pm Y_S+(1-\|Y_S\|_1)\sigma$, with any state $\sigma$, realize any such Hermitian $Y_S$. With $Y_R$ defined likewise, $y_j=y_j^S-y_j^R=\Tr(Y_SX_j)-\Tr(Y_RX_j^R)$.

The largest response to $\omega_a$ without response to the other parameters is
\begin{equation}
	\max_{y_j=0\ \forall\theta_j\neq\omega_a}y_{\omega_a}=\min_{\bm v}\bigl[d_a^S(\bm v)+d_a^R(\bm v)\bigr],
	\label{eq:joint_code_dual_em}
\end{equation}
where $d_a^S(\bm v)$ and $d_a^R(\bm v)$ are defined through Eq.~\eqref{eq:dav} and the maximum is over Hermitian $Y_S\perp\mathcal{S}$ and $Y_R\perp\mathcal{S}_R$ with $\|Y_S\|_1,\|Y_R\|_1\leq1$. With multipliers $v_j$ for the constraints and $v_{\omega_a}=1$, the Lagrangian is $\Tr(Y_S\sum_jv_jX_j)-\Tr(Y_R\sum_jv_jX_j^R)$, whose maximum is $d_a^S(\bm v)+d_a^R(\bm v)$ by Eq.~\eqref{eq:optimal_code_dual_em} with $\mathcal{N}_a$ replaced by $\mathcal{S}$ and $\mathcal{S}_R$. Strong duality holds since $Y_S=Y_R=0$ is strictly feasible, and Eq.~\eqref{eq:dSR} identifies the optimum as $d_a^{SR}$.

For the case of Eq.~\eqref{eq:single}, the same duality gives operators $W_S\perp\mathcal{S}$ and $W_R\perp\mathcal{S}_R$ with trace norms at most one, $\Tr(W_SC_{\eta_b})=a_S$, and $\Tr(W_RC_{\eta_b}^R)=a_R$. Since $\partial_{\omega_a}H+\lambda_SC_{\eta_b}\in\mathcal{S}$ and $\partial_{\omega_a}H_R+\lambda_RC_{\eta_b}^R\in\mathcal{S}_R$, the choice $Y_S=q_SW_S$ and $Y_R=q_RW_R$ with $q_S,q_R\in[0,1]$ gives
\begin{equation}
	y_{\eta_b}=q_Sa_S-q_Ra_R,\qquad y_{\omega_a}=\lambda_Rq_Ra_R-\lambda_Sq_Sa_S.
	\label{eq:single_responses_em}
\end{equation}
Equal responses to $\eta_b$, $q_Sa_S=q_Ra_R=\min\{a_S,a_R\}$, then give $y_{\omega_a}=\min\{a_S,a_R\}(\lambda_R-\lambda_S)$, whose magnitude is $d_a^{SR}$ in Eq.~\eqref{eq:single}.
\par\endgroup

\par\onecolumngrid\clearpage
\makeatletter
\do@output@MVL{\setcounter{page}{1}}
\makeatother
\setcounter{section}{0}\setcounter{subsection}{0}\setcounter{subsubsection}{0}
\setcounter{equation}{0}\setcounter{figure}{0}\setcounter{table}{0}
\counterwithout*{equation}{section}
\setlength{\lineskiplimit}{0pt}
\fussy
\setlength{\emergencystretch}{\suppEmergencyStretch}
\makeatletter
\let\addtocontents\savedAddToContents
\let\set@listindent\set@listindent@parshape
\let\close@column\newpage
\let\set@footnotewidth\set@footnotewidth@one
\let\compose@footnotes\compose@footnotes@one
\def\title@column#1{\minipagefootnote@init#1\minipagefootnote@foot}
\@booleanfalse\twocolumn@sw
\def\@extra@b@citeb{.supp}
\immediate\write\@auxout{\string\gdef\string\@extra@binfo{.supp}}
\patchcmd{\NAT@bibsetnum}{LastBibItem}{LastSupplementBibItem}{}{\errmessage{Last bib label width patch failed}}
\patchcmd{\endthebibliography}{LastBibItem}{LastSupplementBibItem}{}{\errmessage{Last bib label patch failed}}
\patchcmd{\titleblock@produce}{\label{FirstPage}}{\label{FirstSupplementPage}}{}{\errmessage{First page patch failed}}
\patchcmd{\frontmatter@footnotemark}{frontmatter.}{supp.frontmatter.}{}{\errmessage{Footnote link patch failed}}
\patchcmd{\frontmatter@makefntext}{frontmatter.}{supp.frontmatter.}{}{\errmessage{Footnote anchor patch failed}}
\patchcmd{\present@bibnote}{frontmatter.}{supp.frontmatter.}{}{\errmessage{Supplement email anchor patch failed}}
\frontmatter@init
\makeatother
\renewcommand{\thesection}{S\arabic{section}}
\renewcommand{\thesubsection}{\thesection.\Alph{subsection}}
\renewcommand{\theequation}{S\arabic{equation}}
\renewcommand{\thefigure}{S\arabic{figure}}
\renewcommand{\thetable}{S\arabic{table}}

\setcounter{secnumdepth}{2}
\setcounter{tocdepth}{1}

\makeatletter
\renewcommand{\p@subsection}{}
\renewcommand{\l@subsection}[2]{}
\renewcommand{\l@subsubsection}[2]{}
\renewcommand{\l@f@section}{\addpenalty{\@secpenalty}\addvspace{0.15em}}
\makeatother

\begin{bibunit}[apsrev4-2]
\makeatletter
\renewcommand{\@bibunitname}{supplemental}
\makeatother

\title{Supplemental Material for ``Restoring Heisenberg Scaling in Hamiltonian Estimation despite Noise Uncertainty''}

\author{Yan Luo}
\affiliation{Institute of Quantum Precision Measurement, State Key Laboratory of Radio Frequency Heterogeneous Integration, College of Physics and Optoelectronic Engineering, Shenzhen University, Shenzhen, China}

\author{Lingna Wang}
\affiliation{Department of Mechanical and Automation Engineering, The Chinese University of Hong Kong, Shatin, Hong Kong SAR, China}

\author{Hongzhen Chen}
\email{hzchen@szu.edu.cn}
\affiliation{Institute of Quantum Precision Measurement, State Key Laboratory of Radio Frequency Heterogeneous Integration, College of Physics and Optoelectronic Engineering, Shenzhen University, Shenzhen, China}

\maketitle

\tableofcontents

\section{Joint estimation in a spin model with transverse noise}\label{sec:finite}
Consider a spin-$1/2$ sensor governed by the Markovian master equation in Eq.~\eqref{eq:toy} of the main text,
\begin{equation}
	\frac{d\rho}{dt}=-i\frac{\omega}{2}[\sigma_z,\rho]
	+\kappa(\sigma_\phi\rho\sigma_\phi-\rho).
	\label{eq:spin_master_explicit}
\end{equation}
Here $\omega$ is the frequency to be estimated, and $\sigma_\phi=\cos\phi\,\sigma_x+\sin\phi\,\sigma_y$ specifies the transverse noise direction. The noise rate $\kappa>0$ is known, except in the last paragraph of Sec.~\ref{sec:toybridge}. Using $\nu$ independent interrogations of duration $t$, we compare estimation of $\omega$ for a known noise direction $\phi_0$ with joint estimation of $(\omega,\phi)$, first without control (Sec.~\ref{sec:toy_uncontrolled}) and then with the single-sensor error-correction protocol of the main text (Sec.~\ref{sec:toybridge}). Section~\ref{sec:toy_simulation} tests the predicted scalings numerically. Using the general bounds of Sec.~\ref{sec:upper}, Sec.~\ref{sec:spin_optimality} proves that the single-sensor protocol is asymptotically optimal under arbitrary control and that, at fixed total time, calibrating $\phi$ first does not improve the leading term of the frequency variance at long times.

\subsection{Frequency estimation without control}\label{sec:toy_uncontrolled}

Without control, each error (a spin flip) reverses the sign of the subsequent phase accumulation. For known $\phi=\phi_0$, let the sensor and a noiseless ancilla $A$ start in an arbitrary pure state $\ket{\psi_0}$. In the quantum-jump unraveling of Eq.~\eqref{eq:spin_master_explicit}~\cite{Dalibard1992WaveFunction}, errors occur at the rate $\kappa\Tr[(\sigma_{\phi_0}^\dagger\sigma_{\phi_0}\otimes I_A)\rho]=\kappa$ for any sensor--ancilla state $\rho$. We denote the record of error times by $\mathsf r$. The number $N(s)$ of errors up to time $s$ is a Poisson process with rate $\kappa$, and the distribution of $\mathsf r$ is independent of $\omega$. Between errors, the unnormalized state is propagated over a time $s$ by $e^{-\kappa s/2}U_0(s)$, where $U_0(s)=e^{-i\omega s\sigma_z/2}$.

For $m=N(t)$ errors at $0<t_1<\cdots<t_m<t$, put $t_0=0$ and $t_{m+1}=t$. The relation $\sigma_{\phi_0}U_0(s)=U_0(-s)\sigma_{\phi_0}$ gives the normalized evolution conditional on this record
\begin{equation}
	\begin{aligned}
		U_{\mathsf r}(\omega)
		&=\left[U_0(t-t_m)\sigma_{\phi_0}U_0(t_m-t_{m-1})\cdots\sigma_{\phi_0}U_0(t_1)\right]\otimes I_A\\
		&=(\sigma_{\phi_0}^{m}\otimes I_A)\exp\!\left[-\frac{i\omega T_{\mathsf r}(t)}{2}(\sigma_z\otimes I_A)\right],
	\end{aligned}
	\label{eq:spin_trajectory_unitary}
\end{equation}
where the net phase-accumulation time is \mbox{$T_{\mathsf r}(t)=\sum_{j=0}^{m}(-1)^j(t_{j+1}-t_j)=\int_0^t(-1)^{N(s)}\,ds$}.
The conditional state $\ket{\psi_{\omega|\mathsf r}}=U_{\mathsf r}(\omega)\ket{\psi_0}$ satisfies
\begin{equation}
	\partial_\omega\ket{\psi_{\omega|\mathsf r}}=-iU_{\mathsf r}(\omega)\frac{T_{\mathsf r}(t)}{2}(\sigma_z\otimes I_A)\ket{\psi_0}.
	\label{eq:spin_trajectory_derivative}
\end{equation}
Its quantum Fisher information (QFI) is
\begin{equation}
	\begin{aligned}
		J_{Q|\mathsf r}
		&=4\left(\langle\partial_\omega\psi_{\omega|\mathsf r}|\partial_\omega\psi_{\omega|\mathsf r}\rangle-\left|\langle\psi_{\omega|\mathsf r}|\partial_\omega\psi_{\omega|\mathsf r}\rangle\right|^2\right)\\
		&=4\Var_{\psi_0}\!\left[\frac{T_{\mathsf r}(t)}{2}(\sigma_z\otimes I_A)\right]
		=T_{\mathsf r}(t)^2\left[1-\langle\sigma_z\otimes I_A\rangle_{\psi_0}^2\right]\leq T_{\mathsf r}(t)^2.
	\end{aligned}
	\label{eq:spin_pure_qfi}
\end{equation}
Averaging the conditional states gives the output state of the sensor and ancilla, \mbox{$\rho_\omega(t)=\EE_{\mathsf r}\!\left[\ket{\psi_{\omega|\mathsf r}}\bra{\psi_{\omega|\mathsf r}}\right]$}. Since the distribution of $\mathsf r$ is independent of $\omega$, convexity of the QFI yields~\cite{Albarelli2018Monitoring}
\begin{equation}
	J_Q(t)\leq\EE[J_{Q|\mathsf r}]\leq\EE[T_{\mathsf r}(t)^2]=2\int_0^t ds\int_0^s du\,e^{-2\kappa(s-u)}=\frac{t}{\kappa}-\frac{1-e^{-2\kappa t}}{2\kappa^2}\leq\frac{t}{\kappa},
	\label{eq:toy_uncontrolled_bound}
\end{equation}
where we used $\EE[(-1)^{N(s)+N(u)}]=\EE[(-1)^{N(s)-N(u)}]=e^{-2\kappa(s-u)}$ for $s\geq u$, which follows from independent Poisson increments. If the error times are recorded, the QFI of the output state together with the record equals $\EE[J_{Q|\mathsf r}]$, so monitoring the errors without feedback cannot exceed this bound. Purification extends Eq.~\eqref{eq:toy_uncontrolled_bound} to mixed initial states of the sensor and ancilla. Treating $\phi$ as unknown cannot improve the precision, so every locally unbiased estimator of $\omega$ in joint estimation of $(\omega,\phi)$ satisfies $\nu\Var(\hat\omega)\geq[t/\kappa-(1-e^{-2\kappa t})/(2\kappa^2)]^{-1}$.

\subsection{Joint estimation with quantum error correction}\label{sec:toybridge}
Quantum error correction (QEC) with a noiseless ancilla allows transverse spin flips to be detected without measuring the accumulated signal phase~\cite{Kessler2014QEC,Zhou2018Heisenberg}. We use a recovery independent of $\phi$ and retain the error count to estimate the frequency and noise direction jointly.

Prepare the sensor and a noiseless ancilla qubit in $(\ket{C_0}+\ket{C_1})/\sqrt2$, where $\ket{C_0}=\ket{00}$ and $\ket{C_1}=\ket{11}$, with the sensor as the first qubit. During the interrogation, repeatedly measure $\sigma_z\otimes\sigma_z$. For each outcome $-1$, apply $\sigma_x$ to the sensor and record the flip. This is the single-sensor protocol of the main text.

Each error followed by recovery applies the net rotation $\sigma_x\sigma_\phi=e^{i\phi\sigma_z}$ to the sensor, that is, the logical rotation $e^{i\phi\sigma_z^C}$ with $\sigma_z^C=\ket{C_0}\bra{C_0}-\ket{C_1}\bra{C_1}$. In the rapid-correction limit, with the correction interval tending to zero at fixed $t$, the recorded error count $N$ has probability $p_N=e^{-\kappa t}(\kappa t)^N/N!$. Conditional on $N$, the logical state carries the relative phase $\Phi_N$ of the main text,
\begin{equation}
	\ket{\psi_N(\omega,\phi)}
	=\frac{\ket{C_0}+e^{i\Phi_N}\ket{C_1}}{\sqrt2},\qquad
	\Phi_N=\omega t-2N\phi.
	\label{eq:toy_conditional_state}
\end{equation}
For known $\phi=\phi_0$, the count-dependent logical rotation $e^{-iN\phi_0\sigma_z^C}$ compensates the error phase $-2N\phi_0$ and leaves the signal phase $\omega t$. An $N$-independent phase measurement then attains the QFI $t^2$ for $\omega$, giving $\Var(\hat\omega)=1/(\nu t^2)$.

\paragraph{Joint estimation of frequency and noise direction.}
For unknown $\phi$, write $\Omega=\omega-2\kappa\phi$ and $\tilde\Omega=\tilde\omega-2\kappa\tilde\phi$, where the prior estimates $(\tilde\omega,\tilde\phi)$ serve as the local reference point. The phase is $\Phi_N=t\Omega-2(N-\kappa t)\phi$: its mean determines $\Omega$, whereas $\phi$ enters only through the count fluctuation $N-\kappa t$. Recording $N$ in a classical register gives the output state $\rho_{\omega,\phi}^{\mathrm{out}}=\sum_Np_N\ket N\bra N\otimes\ket{\psi_N}\bra{\psi_N}$. The count probabilities are independent of both estimated parameters, and the QFI of each $\ket{\psi_N}$ with respect to $\Phi_N$ is 1. 
With the gradient $\bm g_N=\nabla\Phi_N=(t,-2N)^{\mathsf T}$ of $\Phi_N$ with respect to $(\omega,\phi)$, the quantum Fisher information matrix (QFIM) per interrogation is therefore
\begin{equation}
	\JQ=\EE\bigl[\bm g_N\bm g_N^{\mathsf T}\bigr]
	=\EE\begin{pmatrix}t^2&-2tN\\-2tN&4N^2\end{pmatrix}
	=\begin{pmatrix}t^2&-2\kappa t^2\\-2\kappa t^2&4\kappa t(1+\kappa t)\end{pmatrix},
	\label{eq:toy-QFIM}
\end{equation}
where the last step uses $\EE N=\kappa t$ and $\EE N^2=\kappa t(1+\kappa t)$.

We measure the logical qubit in the count-dependent basis $(\ket{C_0}\pm i e^{i(\tilde\omega t-2N\tilde\phi)}\ket{C_1})/\sqrt2$, which is optimal at the reference point $(\tilde\omega,\tilde\phi)$. This basis and the recovery are held fixed when we differentiate the outcome probabilities. The count $N$ and the outcome $z=\pm1$ occur with joint probability
\begin{equation}
	P(N,z\mid\omega,\phi)
	=p_N\frac{1+z\sin[t(\omega-\tilde\omega)-2N(\phi-\tilde\phi)]}{2}.
	\label{eq:toy-observed-prob}
\end{equation}
At $(\tilde\omega,\tilde\phi)$, Eq.~\eqref{eq:toy-observed-prob} gives $P=p_N/2$ and $\nabla P=(p_Nz/2)\,\bm g_N$. The classical Fisher information matrix (CFIM) is therefore
\begin{equation}
	\JC=\sum_N\sum_{z=\pm1}\frac{\nabla P\,(\nabla P)^{\mathsf T}}{P}
	=\sum_N\sum_{z=\pm1}\frac{p_Nz^2}{2}\,\bm g_N\bm g_N^{\mathsf T}
	=\EE\bigl[\bm g_N\bm g_N^{\mathsf T}\bigr]
	=\JQ.
	\label{eq:toy-CFIM}
\end{equation}
The Fisher information for this fixed recovery and the complete measurement record was also derived in Ref.~\cite{Tanggara2026Sequential}. For locally unbiased estimators from $\nu$ independent interrogations, the classical Cram\'er--Rao bound is $\Cov(\hat{\bm\theta})\succeq\JC^{-1}/\nu$, with $\bm\theta=(\omega,\phi)^{\mathsf T}$. The inverse of the matrix in Eq.~\eqref{eq:toy-QFIM} is
\begin{equation}
	\JC^{-1}
	=\begin{pmatrix}t^{-2}+\kappa/t&1/(2t)\\1/(2t)&1/(4\kappa t)\end{pmatrix}.
	\label{eq:toy-CFIM-inverse}
\end{equation}
To attain this bound, we estimate $\Omega$ and $\phi$ separately. In the coordinates $(\Omega,\phi)$, the gradient of $\Phi_N=t\Omega-2(N-\kappa t)\phi$ is $(t,-2(N-\kappa t))^{\mathsf T}$, and Eq.~\eqref{eq:toy-CFIM} with this gradient in place of $\bm g_N$ gives the CFIM $\diag(t^2,4\kappa t)$, since $N-\kappa t$ has zero mean and variance $\kappa t$. With bars denoting means over the $\nu$ interrogations, the estimators $\hat\Omega=\tilde\Omega+\bar z/t$ and $\hat\phi=\tilde\phi-\overline{(N-\kappa t)z}/(2\kappa t)$ are therefore locally unbiased and uncorrelated at $(\tilde\omega,\tilde\phi)$, with ${\Var(\hat\Omega)=1/(\nu t^2)}$. The estimators $\hat\omega=\hat\Omega+2\kappa\hat\phi$ and $\hat\phi$ of the original parameters then have
\begin{equation}
	\Var(\hat\omega)=\frac{1}{\nu t^2}+\frac{\kappa}{\nu t},\qquad
	\Var(\hat\phi)=\frac{1}{4\nu\kappa t},\qquad
	\Cov(\hat\omega,\hat\phi)=\frac{1}{2\nu t}.
	\label{eq:toy_error_propagation}
\end{equation}
These are the entries of $\JC^{-1}/\nu$, so the count-dependent readout attains the Cram\'er--Rao bound for both parameters. The unknown noise direction adds $\kappa/(\nu t)$ to the variance $1/(\nu t^2)$ for known $\phi$. Recording $N$ is essential. If it is discarded after the readout, $P$ is replaced by its sum over $N$, which equals $1/2$ at the reference point and has gradient $(z/2)\,\EE\bm g_N$. The CFIM then becomes the rank-one matrix $(\EE\bm g_N)(\EE\bm g_N)^{\mathsf T}$ with $\EE\bm g_N=(t,-2\kappa t)^{\mathsf T}$, so only $\Omega$ remains locally identifiable.

\paragraph{Joint estimation of frequency and noise rate.}
An unknown noise rate has a different effect: $\kappa$ enters only the count distribution, not the corrected logical phase. For joint estimation of $(\omega,\kappa)$ with known $\phi=\phi_0$, the phase compensation described above gives
\begin{equation}
	P(N,z\mid\omega,\kappa)
	=e^{-\kappa t}\frac{(\kappa t)^N}{N!}
	\frac{1+z\sin[t(\omega-\tilde\omega)]}{2}.
	\label{eq:rate_joint_likelihood}
\end{equation}
At $\omega=\tilde\omega$, Eq.~\eqref{eq:rate_joint_likelihood} gives $P=p_N/2$, and the gradient of $P$ with respect to $(\omega,\kappa)$ is $\nabla P=(p_N/2)\,(zt,N/\kappa-t)^{\mathsf T}$. The terms odd in $z$ cancel in the sum over $z=\pm1$, and $\EE[(N/\kappa-t)^2]=t/\kappa$, so the CFIM is
\begin{equation}
	\JC=\sum_N\sum_{z=\pm1}\frac{\nabla P\,(\nabla P)^{\mathsf T}}{P}
	=\sum_Np_N\begin{pmatrix}t^2&0\\0&(N/\kappa-t)^2\end{pmatrix}
	=\begin{pmatrix}t^2&0\\0&t/\kappa\end{pmatrix}.
	\label{eq:rate_joint_cfim}
\end{equation}
The estimators $\hat\omega=\tilde\omega+\bar z/t$ and $\hat\kappa=\bar N/t$ are jointly locally unbiased, with covariance
\begin{equation}
	\Cov\!\left[\begin{pmatrix}\hat\omega\\\hat\kappa\end{pmatrix}\right]
	=\frac1\nu\begin{pmatrix}t^{-2}&0\\0&\kappa/t\end{pmatrix}.
	\label{eq:rate_joint_covariance}
\end{equation}
Section~\ref{sec:spin_optimality} shows that both variances are optimal under arbitrary control.

\subsection{Numerical simulations of adaptive estimation}\label{sec:toy_simulation}
\begin{figure}[!htbp]
	\centering
	\includegraphics[width=.88\textwidth]{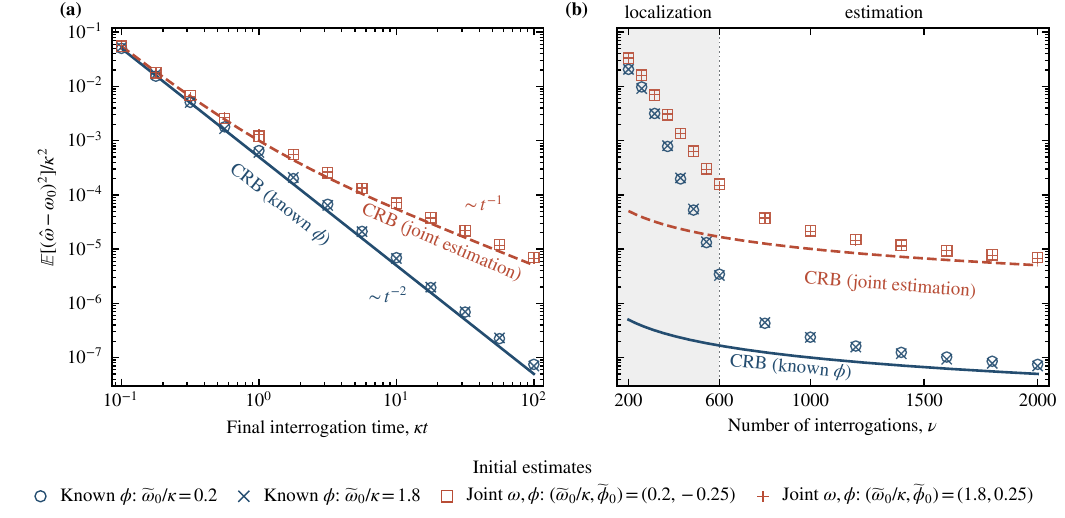}
	\caption{Adaptive frequency estimation with true values $\omega_0/\kappa=1$ and $\phi_0=0.08$. The joint fits start from $(\tilde\omega_0/\kappa,\tilde\phi_0)=(0.2,-0.25)$ or $(1.8,0.25)$, and the fits for known $\phi$ use the corresponding initial frequency. Symbols show the normalized frequency mean-square error $\mathbb E[(\hat\omega-\omega_0)^2]/\kappa^2$ versus (a) the final interrogation time $\kappa t$ for $\nu=2000$ interrogations and (b) the cumulative number of interrogations $\nu$ at $\kappa t=100$. The first 600 interrogations localize the phase (gray region in (b)), followed by 1400 at duration $t$. Curves show the ideal Cram\'er--Rao bounds (CRBs) for $\nu$ interrogations all lasting $t$: $1/[\nu(\kappa t)^2]$ for known $\phi$ and $1/[\nu(\kappa t)^2]+1/(\nu\kappa t)$ for joint estimation.}
	\label{fig:toy_initialization_precision}
\end{figure}

The estimators of Sec.~\ref{sec:toybridge} are locally unbiased only at their reference point, which must lie close to the unknown true parameters. We therefore test whether adaptive maximum-likelihood estimation from distant initial estimates attains the $t^{-2}$ scaling for known $\phi$ and the bound in Eq.~\eqref{eq:toyvariance} of the main text, counting the interrogations used to locate the parameters.

We use known $\kappa$ and the fixed recovery $\sigma_x$ of Sec.~\ref{sec:toybridge}. Interrogations with the same duration and readout basis form a batch, after which the estimate is updated. Of the $\nu=2000$ interrogations, the first 600 form a localization stage that resolves the phase ambiguity, at durations starting from $\tau_0=\min(t,0.5/\kappa)$ and doubling while the next duration is below $t$. We assign 200 to the first duration and divide 400 as evenly as possible among the longer durations, or use all 600 when there is only one duration. Each localization batch is split as evenly as possible between phase offsets $\alpha=\pm\pi/4$, which together give both quadratures and thus determine the phase modulo $2\pi$. The remaining 1400 interrogations form seven batches at duration $t$, with $\alpha=0$.

In batch $r$, of duration $\tau_r$, the readout basis of Sec.~\ref{sec:toybridge} is set at the current estimate $(\tilde\omega_r,\tilde\phi_r)$ instead of the true parameters $(\omega_0,\phi_0)$, starting from the initial estimate $(\tilde\omega_0,\tilde\phi_0)$. Outcomes are sampled at the true parameters, where the bounds below are also evaluated: $N$ is Poisson distributed with mean $\kappa\tau_r$, and the outcome $z=\pm1$ occurs with probability
\begin{equation}
	p_r(z\mid N,\omega,\phi;\alpha)
	=\frac{1+z\sin[\tau_r(\omega-\tilde\omega_r)-2N(\phi-\tilde\phi_r)+\alpha]}{2}.
	\label{eq:toy_simulation_probability}
\end{equation}

After each batch, we update the estimate by maximizing the likelihood of all outcomes so far, each evaluated with its own duration, phase offset, and readout basis. The first fit starts from a grid search over $0\leq\omega/\kappa\leq2$ and $-0.3\leq\phi\leq0.3$. Each fit is then refined by Newton-type iterations and compared with competing likelihood maxima on a grid within six Cram\'er--Rao standard deviations of the current estimate. For known $\phi=\phi_0$, we fix $\tilde\phi_r=\phi_0$ and fit only $\omega$.

To include the time spent on localization, define $S_1=\sum_{a=1}^{\nu}\tau_a$ and $S_2=\sum_{a=1}^{\nu}\tau_a^2$, with $\tau_a$ the duration of interrogation $a$. For any readout basis and offset, the Fisher information of each readout in Eq.~\eqref{eq:toy_simulation_probability} with respect to the logical phase is 1. Since the gradient of $\log p_r$ for each interrogation has zero mean conditional on the earlier outcomes, which fix the readout basis, Fisher information adds across interrogations. With $\bm g_a=\nabla(\tau_a\omega-2N_a\phi)=(\tau_a,-2N_a)^{\mathsf T}$, averaging over the Poisson counts $N_a$ gives
\begin{equation}
	\JC^{\mathrm{tot}}=\sum_{a=1}^{\nu}\EE[\bm g_a\bm g_a^{\mathsf T}]
	=\begin{pmatrix}
		S_2 & -2\kappa S_2\\
		-2\kappa S_2 & 4\kappa^2S_2+4\kappa S_1
	\end{pmatrix}.
	\label{eq:toy_localization_fisher}
\end{equation}
For locally unbiased estimators, the frequency variance is at least $1/S_2$ for known $\phi$, whereas joint estimation obeys
\begin{equation}
	\Var(\hat\omega)\geq\frac{1}{S_2}+\frac{\kappa}{S_1},
	\label{eq:toy_localization_frequency}
\end{equation}
with $\Var(\hat\phi)\geq1/(4\kappa S_1)$. Since $0.7\nu t\leq S_1\leq\nu t$ and $0.7\nu t^2\leq S_2\leq\nu t^2$, the bounds for known $\phi$ and for joint estimation retain the long-time scalings $t^{-2}$ and $t^{-1}$, respectively.

For each of the two initial estimates in Fig.~\ref{fig:toy_initialization_precision}, we use all 2000 simulation runs at each of 13 logarithmically spaced final interrogation times $0.1\leq\kappa t\leq100$, discarding none. The frequency mean-square errors in panel (a) follow these scalings. Their ratios to the bounds evaluated with the actual $S_1$ and $S_2$ range from $0.92$ to $1.07$, with a Monte Carlo standard error of about $0.03$ per point. Panel (b) shows the convergence as the number of interrogations increases, at $\kappa t=100$. The plotted curves are the ideal bounds with all interrogations lasting $t$. Because the localization interrogations are shorter than $t$, the bounds evaluated with the actual $S_1$ and $S_2$ exceed these curves by up to $40\%$ at $\nu=2000$, which explains why the symbols lie above the curves.

\FloatBarrier

\section{General sequential scheme and precision bounds for joint estimation}\label{sec:upper}
We consider a quantum sensor evolving according to the Markovian master equation
\begin{equation}
	\frac{d\rho}{dt}=\mathcal L_{\bm\theta}(\rho)=-i[H(\bm\theta),\rho]+\sum_k\mathcal D[E_k(\bm\theta)]\rho,
	\label{eq:general_master}
\end{equation}
where $\mathcal L_{\bm\theta}$ is the generator and $\mathcal D[E]\rho=E\rho E^\dagger-\{E^\dagger E,\rho\}/2$ is the Lindblad dissipator. We jointly estimate the parameters $\bm\theta$, which may enter both the Hamiltonian and the noise operators. The model in Eq.~\eqref{eq:lindblad} of the main text has $H=H(\bm\omega)$ and $E_k=E_k(\bm\eta)$, with $\bm\theta=(\bm\omega,\bm\eta)$, $\bm\omega=(\omega_1,\ldots,\omega_r)$, and $\bm\eta=(\eta_1,\ldots,\eta_q)$.

In the general sequential scheme, each of $\nu$ independent interrogations lasts a total time $t$. During this time, the evolution under Eq.~\eqref{eq:general_master} is interleaved with arbitrary parameter-independent operations on the sensor and on noiseless ancillas of any finite dimension, whose dynamics do not depend on $\bm\theta$. These operations include state preparation, unitary control, measurements, and feedback conditioned on earlier outcomes~\cite{Demkowicz2017Adaptive,Sekatski2017FastControl}. The scheme includes the limit of arbitrarily frequent operations at fixed $t$, as well as protocols that alternate between noise characterization and Hamiltonian estimation, with $t$ the total time spent on both stages. The output state $\rho_{\bm\theta}$ of an interrogation includes classical registers that store the measurement records. We denote the true parameter values by $\bm\theta_0$ and the prior estimates by $\tilde{\bm\theta}$, which serve as the reference point for local estimation.

Section~\ref{sec:general_setting} extends Appendix~\ref{sec:joint_direction_variance} of the End Matter: it derives the variance bound for a target parameter, identifies its effective information, and constructs a measurement that attains the bound. Section~\ref{sec:joint_explicit_gauges} extends Appendix~\ref{sec:purification_bound} of the End Matter: it derives a bound on the QFI under arbitrary control and summarizes it in Lemma~\ref{lem:purification_bound}, which yields Eq.~\eqref{eq:target_purification_bound} of the main text and its state-dependent refinement. Section~\ref{sec:spin_optimality} applies the lemma to the spin model of Sec.~\ref{sec:finite}.

\subsection{Variance bound in joint estimation}\label{sec:general_setting}
The precision of joint estimation is characterized by the covariance matrix $\Cov(\hat{\bm\theta})$, with entries $[\Cov(\hat{\bm\theta})]_{jk}=\EE[\hat\theta_j\hat\theta_k]-\EE[\hat\theta_j]\EE[\hat\theta_k]$ and diagonal elements $\Var(\hat\theta_j)$. For one interrogation, a positive-operator-valued measure (POVM) $\{M_y\}$ on $\rho_{\bm\theta}$ gives outcome probabilities $p(y|\bm\theta)=\Tr(\rho_{\bm\theta}M_y)$ and the CFIM $[\JC]_{jk}=\sum_y[\partial_jp(y|\bm\theta)][\partial_kp(y|\bm\theta)]/p(y|\bm\theta)$, with $\partial_j=\partial/\partial\theta_j$. For any measurement, $\JC\preceq\JQ$, where the QFIM has entries $[\JQ]_{jk}=\Reop\Tr(\rho_{\bm\theta}L_jL_k)$ and the symmetric logarithmic derivatives (SLDs) $L_j=L_j^\dagger$ satisfy $\partial_j\rho_{\bm\theta}=(\rho_{\bm\theta}L_j+L_j\rho_{\bm\theta})/2$. For $\nu$ independent interrogations and nonsingular Fisher information matrices, locally unbiased estimators therefore obey the Cram\'er--Rao bounds
\begin{equation}
	\Cov(\hat{\bm\theta})\succeq\frac1\nu\JC^{-1}\succeq\frac1\nu\JQ^{-1}.
	\label{eq:matrix_cramer_rao}
\end{equation}
This is Eq.~\eqref{eq:qcrb} of the main text. Its $ii$ entry gives $\Var(\hat\theta_i)\geq[\JQ^{-1}]_{ii}/\nu$ for a target parameter $\theta_i$, as in Eq.~\eqref{eq:nuisance_bound} of the main text. Below, we express this bound through a reparameterization that leaves $\theta_i$ unchanged and construct a measurement that attains it.

Let $\mathcal J\succ0$ denote either the quantum or classical Fisher information matrix at the reference point. With $\theta_i$ placed first, write
\begin{equation}
	\mathcal J=\begin{pmatrix}
		a&\bm c^{\mathsf T}\\
		\bm c&B
	\end{pmatrix}.
	\label{eq:qfim_block}
\end{equation}
As in Appendix~\ref{sec:joint_direction_variance} of the End Matter, keep $\theta_i$ unchanged and define $\bm\theta'_{-i}=\bm\theta_{-i}+B^{-1}\bm c\,\theta_i$, where $\bm\theta_{-i}$ contains the remaining parameters and $B^{-1}\bm c$ is fixed at the reference point. The chain rule gives the Fisher information matrix in the new coordinates $(\theta_i,\bm\theta'_{-i})$, $\mathcal J'=\diag(\mathcal I_i,B)$, where $\mathcal I_i=a-\bm c^{\mathsf T}B^{-1}\bm c$. For any $\bm v=(1,\bm b^{\mathsf T})^{\mathsf T}$ in this ordering,
\begin{equation}
	\bm v^{\mathsf T}\mathcal J\bm v=\mathcal I_i+(\bm b+B^{-1}\bm c)^{\mathsf T}B(\bm b+B^{-1}\bm c)\geq\mathcal I_i.
	\label{eq:directional_info_decomposition}
\end{equation}
Equality is attained at $\bm b=-B^{-1}\bm c$. Since the reparameterization leaves $\theta_i$ unchanged, $[\mathcal J^{-1}]_{ii}=[\mathcal J'^{-1}]_{ii}=1/\mathcal I_i$. In the original parameter ordering, we therefore obtain~\cite{Gross2021OneFromMany}
\begin{equation}
	\mathcal I_i=\frac{1}{[\mathcal J^{-1}]_{ii}}=\min_{\bm v:v_i=1}\bm v^{\mathsf T}\mathcal J\bm v.
	\label{eq:effinfo}
\end{equation}
We call $\mathcal I_i$ the effective information about $\theta_i$, that is, the information that remains after accounting for the uncertainties of the other unknown parameters. It is also called the partial Fisher information~\cite{Suzuki2020Nuisance}.

Now take $\mathcal J=\JQ$. For an estimator $\hat\theta_i(y)$ based on a single outcome $y$, define $X=\sum_y[\hat\theta_i(y)-\tilde\theta_i]M_y$. Local unbiasedness at $\tilde{\bm\theta}$ requires, with $\rho=\rho_{\tilde{\bm\theta}}$,
\begin{equation}
	\Tr(\rho X)=0,\qquad
	\Tr[(\partial_j\rho)X]=\delta_{ij}.
	\label{eq:unbiased}
\end{equation}
The same reparameterization gives the SLD for $\theta_i$ in the new coordinates,
\begin{equation}
	L'_i=L_i-\bm c^{\mathsf T}B^{-1}\bm L_{-i},
	\label{eq:reparameterized_sld}
\end{equation}
where $\bm L_{-i}$ collects the SLDs for the remaining parameters. Since $\Tr[(\partial_j\rho)L_k]=[\JQ]_{jk}$, the block form of $\JQ$ gives $\Tr[(\partial_i\rho)L'_i]=a-\bm c^{\mathsf T}B^{-1}\bm c=\mathcal I_i$ and $\Tr[(\partial_j\rho)L'_i]=0$ for $j\ne i$. Also, $\Tr(\rho L'_i)=0$ because $\Tr(\rho L_j)=\Tr(\partial_j\rho)=0$. Thus the observable
\begin{equation}
	X_{\mathrm{eff}}=\frac{L'_i}{\mathcal I_i}=\sum_j[\JQ^{-1}]_{ji}L_j
	\label{eq:effective_observable}
\end{equation}
satisfies Eq.~\eqref{eq:unbiased} and has $\Tr(\rho X_{\mathrm{eff}}^2)=1/\mathcal I_i$. Measuring this observable, constructed at $\tilde{\bm\theta}$, and adding $\tilde\theta_i$ to its sample mean over $\nu$ interrogations gives a locally unbiased estimator with variance $1/(\nu\mathcal I_i)=[\JQ^{-1}]_{ii}/\nu$, attaining the bound~\cite{Suzuki2020Nuisance}.

For a singular Fisher information matrix $\mathcal J\succeq0$, positive semidefiniteness implies that $\bm c$ lies in the range of $B$. Replacing $B^{-1}$ by the Moore--Penrose inverse $B^+$ gives the same reparameterization and the decomposition in Eq.~\eqref{eq:directional_info_decomposition}, with $\mathcal I_i=a-\bm c^{\mathsf T}B^+\bm c$. If the $i$th coordinate vector $\bm e_i$ lies in the range of $\mathcal J$, then $\mathcal I_i>0$ and Eq.~\eqref{eq:effinfo} holds with $\mathcal J^{-1}$ replaced by $\mathcal J^+$. Otherwise $\mathcal I_i=0$, and the corresponding quantum or classical model admits no finite-variance locally unbiased estimator of $\theta_i$. When $\theta_i$ is estimable in the quantum model ($\mathcal I_i>0$), $X_{\mathrm{eff}}=\sum_j[\JQ^+]_{ji}L_j$ remains locally optimal.

Let $\JQ(t)$ denote the QFIM of the output state of an interrogation of duration $t$. With $\mathcal J=\JQ(t)$, Eqs.~\eqref{eq:matrix_cramer_rao} and~\eqref{eq:effinfo} give
\begin{equation}
	\frac{1}{\nu\Var(\hat\theta_i)}\leq\mathcal I_i=\min_{\bm v:v_i=1}\bm v^{\mathsf T}\JQ(t)\bm v,
	\label{eq:variance_information_relation}
\end{equation}
which is equivalent to Eq.~\eqref{eq:joint_direction_variance} of the main text. Measuring $X_{\mathrm{eff}}$ attains equality. Let $\Var_{\mathrm{opt}}(\hat\theta_i)$ be the infimum of the variance over all protocols of the general sequential scheme and over locally unbiased estimators of $\theta_i$, without requiring finite-variance estimates of the other parameters. Optimizing the attainable variance over protocols gives
\begin{equation}
	\frac{1}{\nu\Var_{\mathrm{opt}}(\hat\theta_i)}=\sup_{\mathrm{protocols}}\min_{\bm v:v_i=1}\bm v^{\mathsf T}\JQ(t)\bm v.
	\label{eq:optimal_variance_information}
\end{equation}
Since the optimal protocols for different target parameters may differ, these optimal variances need not be attainable simultaneously.

\subsection{Precision bound under arbitrary control}\label{sec:joint_explicit_gauges}
By Eq.~\eqref{eq:variance_information_relation}, any real vector $\bm v_i$ with $(\bm v_i)_i=1$ gives $1/[\nu\Var(\hat\theta_i)]\leq J_{Q,s}(t)$, where $J_{Q,s}(t)=\bm v_i^{\mathsf T}\JQ(t)\bm v_i$ is the QFI with respect to $s$ along the line $\bm\theta(s)=\bm\theta_0+s\bm v_i$ through the true values $\bm\theta_0$. On this line, $\partial_s=\partial_{\bm v_i}=\sum_\mu(\bm v_i)_\mu\partial_\mu$. The reparameterization $\theta'_i=\theta_i$ and $\theta'_\mu=\theta_\mu-(\bm v_i)_\mu\theta_i$ for $\mu\ne i$ makes $\partial_s$ the derivative with respect to $\theta_i$ with the other new coordinates held fixed, as in Sec.~\ref{sec:general_setting}. We now bound $J_{Q,s}(t)$ uniformly over all protocols of the general sequential scheme. The derivation holds for arbitrary real $\bm v_i$.

Purify the initial state and represent the controls, measurements, and feedback by parameter-independent unitaries on the sensor and noiseless ancillas~\cite{Demkowicz2017Adaptive,Kurdzialek2023Adaptiveness}. The measurement outcomes are then stored in ancilla registers, and the output state $\rho_{\bm\theta}$ is a reduced state of the resulting purification $\ket{\Psi_t}$. Its QFI is therefore bounded by that of $\ket{\Psi_t}$~\cite{Escher2011Framework},
\begin{equation}
	J_{Q,s}(t)\leq4\left(\norm{\partial_s\Psi_t}^2-|\langle\Psi_t|\partial_s\Psi_t\rangle|^2\right)\leq4\norm{\partial_s\Psi_t}^2.
	\label{eq:purification_qfi_sm}
\end{equation}
The right-hand side depends on how the evolution is purified. We fix the purification through the Kraus representation of each short-time channel and use this freedom to tighten the bound.

Over a short interval $\delta t$, the evolution generated by $\mathcal L_{\bm\theta}$ has Kraus operators
\begin{equation}
	\begin{aligned}
		K_{\mathrm{ch},0}&=I-\left(iH+\frac12\sum_kE_k^\dagger E_k\right)\delta t+O(\delta t^2),\\
		K_{\mathrm{ch},k}&=E_k\sqrt{\delta t}+O(\delta t^{3/2}),\qquad k\geq1.
	\end{aligned}
	\label{eq:kraus_short_time_sm}
\end{equation}
A Hermitian matrix $g_i$ acting on the Kraus index transforms $\bm K=(K_{\mathrm{ch},0},K_{\mathrm{ch},1},\ldots)^{\mathsf T}$ into an equivalent representation $\widetilde{\bm K}(s)=e^{isg_i}\bm K(s)$ of the same channel~\cite{Escher2011Framework,Demkowicz2017Adaptive}. At $s=0$, define
\begin{equation}
	\alpha_{i,\delta t}=(\partial_s\widetilde{\bm K})^\dagger\partial_s\widetilde{\bm K},\qquad
	\beta_{i,\delta t}=-i(\partial_s\widetilde{\bm K})^\dagger\widetilde{\bm K}.
	\label{eq:kraus_alpha_beta_sm}
\end{equation}
Here $\alpha_{i,\delta t}\succeq0$, and $\beta_{i,\delta t}$ is Hermitian because $\widetilde{\bm K}^\dagger\widetilde{\bm K}=I$ implies $\widetilde{\bm K}^\dagger\partial_s\widetilde{\bm K}=-(\partial_s\widetilde{\bm K})^\dagger\widetilde{\bm K}$. For a pure input $\ket\psi$ and an orthonormal environment basis $\{\ket\ell_E\}$, the channel output has the purification $\ket{\widetilde\Psi_s}=\sum_\ell\widetilde K_{\mathrm{ch},\ell}(s)\ket\psi\otimes\ket\ell_E=(I\otimes e^{isg_i})\sum_\ell K_{\mathrm{ch},\ell}(s)\ket\psi\otimes\ket\ell_E$, with $\norm{\partial_s\widetilde\Psi_s}^2=\bra\psi\alpha_{i,\delta t}\ket\psi$ and $\langle\widetilde\Psi_s|\partial_s\widetilde\Psi_s\rangle=-i\bra\psi\beta_{i,\delta t}\ket\psi$ at $s=0$. The matrix $g_i$ thus acts only on the environment, but it changes $\alpha_{i,\delta t}$ and $\beta_{i,\delta t}$.

For the short-time channel, choose
\begin{equation}
	g_i=\begin{pmatrix}
		a_i\delta t&\sqrt{\delta t}\,\bm b_i^\dagger\\
		\sqrt{\delta t}\,\bm b_i&h_i
	\end{pmatrix},
	\label{eq:kraus_unitary_sm}
\end{equation}
where the coefficients $a_i\in\mathbb R$, $\bm b_i=(b_{i1},b_{i2},\ldots)^{\mathsf T}$, and $h_i=h_i^\dagger$ are finite and independent of $\delta t$. At $s=0$, $\widetilde{\bm K}=\bm K$ and $\partial_s\widetilde{\bm K}=\partial_s\bm K+ig_i\bm K$, so $\beta_{i,\delta t}=-i(\partial_s\bm K)^\dagger\bm K-\bm K^\dagger g_i\bm K$. Differentiating Eq.~\eqref{eq:kraus_short_time_sm} along $\bm\theta(s)$ gives the first term,
\begin{equation}
	-i(\partial_s\bm K)^\dagger\bm K
	=\left[\partial_sH+\frac i2\sum_k\bigl(E_k^\dagger\partial_sE_k-(\partial_sE_k)^\dagger E_k\bigr)\right]\delta t+O(\delta t^2)
	=\sum_\mu(\bm v_i)_\mu\hat H_\mu\,\delta t+O(\delta t^2),
	\label{eq:kraus_coherent_term}
\end{equation}
where
\begin{equation}
	\hat H_\mu=\partial_\mu H+
	\frac{i}{2}\sum_k\left[E_k^\dagger\partial_\mu E_k-(\partial_\mu E_k)^\dagger E_k\right]
	\label{eq:suppHhat}
\end{equation}
is the coherent contribution of $\theta_\mu$~\cite{Chen2024Informative,Das2025Scalings}. For the model in Eq.~\eqref{eq:lindblad} of the main text, $\hat H_{\omega_a}=\partial_{\omega_a}H$ and $\hat H_{\eta_b}=C_{\eta_b}$, as in Eq.~\eqref{eq:Hhat} of the main text. In the second term, $\bm K^\dagger g_i\bm K=S_i\delta t+O(\delta t^2)$, with
\begin{equation}
	S_i=a_iI+\sum_k(b_{ik}^*E_k+b_{ik}E_k^\dagger)+\sum_{kl}(h_i)_{kl}E_k^\dagger E_l\in\mathcal S,
	\label{eq:joint_span_sm}
\end{equation}
where \mbox{$\mathcal S=\operatorname{span}\{I,E_k,E_k^\dagger,E_k^\dagger E_l\}$} is the Lindblad span and $\operatorname{span}$ denotes the complex linear span. Together, the two terms give
\begin{equation}
	\beta_{i,\delta t}=\mathcal R_i\delta t+O(\delta t^2),\qquad
	\mathcal R_i=\sum_\mu(\bm v_i)_\mu\hat H_\mu-S_i.
	\label{eq:kraus_beta_expansion_sm}
\end{equation}
The derivatives of the transformed Kraus operators satisfy $\partial_s\widetilde K_{\mathrm{ch},0}=O(\delta t)$ and $\partial_s\widetilde K_{\mathrm{ch},k}=\sqrt{\delta t}\,F_{i,k}+O(\delta t^{3/2})$ for $k\geq1$, which gives
\begin{equation}
	\alpha_{i,\delta t}=A_i\delta t+O(\delta t^2),\qquad
	A_i=\sum_kF_{i,k}^\dagger F_{i,k},\qquad
	F_{i,k}=\partial_{\bm v_i}E_k+ib_{ik}I+i\sum_l(h_i)_{kl}E_l.
	\label{eq:joint_jump_sm}
\end{equation}
Thus $\mathcal R_i$ generates the coherent part and $A_i$ describes the dissipative part of the parameter shift along $\bm v_i$. Both are fixed by $\bm v_i$ and the coefficients of $g_i$. The coefficient $a_i$ shifts $\mathcal R_i$ by $-a_iI$ but does not affect $A_i$. Conversely, every Hermitian operator in $\mathcal S$ can be written in the form of Eq.~\eqref{eq:joint_span_sm} with finite coefficients, so it can be realized as $S_i$ with finite $A_i$. Different choices of $g_i$ thus give different valid bounds below, and Sec.~\ref{sec:jointscaling} optimizes over them. An exact Kraus representation of the short-time channel may contain additional Kraus operators of higher order in $\delta t$. They can be included by extending $g_i$ with zero rows and columns, and they enter $\alpha_{i,\delta t}$ and $\beta_{i,\delta t}$ only at order $\delta t^2$.

We now combine the short-time channels of a sequential protocol. For the $j$th interval, of duration $\delta t_j$, use a fresh environment $E_j$ and the isometry
\begin{equation}
	V_{\mathrm{ch},j}=\sum_\ell\widetilde K_{\mathrm{ch},\ell,j}\otimes\ket\ell_{E_j}.
	\label{eq:channel_isometry_sm}
\end{equation}
The controls act trivially on these environments. Write $\alpha_{i,j}=\alpha_{i,\delta t_j}$ and $\beta_{i,j}=\beta_{i,\delta t_j}$. Decompose the derivative of $V_{\mathrm{ch},j}$ into a component in its image and an orthogonal component:
\begin{equation}
	\partial_sV_{\mathrm{ch},j}=V_{\mathrm{ch},j}D_j+Z_j,\qquad
	D_j=V_{\mathrm{ch},j}^\dagger\partial_sV_{\mathrm{ch},j}=-i\beta_{i,j},\qquad
	V_{\mathrm{ch},j}^\dagger Z_j=0.
	\label{eq:channel_derivative_split_sm}
\end{equation}
Equation~\eqref{eq:kraus_alpha_beta_sm} gives
\begin{equation}
	Z_j^\dagger Z_j=(\partial_sV_{\mathrm{ch},j})^\dagger\partial_sV_{\mathrm{ch},j}-D_j^\dagger D_j
	=\alpha_{i,j}-\beta_{i,j}^2\succeq0.
	\label{eq:orthogonal_derivative_norm_sm}
\end{equation}
Let $U_j$ be the control following interval $j$, $W_j=U_jV_{\mathrm{ch},j}$, and $W_{>j}$ the product of the subsequent $W_k$. If $\ket{\Psi_{j-1}}$ is the state before interval $j$, differentiating the full sequence gives
\begin{equation}
	\partial_s\ket{\Psi_t}=\sum_jW_{>j}U_j(V_{\mathrm{ch},j}D_j+Z_j)\ket{\Psi_{j-1}}.
	\label{eq:purified_derivative_sum_sm}
\end{equation}
The $Z$ terms of different intervals are orthogonal. For $j<k$, $W_{>j}=W_{>k}W_k\cdots W_{j+1}$, so their overlap contains the factor $W_k^\dagger U_kZ_k=V_{\mathrm{ch},k}^\dagger Z_k=0$. The $Z$ terms therefore add in quadrature, whereas the $D$ terms can add coherently. Since $W_{>j}$, $U_j$, and $V_{\mathrm{ch},j}$ are isometries, the triangle inequality gives
\begin{equation}
	\begin{aligned}
		\norm{\partial_s\Psi_t}
		&\leq\sum_j\norm{D_j\ket{\Psi_{j-1}}}+\Bigl(\sum_j\norm{Z_j\ket{\Psi_{j-1}}}^2\Bigr)^{1/2}\\
		&=\sum_j\sqrt{\Tr(\rho_j\beta_{i,j}^2)}+\sqrt{\sum_j\Tr[\rho_j(\alpha_{i,j}-\beta_{i,j}^2)]},
	\end{aligned}
	\label{eq:purified_derivative_finite_sm}
\end{equation}
where $\rho_j$ is the reduced state of the sensor in $\ket{\Psi_{j-1}}$.

Since the measurement outcomes are stored in ancillas, $\rho_j$ is the record-averaged sensor state $\rho_\tau=\sum_{\bm r}p_\tau(\bm r)\rho_{\tau|\bm r}$ at the start of interval $j$. Here $\bm r$ collects the outcomes up to time $\tau$, $p_\tau(\bm r)$ is their probability, and $\rho_{\tau|\bm r}$ is the conditional sensor state after the corresponding feedback. This average enters only through expectation values of sensor operators, while the estimation still uses the full records. For a fixed protocol, $\rho_\tau$ is evaluated at $\bm\theta_0$ and is independent of $\bm v_i$ and $g_i$, since $e^{isg_i}$ acts only on the environment.

Fix $t$, $\bm v_i$, and the finite coefficients of $g_i$ before taking $\max_j\delta t_j\to0$. Equations~\eqref{eq:kraus_beta_expansion_sm} and~\eqref{eq:joint_jump_sm} give $\beta_{i,j}=\mathcal R_i\delta t_j+O(\delta t_j^2)$ and $\alpha_{i,j}-\beta_{i,j}^2=A_i\delta t_j+O(\delta t_j^2)$, so Eq.~\eqref{eq:purified_derivative_finite_sm} becomes
\begin{equation}
	\norm{\partial_s\Psi_t}\leq\int_0^t\sqrt{\Tr(\rho_\tau\mathcal R_i^2)}\,d\tau
	+\sqrt{\int_0^t\Tr(\rho_\tau A_i)\,d\tau}.
	\label{eq:purified_derivative_integral_sm}
\end{equation}
This limit is uniform over the controls. Replacing $\beta_{i,j}$ by $\mathcal R_i\delta t_j$ and $\alpha_{i,j}-\beta_{i,j}^2$ by $A_i\delta t_j$ changes the first sum in Eq.~\eqref{eq:purified_derivative_finite_sm} and the quantity under its second square root by $O(\sum_j\delta t_j^2)=O(t\max_j\delta t_j)$, and $|\sqrt{x}-\sqrt{y}|\leq\sqrt{|x-y|}$ bounds the resulting change of the square root. Since $\rho_\tau$ changes at a bounded rate between control operations, replacing the sums by integrals costs $O(t\sqrt{\max_j\delta t_j})$ in the first term and $O(t\max_j\delta t_j)$ under the second square root. The constants depend on $\bm v_i$, the coefficients of $g_i$, and the values and derivatives of $H$ and $E_k$ at $\bm\theta_0$, but not on the controls. Refining each free-evolution interval and taking limits of finite sequential protocols then extends Eq.~\eqref{eq:purified_derivative_integral_sm} to every protocol of the general sequential scheme.

Finally, with the time-averaged sensor state $\bar\rho=t^{-1}\int_0^t\rho_\tau\,d\tau$, the Cauchy--Schwarz inequality gives
\begin{equation}
	\int_0^t\sqrt{\Tr(\rho_\tau\mathcal R_i^2)}\,d\tau
	\leq\sqrt{t\int_0^t\Tr(\rho_\tau\mathcal R_i^2)\,d\tau}
	=t\sqrt{\Tr(\bar\rho\mathcal R_i^2)},
	\label{eq:time_average_cauchy_schwarz}
\end{equation}
and the second term in Eq.~\eqref{eq:purified_derivative_integral_sm} equals $\sqrt{t\Tr(\bar\rho A_i)}$. Equation~\eqref{eq:purification_qfi_sm} then yields the precision bound
\begin{equation}
	J_{Q,s}(t)\leq4\left[t\sqrt{\Tr(\bar\rho\mathcal R_i^2)}+\sqrt{t\Tr(\bar\rho A_i)}\right]^2.
	\label{eq:directionalstatebound}
\end{equation}
Bounding the expectation values by operator norms, $\Tr(\bar\rho\mathcal R_i^2)\leq\norm{\mathcal R_i}_\infty^2$ and $\Tr(\bar\rho A_i)\leq\norm{A_i}_\infty$, removes the dependence on the protocol,
\begin{equation}
	J_{Q,s}(t)\leq4\left[t\norm{\mathcal R_i}_\infty+\sqrt{t\norm{A_i}_\infty}\right]^2.
	\label{eq:purification_channel_bound}
\end{equation}
The coherent part $\mathcal R_i$ thus contributes information quadratically in $t$, whereas the dissipative part $A_i$ contributes only linearly. We summarize these results as follows.

\begin{lemma}[Precision bound under arbitrary control]\label{lem:purification_bound}
	Fix a real vector $\bm v_i$ and a matrix $g_i$ of the form in Eq.~\eqref{eq:kraus_unitary_sm}, and let $\mathcal R_i$ and $A_i$ be given by Eqs.~\eqref{eq:kraus_beta_expansion_sm} and~\eqref{eq:joint_jump_sm}. For any protocol of the general sequential scheme, the QFI $J_{Q,s}(t)=\bm v_i^{\mathsf T}\JQ(t)\bm v_i$ along $\bm v_i$ obeys Eq.~\eqref{eq:directionalstatebound}, with $\bar\rho$ the time-averaged sensor state of the protocol, and the protocol-independent bound~\eqref{eq:purification_channel_bound}.
\end{lemma}

For the model of the main text, take $\theta_i=\omega_a$ with $(\bm v_i)_{\omega_a}=1$. Then $\mathcal R_i=\partial_{\omega_a}H-K_a(\bm v_i,S_i)$, with $K_a$ defined in Eq.~\eqref{eq:target_K_definition} of the main text, and $A_i$ is the operator $A_a$ defined through Eq.~\eqref{eq:joint_jump_em} of the End Matter. Equation~\eqref{eq:purification_channel_bound} is therefore Eq.~\eqref{eq:target_purification_bound} of the main text, and Eq.~\eqref{eq:directionalstatebound} is the state-dependent refinement mentioned at the end of Appendix~\ref{sec:purification_bound} of the End Matter.

\subsection{Optimality in the spin model}\label{sec:spin_optimality}
We now apply Lemma~\ref{lem:purification_bound} to the spin model of Sec.~\ref{sec:finite} and compare the protocol of Sec.~\ref{sec:toybridge} with all protocols of the general sequential scheme. We show that this protocol attains the optimal covariance matrix in the long-time limit, that calibrating $\phi$ first does not improve it at fixed total time, and that it is also optimal for an unknown noise rate.

With $E=\sqrt\kappa\,\sigma_\phi$ and $\partial_\omega H=\sigma_z/2$, Eq.~\eqref{eq:suppHhat} gives $C_\phi=-\kappa\sigma_z=-2\kappa\,\partial_\omega H$. Changing $\phi$ by $\delta\phi$ thus has the same coherent effect as changing $\omega$ by $-2\kappa\,\delta\phi$, and the two effects cancel along $\bm d=(2\kappa,1)^{\mathsf T}$ in the coordinates $(\omega,\phi)$. For $g_i=0$ in Eq.~\eqref{eq:kraus_unitary_sm}, the operators of Lemma~\ref{lem:purification_bound} along $\bm d$ are
\begin{equation}
	\mathcal R=2\kappa\,\partial_\omega H+C_\phi=0,\qquad
	A=(\partial_\phi E)^\dagger\partial_\phi E=\kappa I,
	\label{eq:spin_direction_operators}
\end{equation}
so Eq.~\eqref{eq:purification_channel_bound} bounds the QFI along $\bm d$ per interrogation under arbitrary control,
\begin{equation}
	\bm d^{\mathsf T}\JQ\bm d\leq4\kappa t.
	\label{eq:spin_qfi_bounds}
\end{equation}
The protocol of Sec.~\ref{sec:toybridge} saturates this bound, since its QFIM in Eq.~\eqref{eq:toy-QFIM} gives $\bm d^{\mathsf T}\JQ\bm d=4\kappa t$.

Equation~\eqref{eq:spin_qfi_bounds} limits the estimation of both parameters. For $\bm v=\bm d/(2\kappa)=(1,1/(2\kappa))^{\mathsf T}$, it gives $\bm v^{\mathsf T}\JQ\bm v\leq t/\kappa$, so Eq.~\eqref{eq:variance_information_relation} yields Eq.~\eqref{eq:toyallcontrol} of the main text, $\nu\Var(\hat\omega)\geq\kappa/t$, which holds even when $\phi$ is not estimated. For jointly locally unbiased estimators of $\bm\theta=(\omega,\phi)^{\mathsf T}$, the Cauchy--Schwarz inequality $(\bm u^{\mathsf T}\bm d)^2\leq(\bm u^{\mathsf T}\JQ^{-1}\bm u)(\bm d^{\mathsf T}\JQ\bm d)$ for every real $\bm u$ gives $\JQ^{-1}\succeq\bm d\bm d^{\mathsf T}/(\bm d^{\mathsf T}\JQ\bm d)$. Equation~\eqref{eq:matrix_cramer_rao} therefore yields
\begin{equation}
	\nu\Cov(\hat{\bm\theta})\succeq\frac{\bm d\bm d^{\mathsf T}}{4\kappa t}
	=\begin{pmatrix}\kappa/t&1/(2t)\\1/(2t)&1/(4\kappa t)\end{pmatrix}.
	\label{eq:spin_covariance_bound}
\end{equation}
The protocol of Sec.~\ref{sec:toybridge} has $\nu\Cov(\hat{\bm\theta})=\JC^{-1}$ [Eq.~\eqref{eq:toy-CFIM-inverse}], which exceeds this bound only in the $\omega\omega$ entry,
\begin{equation}
	\nu\Cov(\hat{\bm\theta})=\frac{\bm d\bm d^{\mathsf T}}{4\kappa t}+\begin{pmatrix}1/t^2&0\\0&0\end{pmatrix}.
	\label{eq:spin_covariance_attained}
\end{equation}
The excess $1/t^2$ is $\nu\Var(\hat\omega)$ for known $\phi$ and is negligible compared with $\kappa/t$ at long times. The protocol therefore attains the optimal covariance matrix in the long-time limit, with $\lim_{t\to\infty}\nu t\Var_{\mathrm{opt}}(\hat\omega)=\kappa$, and its variance $\Var(\hat\phi)=1/(4\nu\kappa t)$ is optimal at every $t$.

\paragraph{Joint estimation with prior calibration.}
Suppose a calibration before sensing provides Fisher information $J_{\mathrm{cal}}$ about $\phi$. Adding $\diag(0,J_{\mathrm{cal}})$ to the CFIM $\nu\JC$ of the joint estimation task in Eqs.~\eqref{eq:toy-QFIM} and~\eqref{eq:toy-CFIM} gives
\begin{equation}
	\Var(\hat\omega)\geq\frac{1}{\nu t^2}+\frac{4\kappa^2}{J_{\mathrm{cal}}+4\kappa\nu t},
	\label{eq:toy_precalibration_variance}
\end{equation}
which approaches the value $1/(\nu t^2)$ for known $\phi$ only if $J_{\mathrm{cal}}\gg4\kappa\nu t$. If the calibration uses the same sensor for a time $T_{\mathrm{cal}}$, however, calibration and sensing form a single sequential protocol of total time $T_{\mathrm{tot}}=T_{\mathrm{cal}}+\nu t$. Applying Eq.~\eqref{eq:spin_qfi_bounds} to the calibration stage and to the whole protocol gives $J_{\mathrm{cal}}\leq4\kappa T_{\mathrm{cal}}$ and, as for Eq.~\eqref{eq:toyallcontrol} of the main text, $\Var(\hat\omega)\geq\kappa/T_{\mathrm{tot}}$. The protocol of Sec.~\ref{sec:toybridge} attains this leading term without calibration, so calibrating $\phi$ first cannot improve it at fixed total time.

\paragraph{Unknown noise rate.}
For joint estimation of $(\omega,\kappa)$ with known $\phi$, the protocol of Sec.~\ref{sec:toybridge} is optimal for both parameters. For $\omega$, $\mathcal R=\partial_\omega H=\sigma_z/2$ and $A=0$, so Lemma~\ref{lem:purification_bound} bounds the QFI for $\omega$ by $t^2$. For $\kappa$, $C_\kappa=0$, so $\mathcal R=0$ and $A=(\partial_\kappa E)^\dagger\partial_\kappa E=I/(4\kappa)$, which bounds the QFI for $\kappa$ by $t/\kappa$. The CFIM in Eq.~\eqref{eq:rate_joint_cfim} attains both bounds. Unlike an unknown direction, an unknown rate therefore leaves the optimal frequency precision unchanged, consistent with Sec.~\ref{sec:fixed_error_space}.

\section{Precision scaling and optimal coefficients (Theorem 1)}\label{sec:jointscaling}
This section proves Theorem~\ref{thm:joint_scaling} of the main text. Section~\ref{sec:general_scaling} proves Lemma~\ref{lem:general_scaling} for a general target parameter, which may be a noise parameter or enter both $H$ and the noise operators, and obtains Theorem~\ref{thm:joint_scaling} as its special case for a Hamiltonian parameter. It also bounds the coefficient of $t$ when Heisenberg scaling is lost. Section~\ref{sec:scaling_applications} shows how unknown noise parameters change the optimal quadratic coefficient.

\subsection{Proof of Theorem 1}\label{sec:general_scaling}
We first prove Lemma~\ref{lem:general_scaling} for a target parameter $\theta_i$ that may enter $H$, the noise operators, or both. Theorem~\ref{thm:joint_scaling} of the main text, restated at the end of this subsection, is its special case for a Hamiltonian parameter. The definitions below involve each parameter $\theta_\mu$ only through its coherent contribution $\hat H_\mu$ in Eq.~\eqref{eq:suppHhat}. For a real vector $\bm v$ with $v_i=1$ and a Hermitian $S$ in the Lindblad span $\mathcal S=\operatorname{span}\{I,E_k,E_k^\dagger,E_k^\dagger E_l\}$, let
\begin{equation}
	K_i(\bm v,S)=S-\sum_{\mu\ne i}v_\mu\hat H_\mu,\qquad
	\mathcal N_i=\mathcal S+\operatorname{span}\{\hat H_\mu\}_{\mu\ne i},\qquad
	d_i=\min_{K=K^\dagger\in\mathcal N_i}\norm{\hat H_i-K}_\infty.
	\label{eq:individualdistance}
\end{equation}

\begin{lemma}\label{lem:general_scaling}
	In the general sequential scheme, a locally unbiased estimator of $\theta_i$ can attain Heisenberg scaling, $\Var(\hat\theta_i)=O(1/(\nu t^2))$, if and only if
	\begin{equation}
		\hat H_i\notin\mathcal N_i.
		\label{eq:jointcriterion}
	\end{equation}
	Otherwise, some $\bm v_i$ with $(\bm v_i)_i=1$ and some $g_i$ in Lemma~\ref{lem:purification_bound} give $\mathcal R_i=0$, and every such choice gives
	\begin{equation}
		\frac{1}{\nu\Var_{\mathrm{opt}}(\hat\theta_i)}\leq4t\norm{A_i}_\infty,
		\label{eq:linear_submodel}
	\end{equation}
	limiting the precision to the scaling of the standard quantum limit (SQL). When Eq.~\eqref{eq:jointcriterion} holds, the optimal variance satisfies
	\begin{equation}
		\lim_{t\to\infty}\nu t^2\Var_{\mathrm{opt}}(\hat\theta_i)=\frac{1}{4d_i^2}.
		\label{eq:general_optimal_variance}
	\end{equation}
	All parameters satisfying Eq.~\eqref{eq:jointcriterion} can attain Heisenberg scaling in one common protocol.
\end{lemma}

\paragraph{Upper bounds.}
By Eq.~\eqref{eq:optimal_variance_information}, every real vector $\bm v$ with $v_i=1$ gives $1/[\nu\Var_{\mathrm{opt}}(\hat\theta_i)]\leq\sup\bm v^{\mathsf T}\JQ(t)\bm v$, where the supremum is over all protocols of the general sequential scheme. To bound the QFI along $\bm v$, we apply Lemma~\ref{lem:purification_bound} with $\bm v_i=\bm v$ and a matrix $g_i$ whose operator $S_i$ in Eq.~\eqref{eq:joint_span_sm} equals a chosen Hermitian $S\in\mathcal S$. Such a $g_i$ exists for every Hermitian $S\in\mathcal S$, as shown in Sec.~\ref{sec:joint_explicit_gauges}. Equation~\eqref{eq:kraus_beta_expansion_sm} then gives
\begin{equation}
	\mathcal R_i=\hat H_i+\sum_{\mu\ne i}v_\mu\hat H_\mu-S=\hat H_i-K_i(\bm v,S),
	\label{eq:residual_target}
\end{equation}
and the protocol-independent bound~\eqref{eq:purification_channel_bound} yields
\begin{equation}
	\frac{1}{\nu\Var_{\mathrm{opt}}(\hat\theta_i)}\leq4\left[t\norm{\hat H_i-K_i}_\infty+\sqrt{t\norm{A_i}_\infty}\right]^2
	\label{eq:theorem1_upper}
\end{equation}
for every $\bm v$ and $S$. As $\bm v$ and $S$ vary, $K_i$ ranges over all Hermitian operators in $\mathcal N_i$. Indeed, a Hermitian $K\in\mathcal N_i$ equals its Hermitian part, which has the form of $K_i$ because $\mathcal S$ is spanned by Hermitian operators and the operators $\hat H_\mu$ are Hermitian.

If $\hat H_i\in\mathcal N_i$, some $\bm v$ and $S$ give $K_i=\hat H_i$, that is, $\mathcal R_i=0$. Equation~\eqref{eq:theorem1_upper} then reduces to Eq.~\eqref{eq:linear_submodel}, which rules out Heisenberg scaling. If $\hat H_i\notin\mathcal N_i$, the smallest value of $\norm{\hat H_i-K_i}_\infty$ over $\bm v$ and $S$ is $d_i>0$. This minimum is attained because $\mathcal N_i$ is finite-dimensional: a minimizing sequence is bounded and, after linearly dependent operators are removed from the spanning set, has a limit with finite coefficients. Inserting the minimizing $\bm v$ and $S$ into Eq.~\eqref{eq:theorem1_upper} and taking $t\to\infty$ gives
\begin{equation}
	\limsup_{t\to\infty}\frac{1}{\nu t^2\Var_{\mathrm{opt}}(\hat\theta_i)}\leq4d_i^2.
	\label{eq:single_coefficient_upper}
\end{equation}

\paragraph{Attainability.}
For $d_i>0$, the protocol of Appendix~\ref{sec:optimal_coefficient} of the End Matter attains the bound~\eqref{eq:single_coefficient_upper}, with $\theta_i$, $\hat H_i$, and $\mathcal N_i$ in place of $\omega_a$, $\partial_{\omega_a}H$, and $\mathcal N_a$ there and in Eq.~\eqref{eq:code_conditions} of the main text. Following that section, we first derive the logical dynamics of a code that corrects the noise, for later use in Secs.~\ref{sec:noise_reference} and~\ref{sec:noise_modulation}, and then choose the code.

Let $V=\ket{C_0}\bra0+\ket{C_1}\bra1$ encode a logical qubit in the sensor and a noiseless ancilla, with projector $P=VV^\dagger$ and state $\rho=V\rho_CV^\dagger$. The code, the recovery, and the control are fixed at the prior estimate $\tilde{\bm\theta}$, where the code satisfies the Knill--Laflamme conditions
\begin{equation}
	V^\dagger E_kV=\epsilon_kI_C,\qquad V^\dagger E_k^\dagger E_lV=\Lambda_{kl}I_C.
	\label{eq:code_kl_explicit}
\end{equation}
Here $I_C$ is the logical identity. The errors $M_k=(I-P)E_kP$ then obey $M_k^\dagger M_l=\Gamma_{kl}P$ with $\Gamma_{kl}=\Lambda_{kl}-\epsilon_k^*\epsilon_l$. A fixed unitary mixing $E_k\mapsto\sum_lu_{lk}E_l$ with $u^\dagger\Gamma u=\diag(\chi_k)$ leaves the master equation, $\mathcal S$, and every $\hat H_\mu$ unchanged and, with the same notation for the transformed operators, gives $M_k^\dagger M_l=\chi_k\delta_{kl}P$. The recovery $\mathcal R_{\mathrm{rec}}$ has Kraus operators $R_0=P$ and $R_k=PM_k^\dagger/\sqrt{\chi_k}$ for $\chi_k>0$, completed to a trace-preserving map into the code space by operators that vanish on the code space and on the ranges of the $M_k$. Hence all Kraus operators except $R_0$ annihilate the code space, the completing ones annihilate every $E_kV$, and
\begin{equation}
	R_0E_kV=\epsilon_kV,\qquad R_jE_kV=\delta_{jk}\sqrt{\chi_k}V\quad(\chi_j>0).
	\label{eq:recovery_identities}
\end{equation}

A correction cycle evolves $\rho$ for a time $\delta t$ under Eq.~\eqref{eq:general_master}, with $H$ and $E_k$ at their true values and the control Hamiltonian $-H(\tilde{\bm\theta})$ added, and then applies $\mathcal R_{\mathrm{rec}}$. Since only $R_0=P$ contributes to the commutator and the anticommutators, one cycle gives $\rho_C(\tau+\delta t)=\rho_C(\tau)+\delta t\,\mathcal L^C(\bm\theta)(\rho_C)+O(\delta t^2)$ with
\begin{equation}
	\mathcal L^C(\bm\theta)(\rho_C)=-i\bigl[V^\dagger(H-H(\tilde{\bm\theta}))V,\rho_C\bigr]+\sum_k\Bigl(V^\dagger\mathcal R_{\mathrm{rec}}[E_k\rho E_k^\dagger]V-\frac12\bigl\{V^\dagger E_k^\dagger E_kV,\rho_C\bigr\}\Bigr).
	\label{eq:logical_generator}
\end{equation}
In the rapid-correction limit $\delta t\to0$ at fixed $t$, $\mathcal L^C$ generates the logical evolution. At $\tilde{\bm\theta}$, the commutator vanishes and Eq.~\eqref{eq:recovery_identities} gives $V^\dagger\mathcal R_{\mathrm{rec}}[E_k\rho E_k^\dagger]V=(|\epsilon_k|^2+\chi_k)\rho_C=\Lambda_{kk}\rho_C$, which cancels the anticommutator, so $\mathcal L^C(\tilde{\bm\theta})=0$.

For any parameter $\theta_\mu$, let $D_{\mu,k}=\partial_\mu E_k$ and $Q_\mu=V^\dagger\sum_kE_k^\dagger D_{\mu,k}V$ at $\tilde{\bm\theta}$. Since $\epsilon_k^*V^\dagger=V^\dagger E_k^\dagger P$ and $(I-P)E_kV=0$ whenever $\chi_k=0$, the contributions of $R_0$ and $R_k$ in Eq.~\eqref{eq:recovery_identities} combine into
\begin{equation}
	V^\dagger\mathcal R_{\mathrm{rec}}\Bigl[\sum_kD_{\mu,k}\rho E_k^\dagger\Bigr]V=\sum_kV^\dagger E_k^\dagger\bigl[P+(I-P)\bigr]D_{\mu,k}V\rho_C=Q_\mu\rho_C.
	\label{eq:recovered_noise_terms}
\end{equation}
The conjugate terms give $\rho_CQ_\mu^\dagger$, and the anticommutators give $-\{Q_\mu+Q_\mu^\dagger,\rho_C\}/2$. Adding the Hamiltonian derivative and using Eq.~\eqref{eq:suppHhat}, we obtain
\begin{equation}
	\partial_\mu\mathcal L^C(\rho_C)=-i[V^\dagger\partial_\mu HV,\rho_C]+\Bigl[\frac{Q_\mu-Q_\mu^\dagger}{2},\rho_C\Bigr]=-i[V^\dagger\hat H_\mu V,\rho_C].
	\label{eq:recoverderivative}
\end{equation}
To first order in $\delta\bm\theta=\bm\theta-\tilde{\bm\theta}$, the logical state thus evolves under the Hamiltonian $\sum_\mu\delta\theta_\mu V^\dagger\hat H_\mu V$, as in Eq.~\eqref{eq:corrected_local_generator_em} of the End Matter.

We now choose the code. Let $\rho_0$ and $\rho_1$ be the reduced states of the codewords on the sensor, so that $\langle O\rangle_0=\Tr(\rho_0O)$ and $\langle O\rangle_1=\Tr(\rho_1O)$, and let $Y=(\rho_0-\rho_1)/2$, with trace norm $\norm{Y}_1\leq1$. The first condition in Eq.~\eqref{eq:code_conditions} of the main text requires $Y\perp\mathcal N_i$, that is, $\Tr(YO)=0$ for all $O\in\mathcal N_i$, so for every Hermitian $K\in\mathcal N_i$,
\begin{equation}
	\langle\hat H_i\rangle_0-\langle\hat H_i\rangle_1=2\Tr[Y(\hat H_i-K)]\leq2\norm{\hat H_i-K}_\infty,
	\label{eq:splitting_bound}
\end{equation}
and this difference is at most $2d_i$. Norm duality shows that this bound is tight. For a space $\mathcal N$ of operators that contains $I$ and is spanned by Hermitian operators, and a Hermitian operator $X$,
\begin{equation}
	\min_{K=K^\dagger\in\mathcal N}\norm{X-K}_\infty=\max_{\substack{Y=Y^\dagger,\ \norm{Y}_1\leq1\\Y\perp\mathcal N}}\Tr(YX),
	\label{eq:dual}
\end{equation}
where the inequality $\geq$ follows as in Eq.~\eqref{eq:splitting_bound} and equality holds by strong duality, since $-dI\prec X-K\prec dI$ is strictly feasible for sufficiently large $d$. For $\mathcal N=\mathcal N_i$ and $X=\hat H_i$, an optimizer $Y_\star$ has $\Tr(Y_\star\hat H_i)=d_i$ and, since $d_i>0$ and $I\in\mathcal N_i$, $\norm{Y_\star}_1=1$ and $\Tr Y_\star=0$. Purifying the density matrices $\rho_\pm=|Y_\star|\pm Y_\star$ in orthogonal subspaces of a noiseless ancilla gives codewords with $\rho_0=\rho_+$ and $\rho_1=\rho_-$, whose encoding obeys
\begin{equation}
	V^\dagger OV=\Tr(|Y_\star|O)I_C+\Tr(Y_\star O)\sigma_z^C
	\label{eq:optimal_code_compression}
\end{equation}
for any sensor operator $O$. Since $Y_\star\perp\mathcal N_i\supseteq\mathcal S$ and $\Tr(Y_\star\hat H_i)=d_i$, this code satisfies Eq.~\eqref{eq:code_kl_explicit} and Eq.~\eqref{eq:code_conditions} of the main text. Equation~\eqref{eq:recoverderivative} then gives the logical Hamiltonian $d_i\,\delta\theta_i\sigma_z^C$ up to a multiple of $I_C$, as in Eq.~\eqref{eq:logical_evolution} of the main text.

Prepare $(\ket{C_0}+\ket{C_1})/\sqrt2$ and measure $\sigma_y^C$ after time $t$. The outcomes $z=\pm1$ occur with probabilities
\begin{equation}
	p(z|\tilde{\bm\theta}+\delta\bm\theta)=\frac12[1+2zd_it\,\delta\theta_i]+O(\norm{\delta\bm\theta}^2),
	\label{eq:hs_probabilities}
\end{equation}
which depend on the other parameters only at second order. The estimator $\hat\theta_i=\tilde\theta_i+\bar z/(2d_it)$, with $\bar z$ the mean of the $\nu$ outcomes, is therefore locally unbiased and has variance $1/(4\nu d_i^2t^2)$. This saturates Eq.~\eqref{eq:single_coefficient_upper} and proves Eq.~\eqref{eq:general_optimal_variance}. The rapid-correction limit and the limit of many interrogations, in which the prior estimates approach $\bm\theta_0$, are taken at fixed $t$, before $t\to\infty$. The prior estimates can be refined adaptively with a vanishing fraction of the interrogations, for example with the auxiliary experiment described below, and Sec.~\ref{sec:toy_simulation} demonstrates such adaptive localization in the spin model.

\paragraph{Simultaneous Heisenberg scaling.}
We now prove the last statement of Lemma~\ref{lem:general_scaling}. Let $\mathcal T$ be the set of parameters satisfying Eq.~\eqref{eq:jointcriterion}, and use the optimal code for each $\theta_i$, $i\in\mathcal T$, in a fraction $w_i>0$ of the interrogations. The resulting estimates come from independent data and have covariance
\begin{equation}
	\Cov(\hat{\bm\theta}_{\mathcal T})
	=\frac{1}{4\nu t^2}\operatorname{diag}_{i\in\mathcal T}
	\left(\frac{1}{w_i d_i^2}\right)+o((\nu t^2)^{-1}).
	\label{eq:jointscalingconstruction}
\end{equation}
Since every $w_i>0$, all parameters in $\mathcal T$ attain Heisenberg scaling in this common protocol.

\paragraph{Remaining parameters.}
As stated in Appendix~\ref{sec:optimal_coefficient} of the End Matter, the remaining parameters can be estimated without changing the optimal coefficient. Suppose that an auxiliary experiment of fixed duration $\tau$ within the general sequential scheme has outcome probabilities $p_0(y|\bm\theta)$ whose CFIM $\JC^{\mathrm{aux}}$ for the full vector $\bm\theta$ is nonsingular. Reserve a fraction $w_0>0$ of the interrogations for it, with $w_0+\sum_{i\in\mathcal T}w_i=1$, and repeat it $m=\lfloor t/\tau\rfloor$ times within each interrogation, resetting the sensor in between. The outcomes $y_n$ of these $\nu_0m$ runs, with $\nu_0=w_0\nu$, give the local estimator
\begin{equation}
	\hat{\bm\theta}_0=\tilde{\bm\theta}+
	\frac{(\JC^{\mathrm{aux}})^{-1}}{\nu_0m}\sum_{n=1}^{\nu_0m}\nabla_{\bm\theta}\log p_0(y_n|\tilde{\bm\theta}).
	\label{eq:regular_score_estimator}
\end{equation}
At $\tilde{\bm\theta}$, the gradient $\nabla_{\bm\theta}\log p_0(y|\tilde{\bm\theta})$ has zero mean and covariance $\JC^{\mathrm{aux}}$, and the derivative of its mean with respect to $\bm\theta$ is $\JC^{\mathrm{aux}}$. Hence $\hat{\bm\theta}_0$ is locally unbiased with covariance $(\JC^{\mathrm{aux}})^{-1}/(\nu_0m)=O(1/(\nu t))$. We take $\hat\theta_i$ from the codes for $i\in\mathcal T$ and from $\hat{\bm\theta}_0$ otherwise. For a parameter that violates Eq.~\eqref{eq:jointcriterion}, the scaling $O(1/(\nu t))$ is optimal by Eq.~\eqref{eq:linear_submodel}.

For a single target parameter $\theta_i$, take $w_i=1-w_0$. Then
\begin{equation}
	\Var(\hat\theta_i)=\frac{1}{4\nu(1-w_0)d_i^2t^2},
	\label{eq:scalar_auxiliary_variance}
\end{equation}
and the other parameters, estimated with Eq.~\eqref{eq:regular_score_estimator}, have
\begin{equation}
	\Cov(\hat{\bm\theta}_{-i})=\frac{[(\JC^{\mathrm{aux}})^{-1}]_{-i,-i}}{\nu w_0\lfloor t/\tau\rfloor},
	\label{eq:scalar_auxiliary_covariance}
\end{equation}
where $-i$ denotes all parameters other than $\theta_i$. Their covariance with $\hat\theta_i$ vanishes because the data sets are independent. A fixed $w_0>0$ thus gives $\Cov(\hat{\bm\theta}_{-i})=O(1/(\nu t))$ but increases $\Var(\hat\theta_i)$ by the factor $1/(1-w_0)$. Taking $w_0(t)\to0$ with $w_0(t)t\to\infty$ gives $\nu t^2\Var(\hat\theta_i)\to1/(4d_i^2)$, the optimum in Eq.~\eqref{eq:general_optimal_variance}, while $\Cov(\hat{\bm\theta}_{-i})=O(1/(\nu w_0(t)t))$ still vanishes. For example, $w_0(t)=(t/\tau)^{-\alpha}$ with $0<\alpha<1$ gives $\Cov(\hat{\bm\theta}_{-i})=O(1/(\nu t^{1-\alpha}))$.

\paragraph{Precision at the SQL.}
When $d_i=0$, some $\bm v_i$ and $g_i$ cancel the coherent part, $\mathcal R_i=0$, and only the dissipative part $A_i$ limits the precision. Minimizing Eq.~\eqref{eq:linear_submodel} over $\bm v_i$ and $g_i$ gives, for every $t$,
\begin{equation}
	\nu t\Var_{\mathrm{opt}}(\hat\theta_i)\geq\frac1{\mathcal C_i},\qquad
	\mathcal C_i=4\inf_{\bm v_i,\,g_i:\ (\bm v_i)_i=1,\ \mathcal R_i=0}\norm{A_i}_\infty.
	\label{eq:optimal_sql_coefficient}
\end{equation}
In the spin model of Sec.~\ref{sec:finite}, $\mathcal C_\omega=1/\kappa$, and the protocol of Sec.~\ref{sec:toybridge} attains this bound at long times (Sec.~\ref{sec:spin_optimality}). With known noise, approximate QEC attains the corresponding bound for a single Hamiltonian parameter~\cite{Zhou2020Approximate}, also when other Hamiltonian parameters are unknown~\cite{Abbasgholinejad2026Function}.

\paragraph{Theorem 1 of the main text.}
For the model in Eq.~\eqref{eq:lindblad} of the main text, with $H=H(\bm\omega)$ and $E_k=E_k(\bm\eta)$, Eq.~\eqref{eq:suppHhat} gives $\hat H_{\omega_a}=\partial_{\omega_a}H$ and
\begin{equation}
	\hat H_{\eta_b}=C_{\eta_b}=\frac i2\sum_k\left[E_k^\dagger\partial_{\eta_b}E_k
	-(\partial_{\eta_b}E_k)^\dagger E_k\right],
	\label{eq:physical_noise_response}
\end{equation}
as in Eq.~\eqref{eq:Hhat} of the main text. For the target $\theta_i=\omega_a$, the operator $K_i$ in Eq.~\eqref{eq:individualdistance} becomes $K_a$ in Eq.~\eqref{eq:target_K_definition} of the main text, and $\mathcal N_i$ becomes the span in Eq.~\eqref{eq:noise_span} of the main text,
\begin{equation}
	\mathcal N_a=\mathcal S+\operatorname{span}\bigl(\{\partial_{\omega_{a'}}H\}_{a'\ne a}\cup\{C_{\eta_b}\}_b\bigr).
	\label{eq:theorem1_span}
\end{equation}
With these operators, Theorem~\ref{thm:joint_scaling} of the main text reads as follows.

\begin{maintheorem}{\ref{thm:joint_scaling}}
	In the general sequential scheme with unknown noise parameters, a locally unbiased estimator of $\omega_a$ can attain Heisenberg scaling, $\Var(\hat\omega_a)=O(1/(\nu t^2))$, if and only if
	\begin{equation}
		\partial_{\omega_a}H\notin\mathcal N_a.
		\label{eq:theorem1_criterion}
	\end{equation}
	Otherwise, the inverse variance grows at most linearly with the interrogation time, limiting the precision to SQL scaling. If Eq.~\eqref{eq:theorem1_criterion} holds, the optimal variance satisfies
	\begin{equation}
		\lim_{t\to\infty}\nu t^2\Var_{\mathrm{opt}}(\hat\omega_a)=\frac{1}{4d_a^2},
		\label{eq:theorem1_variance}
	\end{equation}
	where the operator-norm distance from $\partial_{\omega_a}H$ to $\mathcal N_a$ is
	\begin{equation}
		d_a=\min_{K=K^\dagger\in\mathcal N_a}\norm{\partial_{\omega_a}H-K}_\infty.
		\label{eq:theorem1_distance}
	\end{equation}
	Here $\norm{X}_\infty$ is the largest absolute eigenvalue of $X$. Equivalently, the optimal quadratic coefficient, i.e., the leading coefficient of $t^2$ in $1/[\nu\Var_{\mathrm{opt}}(\hat\omega_a)]$, is $4d_a^2$.
\end{maintheorem}

The theorem is the case $\theta_i=\omega_a$ of Lemma~\ref{lem:general_scaling}. With $d_i=d_a$, Eqs.~\eqref{eq:jointcriterion}, \eqref{eq:linear_submodel}, and~\eqref{eq:general_optimal_variance} of the lemma give its criterion, the linear bound on the inverse variance, and the optimal variance. For $\theta_i=\eta_b$, the lemma gives the extension to noise parameters stated in the main text.

\paragraph{Semidefinite program for $d_a$.}
The distance $d_a$ in Eq.~\eqref{eq:theorem1_distance} and the optimal code follow from $H$ and the $E_k$ by a semidefinite program, as for known noise~\cite{Zhou2018Heisenberg}. Evaluate $E_k$, $\partial_{\omega_{a'}}H$, and $\partial_{\eta_b}E_k$ at the prior estimates $\tilde{\bm\theta}$, and compute $C_{\eta_b}$ from Eq.~\eqref{eq:physical_noise_response}. The Hermitian operators
\begin{equation}
	I,\qquad E_k+E_k^\dagger,\qquad i(E_k-E_k^\dagger),\qquad E_k^\dagger E_l+E_l^\dagger E_k,\qquad i(E_k^\dagger E_l-E_l^\dagger E_k),
	\label{eq:lindblad_span_hermitian}
\end{equation}
for all $k$ and $l$, labeled $B_m$, span the Hermitian operators in $\mathcal S$ over the reals. The Hermitian operators in $\mathcal N_a$ of Eq.~\eqref{eq:theorem1_span} are then the real linear combinations of the $B_m$, the $\partial_{\omega_{a'}}H$ with $a'\ne a$, and the $C_{\eta_b}$. Since a Hermitian operator $X$ obeys $\norm{X}_\infty\leq r$ if and only if $-rI\preceq X\preceq rI$, Eq.~\eqref{eq:theorem1_distance} becomes
\begin{equation}
	d_a=\min\Bigl\{r:\ -rI\preceq\sum_{a'}v_{\omega_{a'}}\partial_{\omega_{a'}}H+\sum_bv_{\eta_b}C_{\eta_b}-\sum_mx_mB_m\preceq rI\Bigr\},
	\label{eq:distance_sdp}
\end{equation}
where the minimum is over real $r$, $x_m$, and $\bm v$ with $v_{\omega_a}=1$. For fixed $\bm v$, the minimum over $r$ and $x_m$ is $d_a(\bm v)$ in Eq.~\eqref{eq:dav} of the main text. The objective is linear and the constraints are linear matrix inequalities on the sensor, so Eq.~\eqref{eq:distance_sdp} is a semidefinite program. Its dual is the maximization in Eq.~\eqref{eq:dual} with $X=\partial_{\omega_a}H$ and $\mathcal N=\mathcal N_a$, over $Y=Y_+-Y_-$ with $Y_\pm\succeq0$ and $\Tr(Y_++Y_-)\leq1$, subject to $\Tr(YB_m)=0$, $\Tr(Y\partial_{\omega_{a'}}H)=0$ for $a'\ne a$, and $\Tr(YC_{\eta_b})=0$. Standard solvers return $d_a$ together with a dual optimizer $Y_\star$, and purifying $|Y_\star|\pm Y_\star$ in orthogonal subspaces of a noiseless ancilla gives the optimal code of Eq.~\eqref{eq:optimal_code_compression}. With $\hat H_\mu$ of Eq.~\eqref{eq:suppHhat} in place of $\partial_{\omega_{a'}}H$ and $C_{\eta_b}$, the same program gives $d_i$ in Lemma~\ref{lem:general_scaling} for any target $\theta_i$.

\subsection{Effect of unknown noise parameters}\label{sec:scaling_applications}
Unknown noise parameters enter Theorem~\ref{thm:joint_scaling} of the main text only through the operators $C_{\eta_b}$, which can only enlarge $\mathcal N_a$ relative to known noise and therefore never increase $d_a$. In the spin model of Sec.~\ref{sec:finite}, the unknown noise direction contributes $C_\phi=-2\kappa\,\partial_\omega H$ (Sec.~\ref{sec:spin_optimality}), which places $\partial_\omega H$ in $\mathcal N_\omega$ and destroys Heisenberg scaling. The two paragraphs below show milder effects. Heisenberg scaling can survive with a reduced optimal quadratic coefficient, and unknown rates and correlations of fixed noise operators leave this coefficient unchanged.

\paragraph{Heisenberg scaling with a reduced coefficient.}\leavevmode\label{sec:noise_rotation_example}
Consider two spins with unequal signal couplings of the same sign, $H=\omega(\sigma_z^{(1)}+g\sigma_z^{(2)})/2$ with $g>0$ and $g\neq1$, and independent noise described by
\begin{equation}
	E_j(\eta)=\sqrt\kappa\bigl(\cos\eta\,\sigma_x^{(j)}+\sin\eta\,\sigma_y^{(j)}\bigr),\qquad j=1,2.
	\label{eq:two_spin_noise}
\end{equation}
The rate $\kappa>0$ is known, and $(\omega,\eta)$ are estimated jointly near the true value $\eta_0=0$. At $\eta=0$, the Lindblad span is $\mathcal S=\operatorname{span}\{I,\sigma_x^{(1)},\sigma_x^{(2)},\sigma_x^{(1)}\sigma_x^{(2)}\}$. Since $\partial_\eta E_j=\sqrt\kappa\,\sigma_y^{(j)}$, Eq.~\eqref{eq:physical_noise_response} gives
\begin{equation}
	C_\eta=\frac{i\kappa}{2}\sum_{j=1}^2[\sigma_x^{(j)},\sigma_y^{(j)}]
	=-\kappa\bigl(\sigma_z^{(1)}+\sigma_z^{(2)}\bigr).
	\label{eq:two_spin_noise_response}
\end{equation}

With $\eta$ unknown, $\mathcal N_\omega=\mathcal S+\operatorname{span}\{C_\eta\}$, so $d_\omega$ is the minimum of $\norm{\partial_\omega H+v_\eta C_\eta-S}_\infty$ over real $v_\eta$ and Hermitian $S\in\mathcal S$. In the product $\sigma_z$ basis, $\partial_\omega H$ and $C_\eta$ are diagonal, whereas $\sigma_x^{(1)}$, $\sigma_x^{(2)}$, and $\sigma_x^{(1)}\sigma_x^{(2)}$ have no diagonal entries. Discarding off-diagonal entries does not increase the operator norm, so $S$ can be taken proportional to $I$. With $a=2\kappa v_\eta$, this gives
\begin{equation}
	\begin{aligned}
		4d_\omega^2
		&=\min_{a,s\in\mathbb R}\norm{(1-a)\sigma_z^{(1)}+(g-a)\sigma_z^{(2)}-sI}_\infty^2\\
		&=\min_{a\in\mathbb R}\bigl(|1-a|+|g-a|\bigr)^2=(1-g)^2.
	\end{aligned}
	\label{eq:orientation_array_quadratic_optimum}
\end{equation}
Since $(1-a)\sigma_z^{(1)}+(g-a)\sigma_z^{(2)}$ has the eigenvalues $\pm(1-a)\pm(g-a)$, which are symmetric about zero, $s=0$ is optimal, and the minimum over $a$ is attained for $a$ between $1$ and $g$. For known $\eta$, $v_\eta=0$ and thus $a=0$, which gives $4(d_\omega^{\mathrm{known}})^2=(1+g)^2$. Theorem~\ref{thm:joint_scaling} of the main text then gives the optimal quadratic coefficients $(1+g)^2$ and $(1-g)^2$ for known and unknown $\eta$, respectively. Thus Heisenberg scaling survives, but the leading term of the optimal variance increases by the factor $[(1+g)/(1-g)]^2$. For $g=1$, $\partial_\omega H=-C_\eta/(2\kappa)\in\mathcal N_\omega$, and Heisenberg scaling is lost as in Sec.~\ref{sec:finite}.

\paragraph{Unknown rates and correlations of fixed noise operators.}\leavevmode\label{sec:fixed_error_space}
Let $E_k=\sum_lM_{lk}(\bm\eta)E_l^{(0)}$ with fixed operators $E_l^{(0)}$ and an invertible matrix $M(\bm\eta)$, which describes unknown rates and correlations of the noise, as in the main text. Since $E_l^{(0)}=\sum_mE_m(M^{-1})_{ml}$,
\begin{equation}
	\partial_{\eta_b}E_k=\sum_mE_m\Xi^{(b)}_{mk},\qquad \Xi^{(b)}=M^{-1}\partial_{\eta_b}M.
	\label{eq:fixed_error_derivative}
\end{equation}
Equation~\eqref{eq:physical_noise_response} then gives
\begin{equation}
	C_{\eta_b}=\frac i2\sum_{km}\bigl[\Xi^{(b)}_{mk}E_k^\dagger E_m-(\Xi^{(b)}_{mk})^*E_m^\dagger E_k\bigr]\in\operatorname{span}\{E_k^\dagger E_l\}\subseteq\mathcal S.
	\label{eq:fixed_error_response}
\end{equation}
The span in Eq.~\eqref{eq:noise_span} of the main text therefore reduces to $\mathcal N_a=\mathcal S+\operatorname{span}\{\partial_{\omega_{a'}}H\}_{a'\ne a}$, the span for known noise with the same unknown Hamiltonian parameters, so $d_a$ and the optimal quadratic coefficient $4d_a^2$ are unchanged. For unknown rates alone, $E_k=\sqrt{\kappa_k(\bm\eta)}E_k^{(0)}$, so $M=\diag(\sqrt{\kappa_k})$ is real and diagonal and $C_{\eta_b}=0$. More generally, $d_a$ is unchanged whenever all $C_{\eta_b}\in\mathcal S$.

\section{Precision with a reference system (Theorem 2)}\label{sec:noise_reference}
This section proves Theorem~\ref{thm:reference_restoration} of the main text, which states when a noisy reference restores the precision lost to unknown noise parameters. In Sec.~\ref{sec:reference_proof}, we restate the theorem and complete the proof sketched in the main text: a triangle inequality bounds the precision, and the joint codes of the main text attain this bound when the local codes are chosen as in Appendix~\ref{sec:reference_codes} of the End Matter. Section~\ref{sec:reference_special} derives the two special cases discussed after the theorem, and Sec.~\ref{sec:reference_spin} gives the details of the spin model with a reference spin.

\subsection{Proof of Theorem 2}\label{sec:reference_proof}
The reference $R$ evolves alongside the sensor $S$ under its own Hamiltonian $H_R(\bm\omega)$ and Lindblad operators $F_l(\bm\eta)$. The joint system therefore obeys Eq.~\eqref{eq:lindblad} of the main text with the Hamiltonian and Lindblad operators
\begin{equation}
	H_{SR}=H(\bm\omega)\otimes I+I\otimes H_R(\bm\omega),\qquad
	E_k(\bm\eta)\otimes I,\qquad I\otimes F_l(\bm\eta).
	\label{eq:reference_joint_model}
\end{equation}
The operators $H_R$ and $F_l$ may vanish or depend on only some of the parameters. Parameters specific to $R$ are included in $\bm\omega$ and $\bm\eta$, and $H$ and $E_k$ do not depend on them. As in Eq.~\eqref{eq:suppHhat}, the coherent contributions of a parameter $\theta_j$ for $S$ and $R$ are
\begin{equation}
	\begin{aligned}
		\hat H_j&=\partial_{\theta_j}H+\frac i2\sum_k\bigl[E_k^\dagger\partial_{\theta_j}E_k-(\partial_{\theta_j}E_k)^\dagger E_k\bigr],\\
		\hat H_j^R&=\partial_{\theta_j}H_R+\frac i2\sum_l\bigl[F_l^\dagger\partial_{\theta_j}F_l-(\partial_{\theta_j}F_l)^\dagger F_l\bigr],
	\end{aligned}
	\label{eq:reference_local_contributions}
\end{equation}
denoted $X_j$ and $X_j^R$ in the main text. They are $\partial_{\omega_a}H$ and $\partial_{\omega_a}H_R$ for a Hamiltonian parameter $\omega_a$, and $C_{\eta_b}$ and $C_{\eta_b}^R$ for a noise parameter $\eta_b$. Since each Lindblad operator acts on only one system, the coherent contributions of the joint system are
\begin{equation}
	\hat H_j^{SR}=\hat H_j\otimes I+I\otimes\hat H_j^R.
	\label{eq:reference_joint_contributions}
\end{equation}
Every element of the joint Lindblad span $\mathcal S_{SR}$ is a linear combination of products $A\otimes B$ with $A\in\mathcal S$ and $B\in\mathcal S_R$, where $\mathcal S_R$ is the Lindblad span of $R$. Besides $\mathcal S\otimes I$ and $I\otimes\mathcal S_R$, this span contains the cross terms $E_k^\dagger\otimes F_l$ and their Hermitian conjugates.

As in Eq.~\eqref{eq:dav} of the main text, the smallest norms of the coherent parts of $S$ and $R$ along a real vector $\bm v$ with $v_{\omega_a}=1$ are
\begin{equation}
	d_a^S(\bm v)=\min_{S=S^\dagger\in\mathcal S}\Bigl\|\sum_jv_j\hat H_j-S\Bigr\|_\infty,\qquad
	d_a^R(\bm v)=\min_{S'=S'^\dagger\in\mathcal S_R}\Bigl\|\sum_jv_j\hat H_j^R-S'\Bigr\|_\infty.
	\label{eq:reference_directional_distances}
\end{equation}
With this notation, Theorem~\ref{thm:reference_restoration} of the main text reads as follows.

\begin{maintheorem}{\ref{thm:reference_restoration}}
	In the general sequential scheme acting jointly on $S$ and $R$, the parameter $\omega_a$ can attain Heisenberg scaling if and only if no $\bm v$ gives $d_a^S(\bm v)=d_a^R(\bm v)=0$. If so, the optimal variance satisfies $\lim_{t\to\infty}\nu t^2\Var_{\mathrm{opt}}(\hat\omega_a)=1/[4(d_a^{SR})^2]$, where $d_a$ for the joint system reduces to
	\begin{equation}
		d_a^{SR}=\min_{\bm v}\big[d_a^S(\bm v)+d_a^R(\bm v)\big].
		\label{eq:reference_dSR}
	\end{equation}
\end{maintheorem}

The proof follows the proof sketch of the main text and Appendix~\ref{sec:reference_codes} of the End Matter.

\paragraph{Upper bound.}
Fix a real vector $\bm v$ with $v_{\omega_a}=1$ and Hermitian operators $S\in\mathcal S$ and $S'\in\mathcal S_R$. Since $S\otimes I+I\otimes S'$ is a Hermitian element of $\mathcal S_{SR}$, Eq.~\eqref{eq:theorem1_upper} holds for the joint system with the coherent part $\hat H_i-K_i$ given by
\begin{equation}
	\sum_jv_j\hat H_j^{SR}-S\otimes I-I\otimes S'=\Bigl(\sum_jv_j\hat H_j-S\Bigr)\otimes I+I\otimes\Bigl(\sum_jv_j\hat H_j^R-S'\Bigr).
	\label{eq:reference_split}
\end{equation}
The triangle inequality and $\norm{A\otimes I}_\infty=\norm{A}_\infty$ bound its norm by the sum of the norms of the two local terms, which is $d_a^S(\bm v)+d_a^R(\bm v)$ for the minimizing $S$ and $S'$. Taking $t\to\infty$ and minimizing over $\bm v$ gives
\begin{equation}
	\limsup_{t\to\infty}\frac{1}{\nu t^2\Var_{\mathrm{opt}}(\hat\omega_a)}\leq4(d_a^{SR})^2.
	\label{eq:reference_upper}
\end{equation}
If some $\bm v$ gives $d_a^S(\bm v)=d_a^R(\bm v)=0$, the coherent part vanishes for this $\bm v$ and the minimizing $S$ and $S'$. Equation~\eqref{eq:theorem1_upper} then reduces to Eq.~\eqref{eq:linear_submodel}, which rules out Heisenberg scaling.

The bound~\eqref{eq:reference_upper} uses only the elements $S\otimes I+I\otimes S'$ of $\mathcal S_{SR}$. It remains to show why the cross terms cannot tighten it, and how equality is reached in the triangle inequality. Both answers come from the joint codes of the main text, built from two local codes.

\paragraph{Joint codes.}
As in the proof sketch of the main text, each system is encoded in a code that corrects its own noise. The codewords $\ket{C_0}$ and $\ket{C_1}$ of $S$ and a noiseless ancilla satisfy the Knill--Laflamme conditions~\eqref{eq:code_kl_explicit}, and the codewords $\ket{C_0^R}$ and $\ket{C_1^R}$ of $R$ and a second noiseless ancilla satisfy them with $F_l$ in place of $E_k$. The joint codewords are $\ket{C_0^{SR}}=\ket{C_0C_1^R}$ and $\ket{C_1^{SR}}=\ket{C_1C_0^R}$, with the encoding $V_{SR}=\ket{C_0^{SR}}\bra0+\ket{C_1^{SR}}\bra1$. For operators $A$ on $S$ and $B$ on $R$,
\begin{equation}
	\bra{C_i^{SR}}A\otimes B\ket{C_j^{SR}}=\bra{C_i}A\ket{C_j}\bra{C_{1-i}^R}B\ket{C_{1-j}^R}.
	\label{eq:reference_product_elements}
\end{equation}
For $A\in\mathcal S$ and $B\in\mathcal S_R$, the local Knill--Laflamme conditions make both factors proportional to $\delta_{ij}$, with coefficients independent of $i$. Each operator spanning $\mathcal S_{SR}$ is such a product, so the joint code satisfies the Knill--Laflamme conditions for the Lindblad operators $E_k\otimes I$ and $I\otimes F_l$ of the joint system, including the cross terms. For the coherent contributions in Eq.~\eqref{eq:reference_joint_contributions}, Eq.~\eqref{eq:reference_product_elements} with $B=I$ or $A=I$ shows that the off-diagonal elements vanish, because the local codewords are orthogonal, and gives
\begin{equation}
	V_{SR}^\dagger\hat H_j^{SR}V_{SR}=c_jI_C+(y_j^S-y_j^R)\sigma_z^C,\qquad
	y_j^S=\frac{\langle\hat H_j\rangle_0-\langle\hat H_j\rangle_1}{2},\qquad
	y_j^R=\frac{\langle\hat H_j^R\rangle_0^R-\langle\hat H_j^R\rangle_1^R}{2},
	\label{eq:reference_local_responses}
\end{equation}
with real $c_j$, the logical identity $I_C$, and the logical Pauli operator $\sigma_z^C$. Here $\langle\cdot\rangle_i$ and $\langle\cdot\rangle_i^R$ are the expectations in $\ket{C_i}$ and $\ket{C_i^R}$, and $y_j^S$ and $y_j^R$ are the responses of the two codes to $\theta_j$. They enter with opposite signs because the reference codewords enter in reverse order. With the control Hamiltonian $-H_{SR}(\tilde{\bm\omega})$ and a recovery constructed as in Sec.~\ref{sec:general_scaling} for this code, Eq.~\eqref{eq:recoverderivative} gives the first-order logical generator
\begin{equation}
	\delta\mathcal L^C(\rho_C)=-i\sum_j\delta\theta_j(y_j^S-y_j^R)[\sigma_z^C,\rho_C],
	\label{eq:reference_logical_generator}
\end{equation}
which is Eq.~\eqref{eq:jointgen} of the main text.

\paragraph{Optimal codes.}
It remains to choose local codes whose responses to the other parameters cancel, $y_j^S=y_j^R$ for all $\theta_j\ne\omega_a$, and whose response to $\omega_a$ is as large as possible. As in Appendix~\ref{sec:reference_codes} of the End Matter, let $\rho_0$ and $\rho_1$ be the reduced states of the sensor codewords on $S$, and let $Y_S=(\rho_0-\rho_1)/2$, with trace norm $\norm{Y_S}_1=\Tr|Y_S|\leq1$, so that $y_j^S=\Tr(Y_S\hat H_j)$. The Knill--Laflamme conditions require $\Tr(Y_SO)=0$ for all $O\in\mathcal S$, written $Y_S\perp\mathcal S$, so $Y_S$ lies in the set
\begin{equation}
	\mathcal K_S=\bigl\{Y=Y^\dagger:\ Y\perp\mathcal S,\ \norm{Y}_1\leq1\bigr\}.
	\label{eq:reference_dual_set}
\end{equation}
Conversely, every $Y_S\in\mathcal K_S$ is realized. Since $I\in\mathcal S$, $Y_S$ is traceless, so the operators
\begin{equation}
	\rho_\pm^S=|Y_S|\pm Y_S+(1-\norm{Y_S}_1)\sigma,
	\label{eq:reference_padded_states}
\end{equation}
with any state $\sigma$, are density matrices. Purifying them in orthogonal subspaces of a noiseless ancilla gives codewords with $\rho_0=\rho_+^S$ and $\rho_1=\rho_-^S$. Their encoding obeys Eq.~\eqref{eq:optimal_code_compression} with $Y_S$ in place of $Y_\star$ and $(\rho_+^S+\rho_-^S)/2$ in place of $|Y_\star|$, so they satisfy Eq.~\eqref{eq:code_kl_explicit} because $Y_S\perp\mathcal S$. The same holds for $R$, with $\mathcal S_R$, $\mathcal K_R$, $Y_R$, and $y_j^R=\Tr(Y_R\hat H_j^R)$. Choosing the local codes thus amounts to choosing $Y_S\in\mathcal K_S$ and $Y_R\in\mathcal K_R$. As stated in Eq.~\eqref{eq:joint_code_dual_em} of the End Matter, the largest response to $\omega_a$ without response to the other parameters is
\begin{equation}
	\max\bigl\{y_{\omega_a}^S-y_{\omega_a}^R:\ y_j^S=y_j^R\ \ \forall\theta_j\ne\omega_a\bigr\}
	=\min_{\bm v:\,v_{\omega_a}=1}\bigl[d_a^S(\bm v)+d_a^R(\bm v)\bigr]=d_a^{SR},
	\label{eq:reference_program}
\end{equation}
where the maximum is over $Y_S\in\mathcal K_S$ and $Y_R\in\mathcal K_R$ and the last equality is Eq.~\eqref{eq:reference_dSR}. To prove the first equality, we use multipliers $v_j$ for the constraints and set $v_{\omega_a}=1$. The Lagrangian is
\begin{equation}
	\sum_jv_j(y_j^S-y_j^R)=\Tr\Bigl(Y_S\sum_jv_j\hat H_j\Bigr)-\Tr\Bigl(Y_R\sum_jv_j\hat H_j^R\Bigr),
	\label{eq:reference_lagrangian}
\end{equation}
and its maximum over $\mathcal K_S\times\mathcal K_R$ is $d_a^S(\bm v)+d_a^R(\bm v)$, by Eq.~\eqref{eq:dual} with $\mathcal N=\mathcal S$ and $\mathcal N=\mathcal S_R$, because $\mathcal K_R$ is invariant under $Y_R\to-Y_R$. Since $Y_S=Y_R=0$ is strictly feasible, strong duality holds and the minimum over $\bm v$ is attained. The feasible set is compact and thus contains an optimizer $(Y_S,Y_R)$, with
\begin{equation}
	y_j^S=y_j^R\ \ \forall\theta_j\ne\omega_a,\qquad
	y_{\omega_a}^S-y_{\omega_a}^R=d_a^{SR}.
	\label{eq:reference_optimal_responses}
\end{equation}
For the joint code of this optimizer, Eq.~\eqref{eq:reference_logical_generator} reduces to Eq.~\eqref{eq:logical_evolution} of the main text with $d_a^{SR}$ in place of $d_a$. For $d_a^{SR}>0$, the logical state $(\ket{C_0^{SR}}+\ket{C_1^{SR}})/\sqrt2$, the readout of Eq.~\eqref{eq:hs_probabilities}, and the estimator $\hat\omega_a=\tilde\omega_a+\bar z/(2d_a^{SR}t)$ therefore attain
\begin{equation}
	\Var(\hat\omega_a)=\frac{1}{4\nu(d_a^{SR})^2t^2}.
	\label{eq:reference_attainability}
\end{equation}

Equations~\eqref{eq:reference_upper} and~\eqref{eq:reference_attainability} give $\lim_{t\to\infty}\nu t^2\Var_{\mathrm{opt}}(\hat\omega_a)=1/[4(d_a^{SR})^2]$ for $d_a^{SR}>0$. Since the minimum over $\bm v$ is attained, $d_a^{SR}>0$ if and only if no $\bm v$ gives $d_a^S(\bm v)=d_a^R(\bm v)=0$. Heisenberg scaling is then attained, and is otherwise ruled out by the upper bound, as stated in the theorem. Theorem~\ref{thm:joint_scaling} of the main text, applied to the joint system, gives the same criterion and limit with $d_a$ of the joint system in place of $d_a^{SR}$, so $d_a$ for the joint system reduces to $d_a^{SR}$. Since the bound~\eqref{eq:reference_upper} is attained, the cross terms cannot tighten it, and the triangle inequality holds with equality at the minimizing $\bm v$, $S$, and $S'$. The paragraphs of Sec.~\ref{sec:general_scaling} on simultaneous Heisenberg scaling and on the remaining parameters apply to the joint system unchanged.

\paragraph{Semidefinite program for $d_a^{SR}$.}
The distance $d_a^{SR}$ follows from one semidefinite program for the sensor and the reference together. Let $B_m$ be the operators of Eq.~\eqref{eq:lindblad_span_hermitian}, which span the Hermitian operators in $\mathcal S$, and let $B'_m$ be the same operators built from the $F_l$, which span the Hermitian operators in $\mathcal S_R$. With $\hat H_j$ and $\hat H_j^R$ of Eq.~\eqref{eq:reference_local_contributions} evaluated at the prior estimates, Eqs.~\eqref{eq:reference_directional_distances} and~\eqref{eq:reference_dSR} give
\begin{equation}
	\begin{aligned}
		d_a^{SR}=\min\Bigl\{r_S+r_R:\ &-r_SI\preceq\sum_jv_j\hat H_j-\sum_mx_mB_m\preceq r_SI,\\
		&-r_RI\preceq\sum_jv_j\hat H_j^R-\sum_mx'_mB'_m\preceq r_RI\Bigr\},
	\end{aligned}
	\label{eq:reference_sdp}
\end{equation}
where the minimum is over real $r_S$, $r_R$, $x_m$, $x'_m$, and $v_j$ with $v_{\omega_a}=1$. The first pair of matrix inequalities acts on the sensor and the second on the reference, and both contain the same $\bm v$. For fixed $\bm v$, the program splits into two independent programs, over $r_S$ and $x_m$ and over $r_R$ and $x'_m$, whose optimal values are $d_a^S(\bm v)$ and $d_a^R(\bm v)$, and the minimum of their sum over the common $\bm v$ is $d_a^{SR}$. Minimizing the two programs over separate vectors would instead give $\min_{\bm v}d_a^S(\bm v)+\min_{\bm v}d_a^R(\bm v)$, which vanishes for the reference spin of Sec.~\ref{sec:reference_spin} although the reference restores Heisenberg scaling there. As shown above, the dual program is the maximization in Eq.~\eqref{eq:reference_program}, with $\mathcal K_S$ and $\mathcal K_R$ of Eq.~\eqref{eq:reference_dual_set} written with $B_m$ and $B'_m$ as in the dual of Eq.~\eqref{eq:distance_sdp}. Its optimizer $(Y_S,Y_R)$ gives the two local codes through Eq.~\eqref{eq:reference_padded_states}.

\subsection{Special cases of Theorem 2}\label{sec:reference_special}
After Theorem~\ref{thm:reference_restoration}, the main text evaluates $d_a^{SR}$ of Eq.~\eqref{eq:reference_dSR} in two special cases. We derive these results and the codes that attain them, first for a reference that is a copy of the sensor without signal and then for a single unknown noise parameter.

\paragraph{Identical reference without signal.}
Let $R$ be a copy of $S$ with $H_R=0$ and $F_l(\bm\eta)=E_l(\bm\eta)$, where operators on both systems are written in the same basis. Then $\mathcal S_R=\mathcal S$, and Eq.~\eqref{eq:reference_local_contributions} gives $\hat H_{\omega_{a'}}^R=0$ for all $a'$ and $\hat H_{\eta_b}^R=C_{\eta_b}$. For known noise with the same unknown Hamiltonian parameters, Eqs.~\eqref{eq:theorem1_span} and~\eqref{eq:theorem1_distance} reduce to
\begin{equation}
	\mathcal N_a^{\mathrm{known}}=\mathcal S+\operatorname{span}\{\partial_{\omega_{a'}}H\}_{a'\ne a},\qquad
	d_a^{\mathrm{known}}=\min_{K=K^\dagger\in\mathcal N_a^{\mathrm{known}}}\norm{\partial_{\omega_a}H-K}_\infty.
	\label{eq:reference_known_distance}
\end{equation}
For any $\bm v$ with $v_{\omega_a}=1$ and Hermitian $S,S'\in\mathcal S$, the two norms in Eq.~\eqref{eq:reference_directional_distances} obey
\begin{equation}
	\begin{aligned}
		&\Bigl\|\partial_{\omega_a}H+\sum_{a'\ne a}v_{\omega_{a'}}\partial_{\omega_{a'}}H+\sum_bv_{\eta_b}C_{\eta_b}-S\Bigr\|_\infty
		+\Bigl\|\sum_bv_{\eta_b}C_{\eta_b}-S'\Bigr\|_\infty\\
		&\qquad\geq\norm{\partial_{\omega_a}H-K}_\infty\geq d_a^{\mathrm{known}},\qquad
		K=S-S'-\sum_{a'\ne a}v_{\omega_{a'}}\partial_{\omega_{a'}}H.
	\end{aligned}
	\label{eq:reference_known_triangle}
\end{equation}
The noise contributions cancel in the triangle inequality, and $K$ is a Hermitian element of $\mathcal N_a^{\mathrm{known}}$. Both inequalities become equalities for $v_{\eta_b}=0$, $S'=0$, and the $v_{\omega_{a'}}$ and $S$ for which $K$ attains $d_a^{\mathrm{known}}$. Equation~\eqref{eq:reference_dSR} therefore gives
\begin{equation}
	d_a^{SR}=d_a^{\mathrm{known}},
	\label{eq:reference_known_result}
\end{equation}
as stated in the main text. The reference thus restores the optimal quadratic coefficient $4(d_a^{\mathrm{known}})^2$ attainable with known noise.

It remains to show that the joint codewords $\ket{C_0C_1}$ and $\ket{C_1C_0}$, built from the optimal code for known noise, attain this coefficient. For $d_a^{\mathrm{known}}>0$, the codewords of this code purify $|Y|\pm Y$ as in Sec.~\ref{sec:general_scaling}, with $Y$ an optimizer of Eq.~\eqref{eq:dual} for $\mathcal N=\mathcal N_a^{\mathrm{known}}$ and $X=\partial_{\omega_a}H$. Since $Y\perp\mathcal N_a^{\mathrm{known}}\supseteq\mathcal S=\mathcal S_R$ and $\norm{Y}_1=1$, Eq.~\eqref{eq:reference_padded_states} with $Y_S=Y_R=Y$ gives this code on both systems, with the responses
\begin{equation}
	y_{\omega_a}^S-y_{\omega_a}^R=\Tr(Y\partial_{\omega_a}H)=d_a^{SR},\qquad
	y_{\omega_{a'}}^S=y_{\omega_{a'}}^R=0\quad(a'\ne a),\qquad
	y_{\eta_b}^S=y_{\eta_b}^R=\Tr(YC_{\eta_b}).
	\label{eq:reference_known_responses}
\end{equation}
The responses to each noise parameter cancel, so Eq.~\eqref{eq:reference_optimal_responses} holds and the joint code attains the variance~\eqref{eq:reference_attainability}.

\paragraph{Single unknown noise parameter.}
Let a noise parameter $\eta_b$ be the only unknown parameter besides $\omega_a$, and let neither system alone satisfy the criterion~\eqref{eq:theorem1_criterion} for Heisenberg scaling, that is,
\begin{equation}
	\partial_{\omega_a}H+\lambda_SC_{\eta_b}\in\mathcal S,\qquad
	\partial_{\omega_a}H_R+\lambda_RC_{\eta_b}^R\in\mathcal S_R
	\label{eq:reference_single_condition}
\end{equation}
for real $\lambda_S$ and $\lambda_R$. Up to a Hermitian element of $\mathcal S$, the operator $\sum_jv_j\hat H_j=\partial_{\omega_a}H+v_{\eta_b}C_{\eta_b}$ in Eq.~\eqref{eq:reference_directional_distances} equals $(v_{\eta_b}-\lambda_S)C_{\eta_b}$, and likewise for $R$, so that
\begin{equation}
	\begin{aligned}
		d_a^S(\bm v)&=a_S|v_{\eta_b}-\lambda_S|,&\qquad
		a_S&=\min_{S=S^\dagger\in\mathcal S}\norm{C_{\eta_b}-S}_\infty,\\
		d_a^R(\bm v)&=a_R|v_{\eta_b}-\lambda_R|,&\qquad
		a_R&=\min_{S'=S'^\dagger\in\mathcal S_R}\norm{C_{\eta_b}^R-S'}_\infty.
	\end{aligned}
	\label{eq:reference_single_distances}
\end{equation}
The sum $d_a^S(\bm v)+d_a^R(\bm v)$ is piecewise linear in $v_{\eta_b}$ and is smallest at $v_{\eta_b}=\lambda_S$ if $a_S\geq a_R$ and at $v_{\eta_b}=\lambda_R$ otherwise, which gives Eq.~\eqref{eq:single} of the main text,
\begin{equation}
	d_a^{SR}=\min_{v_{\eta_b}}\bigl[d_a^S(\bm v)+d_a^R(\bm v)\bigr]=\min\{a_S,a_R\}\,|\lambda_S-\lambda_R|.
	\label{eq:reference_single_result}
\end{equation}

As in Appendix~\ref{sec:reference_codes} of the End Matter, Eq.~\eqref{eq:dual} gives Hermitian operators $W_S\perp\mathcal S$ and $W_R\perp\mathcal S_R$ with trace norms at most one, $\Tr(W_SC_{\eta_b})=a_S$, and $\Tr(W_RC_{\eta_b}^R)=a_R$. For $d_a^{SR}>0$, we choose the local codes $Y_S=sq_SW_S$ and $Y_R=sq_RW_R$, with $q_Sa_S=q_Ra_R=\min\{a_S,a_R\}$ and $s$ the sign of $\lambda_R-\lambda_S$. By Eq.~\eqref{eq:reference_single_condition}, their responses are
\begin{equation}
	y_{\eta_b}^S=y_{\eta_b}^R=s\min\{a_S,a_R\},\qquad
	y_{\omega_a}^S-y_{\omega_a}^R=\lambda_Ry_{\eta_b}^R-\lambda_Sy_{\eta_b}^S=d_a^{SR},
	\label{eq:reference_single_responses}
\end{equation}
which satisfy Eq.~\eqref{eq:reference_optimal_responses}. For $\lambda_R<\lambda_S$, the sign $s=-1$ interchanges the codewords of each system by Eq.~\eqref{eq:reference_padded_states}, and thus the two joint codewords.

\subsection{Spin model with a reference spin}\label{sec:reference_spin}
This subsection derives the results that the main text quotes for the spin model with a reference spin. The sensor spin $S$ and the reference spin $R$ are flipped independently about the same unknown axis, and only the sensor couples to the signal,
\begin{equation}
	H_{SR}=\frac\omega2\sigma_z^{(S)},\qquad
	E_j=\sqrt{\kappa_j}\,\sigma_\phi^{(j)},\quad j=S,R,
	\label{eq:reference_spin_model}
\end{equation}
with $\sigma_\phi^{(j)}=\cos\phi\,\sigma_x^{(j)}+\sin\phi\,\sigma_y^{(j)}$, known rates $\kappa_S,\kappa_R>0$, and $r=\kappa_R/\kappa_S$. We start from equal rates, then let the rates differ or be unknown, compare the joint encoding with independent encodings of the two spins, and finally allow correlated noise.

\paragraph{Joint encoding and readout.}\leavevmode\label{sec:reference_spin_readout}
As in the main text, each spin is encoded with its own noiseless ancilla in the code $\ket{C_0}=\ket{00}$, $\ket{C_1}=\ket{11}$ of Sec.~\ref{sec:toybridge}, and the joint codewords are
\begin{equation}
	\ket{C_0^{SR}}=\ket{C_0C_1},\qquad\ket{C_1^{SR}}=\ket{C_1C_0},
	\label{eq:reference_spin_code}
\end{equation}
with the sensor and its ancilla first. Prepare $(\ket{C_0^{SR}}+\ket{C_1^{SR}})/\sqrt2$ and correct each spin as in Sec.~\ref{sec:toybridge}, recording the flip counts $N_S$ and $N_R$. One ancilla suffices: the codewords $\ket{010}$ and $\ket{101}$ of $S$, $R$, and a single ancilla $A$, corrected with the parity checks $\sigma_z^{(S)}\sigma_z^{(R)}$ and $\sigma_z^{(R)}\sigma_z^{(A)}$, give the same count statistics and logical state.

On the code space, $\sigma_z^{(S)}=\sigma_z^C$ and $\sigma_z^{(R)}=-\sigma_z^C$, where $\sigma_z^C=\ket{C_0^{SR}}\bra{C_0^{SR}}-\ket{C_1^{SR}}\bra{C_1^{SR}}$ as in the main text, and $\sigma_x^C$ and $\sigma_y^C$ denote the corresponding logical Pauli operators. Since $\sigma_x^{(j)}\sigma_\phi^{(j)}=e^{i\phi \sigma_z^{(j)}}$, corrected flips on $S$ and $R$ shift the logical phase in opposite directions. In the rapid-correction limit at fixed $t$, $N_S$ and $N_R$ are independent Poisson variables with means $\kappa_St$ and $\kappa_Rt$, because $E_j^\dagger E_j=\kappa_jI$. Conditional on the counts, the logical state is, as in Eq.~\eqref{eq:reference_spin_phase} of the main text,
\begin{equation}
	\ket{\psi_D}=\frac{\ket{C_0^{SR}}+e^{i\Phi_D}\ket{C_1^{SR}}}{\sqrt2},
	\qquad \Phi_D=\omega t-2D\phi,\qquad D=N_S-N_R.
	\label{eq:reference_spin_state}
\end{equation}

For equal rates $\kappa_S=\kappa_R=\kappa$, as in the main text, $\EE D=0$ and $\EE D^2=2\kappa t$. We measure $\cos\tilde\Phi_D\,\sigma_y^C-\sin\tilde\Phi_D\,\sigma_x^C$, with $\tilde\Phi_D=\tilde\omega t-2D\tilde\phi$ at the prior estimates $(\tilde\omega,\tilde\phi)$, which is the count-dependent basis of Sec.~\ref{sec:toybridge} with $D$ in place of $N$. As in Eqs.~\eqref{eq:toy-observed-prob} and~\eqref{eq:toy-CFIM}, with $D$ in place of $N$ in the phase, the gradient $\bm g_D=\nabla\Phi_D=(t,-2D)^{\mathsf T}$ in place of $\bm g_N$, and the average taken over both counts, the CFIM is
\begin{equation}
	\JC=\EE\bigl[\bm g_D\bm g_D^{\mathsf T}\bigr]
	=\EE\begin{pmatrix}t^2&-2tD\\-2tD&4D^2\end{pmatrix}
	=\begin{pmatrix}t^2&0\\0&8\kappa t\end{pmatrix}.
	\label{eq:reference_spin_cfim}
\end{equation}
This equals the QFIM $\JQ$ of the main text, so the joint readout attains the quantum Cram\'er--Rao bound. With bars denoting means over the $\nu$ interrogations, the estimators $\hat\omega=\tilde\omega+\bar z/t$ and $\hat\phi=\tilde\phi-\overline{Dz}/(4\kappa t)$ are therefore locally unbiased and uncorrelated at $(\tilde\omega,\tilde\phi)$, with
\begin{equation}
	\Var(\hat\omega)=\frac{1}{\nu t^2},\qquad
	\Var(\hat\phi)=\frac{1}{8\nu\kappa t}.
	\label{eq:reference_spin_variances}
\end{equation}
Compared with the single sensor [Eq.~\eqref{eq:toy_error_propagation}], the reference removes the term $\kappa/(\nu t)$ from $\Var(\hat\omega)$, recovering the variance for known $\phi$, and halves $\Var(\hat\phi)$.

\paragraph{Unequal flip rates.}\leavevmode\label{sec:reference_rates}
For $\kappa_S\neq\kappa_R$, the two counts have different means, so $\phi$ no longer drops out of the mean phase $\EE\Phi_D=\omega t-2(\kappa_S-\kappa_R)t\phi$ in Eq.~\eqref{eq:reference_spin_state}. We first find the optimal quadratic coefficient and then the encoding that attains it. As in Sec.~\ref{sec:spin_optimality}, the coherent contributions of $\phi$ and their distances from the Lindblad spans in Eq.~\eqref{eq:reference_single_distances} are
\begin{equation}
	C_\phi=-\kappa_S\sigma_z^{(S)}=-2\kappa_S\partial_\omega H,\qquad
	C_\phi^R=-\kappa_R\sigma_z^{(R)},\qquad
	a_S=\kappa_S,\qquad a_R=\kappa_R,
	\label{eq:reference_spin_contributions}
\end{equation}
since $\sigma_z^{(j)}$ has operator-norm distance $1$ from $\operatorname{span}\{I,\sigma_\phi^{(j)}\}$. Equation~\eqref{eq:reference_single_result} therefore applies with $\lambda_S=1/(2\kappa_S)$ and $\lambda_R=0$, and gives
\begin{equation}
	d_\omega^{SR}=\frac{\min\{1,r\}}{2},\qquad
	4(d_\omega^{SR})^2=\min\{1,r^2\},
	\label{eq:reference_spin_coefficient}
\end{equation}
the optimal quadratic coefficient shown in Fig.~\ref{fig:reference_rates}(b) of the main text.

To attain it, we keep the joint code of Eq.~\eqref{eq:reference_spin_code} but expose the two spins for different times,
\begin{equation}
	t_S=t\min\{1,r\},\qquad t_R=t\min\{1,1/r\},\qquad \kappa_St_S=\kappa_Rt_R,
	\label{eq:reference_exposure_times}
\end{equation}
which equalize the mean flip counts. At time $\min\{t_S,t_R\}$, the state of the faster-flipping spin is swapped into a further noiseless ancilla, and the other spin is corrected until time $t$. As in Eq.~\eqref{eq:reference_spin_cfim}, with $t_S$ and $\kappa_S$ in place of $t$ and $\kappa$,
\begin{equation}
	\Phi_D=\omega t_S-2D\phi,\qquad \EE D=0,\qquad \EE D^2=2\kappa_St_S,\qquad
	\JC=\begin{pmatrix}t_S^2&0\\0&8\kappa_St_S\end{pmatrix},
	\label{eq:reference_exposure_cfim}
\end{equation}
which attains the optimal quadratic coefficient $t_S^2/t^2=\min\{1,r^2\}$.

For $\kappa_R\to0$, this coefficient vanishes as $r^2$, consistent with Eq.~\eqref{eq:toyallcontrol} of the main text. If the reference also detects the signal, $H_R=\omega\sigma_z^{(R)}/2$, then $\lambda_R=1/(2\kappa_R)$ and Eq.~\eqref{eq:reference_single_result} gives
\begin{equation}
	2d_\omega^{SR}=1-\min\{r,1/r\},
	\label{eq:reference_two_sensors}
\end{equation}
which vanishes for equal rates, $r=1$. Two identical sensors therefore cannot restore Heisenberg scaling, whereas for $r\neq1$ Theorem~\ref{thm:reference_restoration} of the main text gives the optimal quadratic coefficient $(1-\min\{r,1/r\})^2$.

\paragraph{Encoding without knowledge of the rates.}
The exposure times in Eq.~\eqref{eq:reference_exposure_times} require the ratio $r$. If the rates are unknown, we fix the flip counts instead. Each spin is exposed until its $M$th corrected flip, and its state is then swapped into a further noiseless ancilla. Both counts equal $M$, so $\phi$ drops out of the phase,
\begin{equation}
	D=N_S-N_R=0,\qquad \Phi_D=\omega T_S,
	\label{eq:reference_fixed_counts}
\end{equation}
where the time $T_S$ of the $M$th flip of the sensor is recorded and sets the readout basis. We set $M$ below the expected count of the slower-flipping spin by a margin $m_t$,
\begin{equation}
	M=t\min\{\kappa_S,\kappa_R\}-m_t,\qquad \frac{m_t}{\sqrt t}\to\infty,\qquad \frac{m_t}{t}\to0,
	\label{eq:reference_flip_number}
\end{equation}
so that both spins reach $M$ flips before the time $t$ with probability approaching one. The QFI for $\omega$ is $\EE T_S^2$, and the mean $M/\kappa_S$ and variance $M/\kappa_S^2$ of $T_S$ give
\begin{equation}
	\lim_{t\to\infty}\frac{\EE T_S^2}{t^2}=\lim_{t\to\infty}\frac{M(M+1)}{\kappa_S^2t^2}=\min\{1,r^2\},
	\label{eq:reference_fixed_counts_coefficient}
\end{equation}
again the optimal quadratic coefficient. The rates enter only through $M$ and can be estimated from the flip times of earlier interrogations, which are independent of $\omega$ and $\phi$ and thus do not affect the estimation of $\omega$. Since $\phi$ no longer enters, it can be estimated, if needed, with the auxiliary experiment of Sec.~\ref{sec:general_scaling}.

\paragraph{Independent encodings.}\leavevmode\label{sec:reference_independent}
The joint encoding of Eq.~\eqref{eq:reference_spin_code} entangles the sensor and the reference, whereas independent encodings prepare them in a separable state. With the codewords $\ket{C_0}=\ket{00}$ and $\ket{C_1}=\ket{11}$ of each spin and its noiseless ancilla, the two initial states are
\begin{equation}
	\ket{\psi_{\mathrm{joint}}}=\frac{\ket{C_0C_1}+\ket{C_1C_0}}{\sqrt2},\qquad
	\ket{\psi_{\mathrm{ind}}}=\frac{\ket{C_0}+\ket{C_1}}{\sqrt2}\otimes\frac{\ket{C_0}+\ket{C_1}}{\sqrt2}.
	\label{eq:reference_encodings}
\end{equation}
To quantify the gain from the joint encoding, we start from $\ket{\psi_{\mathrm{ind}}}$, run the single-sensor protocol of Sec.~\ref{sec:toybridge} on each spin for the same duration $t$, and measure each logical qubit in its count-dependent basis. Conditional on the counts, the phases of the two logical qubits and their gradients with respect to $(\omega,\phi)$ are
\begin{equation}
	\Phi_S=\omega t-2N_S\phi,\qquad \Phi_R=-2N_R\phi,\qquad
	\bm g_S=(t,-2N_S)^{\mathsf T},\qquad \bm g_R=(0,-2N_R)^{\mathsf T}.
	\label{eq:reference_independent_phases}
\end{equation}
The two logical qubits are independent, so their CFIMs of Eq.~\eqref{eq:toy-CFIM} add. With the Poisson means $\mu_j=\kappa_jt$,
\begin{equation}
	\JC=\EE\bigl[\bm g_S\bm g_S^{\mathsf T}+\bm g_R\bm g_R^{\mathsf T}\bigr]
	=\EE\begin{pmatrix}t^2&-2tN_S\\-2tN_S&4(N_S^2+N_R^2)\end{pmatrix}
	=\begin{pmatrix}
		t^2 & -2t\mu_S\\
		-2t\mu_S & 4(\mu_S^2+\mu_R^2+\mu_S+\mu_R)
	\end{pmatrix}.
	\label{eq:reference_unequal_rates_cfim}
\end{equation}
This CFIM equals the QFIM of the output, including the recorded counts, so no other readout of these encodings does better. Unlike in Eq.~\eqref{eq:reference_spin_cfim}, the off-diagonal entries do not vanish, and eliminating $\phi$ through the Schur complement gives the effective information about $\omega$, normalized by $t^2$,
\begin{equation}
	\frac{1}{t^2[\JC^{-1}]_{\omega\omega}}
	=\frac{\mu_R^2+\mu_S+\mu_R}{\mu_S^2+\mu_R^2+\mu_S+\mu_R}
	\;\xrightarrow{t\to\infty}\;\frac{r^2}{1+r^2}<\min\{1,r^2\}.
	\label{eq:reference_unequal_rates_information}
\end{equation}
The quadratic coefficient $r^2/(1+r^2)$ of independent encodings is thus below the optimum for every $r$, and at equal rates it is only half that for known $\phi$.

\paragraph{Correlated noise.}\leavevmode\label{sec:reference_correlated}
So far, the two spins were flipped independently. If a common fluctuating field flips both, the noise is correlated and is described by the single Lindblad operator
\begin{equation}
	E=\sqrt{\kappa_S}\,\sigma_\phi^{(S)}+s\sqrt{\kappa_R}\,\sigma_\phi^{(R)},\qquad s=\pm1,
	\label{eq:reference_correlated_noise}
\end{equation}
where $s$ is the sign of the relative coupling of the two spins to the field. Theorem~\ref{thm:reference_restoration} of the main text assumes that each Lindblad operator acts on one system, so we apply Theorem~\ref{thm:joint_scaling} to the joint system and then check that the encodings above still work. Since $E$ is Hermitian, its square, the Lindblad span, and the coherent contribution of $\phi$ in Eq.~\eqref{eq:physical_noise_response} are
\begin{equation}
	\begin{aligned}
		E^2&=(\kappa_S+\kappa_R)I+2s\sqrt{\kappa_S\kappa_R}\,\sigma_\phi^{(S)}\sigma_\phi^{(R)},&\qquad
		\mathcal S_{SR}&=\operatorname{span}\bigl\{I,\ E,\ \sigma_\phi^{(S)}\sigma_\phi^{(R)}\bigr\},\\
		C_\phi&=\frac i2[E,\partial_\phi E]=-\kappa_S\sigma_z^{(S)}-\kappa_R\sigma_z^{(R)}.
	\end{aligned}
	\label{eq:correlated_noise_span}
\end{equation}
The cross terms cancel in $C_\phi$ because operators on different spins commute, so $C_\phi$ is as for uncorrelated noise. By Eq.~\eqref{eq:dav} of the main text, the distance of the joint system is
\begin{equation}
	d_\omega^{\mathrm{joint}}=\min_{v_\phi,\,S=S^\dagger\in\mathcal S_{SR}}\norm{\partial_\omega H+v_\phi C_\phi-S}_\infty,\qquad
	\partial_\omega H+v_\phi C_\phi=\frac{(1-u\kappa_S)\sigma_z^{(S)}-u\kappa_R\sigma_z^{(R)}}{2},
	\label{eq:correlated_noise_operator}
\end{equation}
with $u=2v_\phi$. This operator is diagonal in the product $\sigma_z$ basis, whereas $E$ and $\sigma_\phi^{(S)}\sigma_\phi^{(R)}$ have no diagonal entries. As in Sec.~\ref{sec:noise_rotation_example}, discarding off-diagonal entries does not increase the operator norm, so $S$ can be taken proportional to $I$, and $S=0$ is optimal because the eigenvalues $[\pm(1-u\kappa_S)\pm u\kappa_R]/2$ are symmetric about zero. Hence
\begin{equation}
	2d_\omega^{\mathrm{joint}}=\min_{u\in\mathbb R}\bigl(|1-u\kappa_S|+|u|\kappa_R\bigr)
	=\min\{1,r\},\qquad
	4(d_\omega^{\mathrm{joint}})^2=\min\{1,r^2\},
	\label{eq:reference_spin_distance}
\end{equation}
and, as stated in the main text, Theorem~\ref{thm:joint_scaling} gives the same optimal quadratic coefficient as for uncorrelated noise.

It remains to check that the joint code of Eq.~\eqref{eq:reference_spin_code} attains this coefficient. Each term of $E$ flips one spin and $\sigma_\phi^{(S)}\sigma_\phi^{(R)}$ flips both, which takes both codewords out of the code space, so the encoding $V_{SR}=\ket{C_0^{SR}}\bra0+\ket{C_1^{SR}}\bra1$ satisfies the Knill--Laflamme conditions~\eqref{eq:code_kl_explicit},
\begin{equation}
	V_{SR}^\dagger EV_{SR}=0,\qquad V_{SR}^\dagger E^\dagger EV_{SR}=(\kappa_S+\kappa_R)I_C,
	\label{eq:correlated_noise_kl}
\end{equation}
and jumps occur at the rate $\kappa_S+\kappa_R$. A jump acts on a code state $\ket\psi$ as
\begin{equation}
	E\ket\psi=\sqrt{\kappa_S}\,\sigma_\phi^{(S)}\ket\psi+s\sqrt{\kappa_R}\,\sigma_\phi^{(R)}\ket\psi,
	\label{eq:correlated_noise_jump}
\end{equation}
a superposition of flips on $S$ and on $R$ with different syndromes, of which the syndrome measurement selects the flip on spin $j$ with probability $\kappa_j/(\kappa_S+\kappa_R)$. The corrected flips on $S$ and $R$ are thus independent Poisson processes with rates $\kappa_S$ and $\kappa_R$, as for uncorrelated noise. This remains true after one spin is swapped into a further noiseless ancilla, since its later flips leave the logical state unchanged. Equations~\eqref{eq:reference_spin_state}--\eqref{eq:reference_spin_variances} and the encodings with the exposure times in Eq.~\eqref{eq:reference_exposure_times} or with $M$ flips per spin therefore remain valid.

\section{Precision of a single sensor with noise modulation}\label{sec:noise_modulation}
This section expands the remark in the main text that a single sensor can act as its own reference. Without a reference system, the sensor switches between two noise strengths and interchanges its codewords at the switch, so the two stages play the roles of the sensor and the reference with opposite codewords. Section~\ref{sec:modulation_protocol} constructs a protocol in which the phases due to the noise parameters cancel between the two stages, Sec.~\ref{sec:modulation_optimality} shows that no switching sequence attains a larger quadratic coefficient, and Sec.~\ref{sec:modulation_spin} applies the protocol to the spin model.

\subsection{Protocol with two noise strengths}\label{sec:modulation_protocol}
We consider a single unknown Hamiltonian parameter $\omega$ and unknown noise parameters $\eta_b$. A classical control, for example engineered noise, sets the noise strength $\gamma$ without changing $H$. It rescales the Lindblad operators and the coherent contributions in Eq.~\eqref{eq:physical_noise_response} as
\begin{equation}
	E_k\to\sqrt\gamma\,E_k,\qquad C_{\eta_b}\to\gamma\,C_{\eta_b},\qquad 0<\gamma_-\leq\gamma\leq\gamma_+,
	\label{eq:modulation_scaling}
\end{equation}
and leaves the Lindblad span $\mathcal S$ unchanged. Let $\omega$ fail the criterion~\eqref{eq:theorem1_criterion} for Heisenberg scaling at every fixed strength, that is,
\begin{equation}
	\partial_\omega H+\sum_b\lambda_bC_{\eta_b}\in\mathcal S
	\label{eq:modulation_condition}
\end{equation}
for real $\lambda_b$, with $C_{\eta_b}$ evaluated at $\gamma=1$ here and below. For known noise, the distance of $\partial_\omega H$ from the Lindblad span is
\begin{equation}
	d^{\mathrm{known}}=\min_{S=S^\dagger\in\mathcal S}\norm{\partial_\omega H-S}_\infty,
	\label{eq:modulation_known_distance}
\end{equation}
which is Eq.~\eqref{eq:reference_known_distance} for a single Hamiltonian parameter.

For $d^{\mathrm{known}}>0$, let $Y$ be an optimizer of the dual problem~\eqref{eq:dual} with $\mathcal N=\mathcal S$ and $X=\partial_\omega H$, and build the codewords $\ket{C_0}$ and $\ket{C_1}$ from $Y$ as in Sec.~\ref{sec:general_scaling}. Since $\mathcal S$ does not change, the code satisfies the Knill--Laflamme conditions~\eqref{eq:code_kl_explicit} at every strength. The errors $M_k=(I-P)E_kP$ and the recovery operators of Sec.~\ref{sec:general_scaling} scale as
\begin{equation}
	M_k\to\sqrt\gamma\,M_k,\qquad \chi_k\to\gamma\chi_k,\qquad R_k=\frac{PM_k^\dagger}{\sqrt{\chi_k}}\to R_k,
	\label{eq:modulation_recovery}
\end{equation}
so the recovery is also the same at every strength. With the control Hamiltonian $-H(\tilde{\bm\omega})$, Eqs.~\eqref{eq:recoverderivative} and~\eqref{eq:optimal_code_compression} give the first-order logical generator at strength $\gamma$,
\begin{equation}
	\delta\mathcal L^C_\gamma(\rho_C)=-i\Bigl(d^{\mathrm{known}}\delta\omega+\gamma\sum_by_{\eta_b}\delta\eta_b\Bigr)[\sigma_z^C,\rho_C],\qquad y_{\eta_b}=\Tr(YC_{\eta_b}).
	\label{eq:modulation_generator}
\end{equation}
The noise parameters thus enter through a phase that scales with $\gamma$, whereas the signal phase does not.

The protocol has two stages. The strength is set to $\gamma_-$ for a time $t_-$, the codewords are interchanged, and the strength is set to $\gamma_+$ for a time $t_+$, with
\begin{equation}
	t_-=\frac{\gamma_+t}{\gamma_++\gamma_-},\qquad t_+=\frac{\gamma_-t}{\gamma_++\gamma_-},\qquad \gamma_-t_-=\gamma_+t_+.
	\label{eq:modulation_times}
\end{equation}
The interchange reverses the sign of the relative phase accumulated afterward. By Eq.~\eqref{eq:modulation_generator}, the relative phase of the two codewords after both stages is therefore
\begin{equation}
	2\Bigl(d^{\mathrm{known}}\delta\omega+\gamma_-\sum_by_{\eta_b}\delta\eta_b\Bigr)t_-
	-2\Bigl(d^{\mathrm{known}}\delta\omega+\gamma_+\sum_by_{\eta_b}\delta\eta_b\Bigr)t_+
	=2d^{\mathrm{known}}\delta\omega\,(t_--t_+),
	\label{eq:modulation_phase}
\end{equation}
where the phases due to all $\eta_b$ cancel between the two stages because $\gamma_-t_-=\gamma_+t_+$. The readout of Eq.~\eqref{eq:hs_probabilities}, with $d_it\,\delta\theta_i$ replaced by $d^{\mathrm{known}}(t_--t_+)\delta\omega$, then gives the quadratic coefficient
\begin{equation}
	4(d^{\mathrm{known}})^2\frac{(t_--t_+)^2}{t^2}=4(d^{\mathrm{known}})^2\Bigl(\frac{\gamma_+-\gamma_-}{\gamma_++\gamma_-}\Bigr)^2.
	\label{eq:modulation_coefficient}
\end{equation}
This is the coefficient $4(d^{\mathrm{known}})^2$ for known noise, reduced by a factor that approaches one when $\gamma_+\gg\gamma_-$.

\subsection{Precision bound under arbitrary noise modulation}\label{sec:modulation_optimality}
We now show that the coefficient in Eq.~\eqref{eq:modulation_coefficient} is optimal. No protocol of the general sequential scheme exceeds it when the strength in each interval lies in $[\gamma_-,\gamma_+]$, even if the strength is chosen by feedback. To this end, we apply Lemma~\ref{lem:purification_bound} along the direction $\bm v$ whose coherent part vanishes at the mean strength $\bar\gamma$,
\begin{equation}
	v_\omega=1,\qquad v_{\eta_b}=\frac{\lambda_b}{\bar\gamma},\qquad \bar\gamma=\frac{\gamma_-+\gamma_+}{2}.
	\label{eq:modulation_direction}
\end{equation}
At strength $\gamma$, the operator $\mathcal R_i=\sum_\mu v_\mu\hat H_\mu-S_i$ of the lemma contains
\begin{equation}
	\partial_\omega H+\gamma\sum_bv_{\eta_b}C_{\eta_b}
	=\Bigl(\frac{\gamma}{\bar\gamma}-1\Bigr)\sum_b\lambda_bC_{\eta_b}+\Bigl(\partial_\omega H+\sum_b\lambda_bC_{\eta_b}\Bigr),
	\label{eq:modulation_split}
\end{equation}
from which $g_i$ subtracts the Hermitian element $S_i\in\mathcal S$ of Eq.~\eqref{eq:joint_span_sm}. By Eq.~\eqref{eq:modulation_condition}, the last term lies in $\mathcal S$, and $\sum_b\lambda_bC_{\eta_b}$ has the same distance $d^{\mathrm{known}}$ from $\mathcal S$ as $\partial_\omega H$. Choosing $g_i$ for the strength in each interval therefore gives
\begin{equation}
	\norm{\mathcal R_i}_\infty=\Bigl|\frac{\gamma}{\bar\gamma}-1\Bigr|\,d^{\mathrm{known}}\leq\frac{\gamma_+-\gamma_-}{\gamma_++\gamma_-}\,d^{\mathrm{known}}.
	\label{eq:modulation_residual}
\end{equation}
The coefficients of $g_i$ that realize a fixed element of $\mathcal S$ scale as $\gamma^{-1/2}$ or $\gamma^{-1}$, because the Kraus operators contain $\sqrt\gamma E_k$. They remain bounded for $\gamma\geq\gamma_->0$, and so does $\norm{A_i}_\infty$.

It remains to extend the bound of Lemma~\ref{lem:purification_bound} to a strength that changes during the protocol. Its derivation in Sec.~\ref{sec:joint_explicit_gauges} applies interval by interval. Equation~\eqref{eq:purified_derivative_finite_sm} holds when the channel changes between intervals, and also when the strength is chosen by feedback, since a channel conditioned on a classical register decomposes as in Eq.~\eqref{eq:channel_derivative_split_sm} for each value of the register. The integral bound~\eqref{eq:purified_derivative_integral_sm}, with $\mathcal R_i$ and $A_i$ evaluated at the strength in use, then gives
\begin{equation}
	\norm{\partial_s\Psi_t}\leq t\,d^{\mathrm{known}}\,\frac{\gamma_+-\gamma_-}{\gamma_++\gamma_-}+O(\sqrt t).
	\label{eq:modulation_derivative_bound}
\end{equation}
Together with $J_{Q,s}(t)\leq4\norm{\partial_s\Psi_t}^2$ and Eq.~\eqref{eq:optimal_variance_information}, with the optimum now taken also over the strength in each interval, this gives
\begin{equation}
	\limsup_{t\to\infty}\frac{1}{\nu t^2\Var_{\mathrm{opt}}(\hat\omega)}\leq4(d^{\mathrm{known}})^2\Bigl(\frac{\gamma_+-\gamma_-}{\gamma_++\gamma_-}\Bigr)^2,
	\label{eq:modulation_upper}
\end{equation}
which the protocol of Sec.~\ref{sec:modulation_protocol} attains.

\subsection{Noise modulation in the spin model}\label{sec:modulation_spin}
We now apply the protocol to the spin model of Sec.~\ref{sec:finite}, with the flip rate switched between $\gamma_-\kappa$ and $\gamma_+\kappa$. For $H=\omega\sigma_z/2$ and $E=\sqrt{\gamma\kappa}\,\sigma_\phi$, the condition~\eqref{eq:modulation_condition} and the distance~\eqref{eq:modulation_known_distance} read
\begin{equation}
	\partial_\omega H+\lambda_\phi C_\phi=0,\qquad C_\phi=-\kappa\sigma_z,\qquad \lambda_\phi=\frac{1}{2\kappa},\qquad d^{\mathrm{known}}=\frac12.
	\label{eq:modulation_spin_model}
\end{equation}
The protocol uses the code $\ket{C_0}=\ket{00}$, $\ket{C_1}=\ket{11}$ of Eq.~\eqref{eq:toystate} of the main text and the times of Eq.~\eqref{eq:modulation_times}. Conditional on the numbers $N_-$ and $N_+$ of corrected flips in the two stages, the logical phase is
\begin{equation}
	\Phi=\omega(t_--t_+)-2(N_--N_+)\phi.
	\label{eq:modulation_spin_phase}
\end{equation}
The counts are independent Poisson variables with equal means $\gamma_-\kappa t_-=\gamma_+\kappa t_+$. As in Eq.~\eqref{eq:toyqfim} of the main text, the QFIM per interrogation is therefore
\begin{equation}
	\EE(N_--N_+)=0,\qquad \EE(N_--N_+)^2=2\gamma_-\kappa t_-,\qquad
	\JQ=\begin{pmatrix}(t_--t_+)^2&0\\0&8\gamma_-\kappa t_-\end{pmatrix}.
	\label{eq:modulation_spin_qfim}
\end{equation}
Its $\omega\omega$ entry gives the quadratic coefficient $(t_--t_+)^2/t^2=[(\gamma_+-\gamma_-)/(\gamma_++\gamma_-)]^2$, in agreement with Eq.~\eqref{eq:modulation_coefficient} for $d^{\mathrm{known}}=1/2$, and the count-dependent readout of Sec.~\ref{sec:toybridge} attains this QFIM.

The times in Eq.~\eqref{eq:modulation_times} require the ratio $\gamma_+/\gamma_-$. If it is not known in advance, we fix the flip counts instead, as in Eq.~\eqref{eq:reference_fixed_counts}. Each stage is stopped after the same number $M$ of corrected flips, so that $\phi$ drops out of the phase,
\begin{equation}
	N_-=N_+=M,\qquad \Phi=\omega(T_--T_+),
	\label{eq:modulation_fixed_counts}
\end{equation}
where the recorded stage durations $T_\pm$ replace $t_\pm$ and set the readout basis. We set $M$ below the expected count of each stage by the margin $m_t$ of Eq.~\eqref{eq:reference_flip_number},
\begin{equation}
	M=\gamma_-\kappa t_--m_t=\gamma_+\kappa t_+-m_t,
	\label{eq:modulation_flip_number}
\end{equation}
so that both stages fit within the time $t$ with probability approaching one. The QFI for $\omega$ is $\EE(T_--T_+)^2$, and the means $M/(\gamma_\pm\kappa)$ and variances $M/(\gamma_\pm\kappa)^2$ of $T_\pm$ give
\begin{equation}
	\lim_{t\to\infty}\frac{\EE(T_--T_+)^2}{t^2}=\Bigl(\frac{\gamma_+-\gamma_-}{\gamma_++\gamma_-}\Bigr)^2,
	\label{eq:modulation_fixed_counts_coefficient}
\end{equation}
the same coefficient. The two flip rates enter only through $M$ and can be estimated from the flip times of earlier interrogations, which are independent of $\omega$ and $\phi$.

\FloatBarrier
\putbib[\jobname Notes,reference]
\makeatletter
\immediate\write\@bibunitaux{\string\citation{REVTEX42Control,apsrev42Control}}
\makeatother
\end{bibunit}

\makeatletter
\immediate\write\@auxout{\string\gdef\string\@extra@binfo{}}
\makeatother
\end{document}